\documentclass{aa}

\usepackage{graphicx}
\usepackage{txfonts}
\usepackage{placeins}
\usepackage{mathrsfs}

\begin{document}

\title{Methodological refinement of the\\ submillimeter galaxy magnification bias \\ II. Cosmological analysis with tomography}
\titlerunning{Methodological refinement of the submillimeter galaxy magnification bias. Paper II.}
\authorrunning{Bonavera L. et al.}

   \author{L. Bonavera\inst{1,2}, M. M. Cueli\inst{3,4}, J. Gonz{\'a}lez-Nuevo\inst{1,2}, J. M. Casas\inst{1,2}, D. Crespo\inst{1,2}}

  \institute{
  $^1$Departamento de Fisica, Universidad de Oviedo, C. Federico Garcia Lorca 18, 33007 Oviedo, Spain\\
    $^2$Instituto Universitario de Ciencias y Tecnologías Espaciales de Asturias (ICTEA), C. Independencia 13, 33004 Oviedo, Spain\\
  $^3$SISSA, Via Bonomea 265, 34136 Trieste, Italy\\
    $^4$IFPU - Institute for fundamental physics of the Universe, Via Beirut 2, 34014 Trieste, Italy}
\date{}
\abstract
{This work focuses on the submillimeter galaxy magnification bias, specifically in the tomographic scenario. It builds upon previous works, while utilising updated data to refine the methodology employed in constraining the free parameters of the halo occupation distribution model and cosmological parameters within a flat $\Lambda$CDM model.}
{This work aims to optimise CPU time and explore strategies for analysing different redshift bins, while maintaining measurement precision. Additionally, it seeks to examine the impact of excluding the GAMA15 field, one of the H-ATLAS fields that was found to have an anomalous strong cross-correlation signal, and increasing the number of redshift bins on the results.}
{The study uses a tomographic approach, dividing the redshift range into a different number of bins and analysing cross-correlation measurements between H-ATLAS submillimeter galaxies with photometric redshifts in the range $1.2<z<4.0$ and foreground GAMA galaxies with spectroscopic redshifts in the range $0.01<z<0.9$. Interpreting the weak lensing signal within the halo model formalism and carrying out a Markov chain Monte Carlo algorithm, we obtain the posterior distribution of both halo occupation distribution and cosmological parameters within a flat $\Lambda$CDM model. Comparative analyses are conducted between different scenarios, including different combinations of redshift bins and the inclusion or exclusion of the GAMA15 field.}
{The mean-redshift approximation employed in the "base case" yields results that are in good agreement with the more computationally intensive "full model" case. Marginalised posterior distributions confirm a systematic increase in the minimum mass of the lenses with increasing redshift. The inferred cosmological parameters show narrower posterior distributions compared to previous studies on the same topic, indicating reduced measurement uncertainties. Excluding the GAMA15 field  demonstrates a reduction in the cross-correlation signal, particularly in two of the redshift bins, suggesting a sample variance within the large-scale structure along the line of sight. Moreover, extending the redshift range improves the robustness against the sample variance issue and produces similar, but tighter constraints compared to excluding the GAMA15 field.}
{The study emphasises the importance of considering sample variance and redshift binning in tomographic analyses. Increasing the number of independent fields and the number of redshift bins can minimise both the spatial and redshift sample variance, resulting in more robust measurements. The adoption of additional wide area field observed by \textit{Herschel} and of updated foreground catalogues, such as the Dark Energy Survey or the future \textit{Euclid} mission, is important for implementing these approaches effectively.}


\maketitle

\section{Introduction}

The excess of high-redshift sources near low-redshift mass structures, known as magnification bias, has the potential to serve as a valuable tool for investigating the large-scale structure of the universe and the distribution of dark matter when applied to background submillimeter galaxies \citep[][and references therein]{BON22}. This can be done by providing independent constraints on cosmological parameters, whereby measurements of the magnification bias can complement other cosmological probes. The phenomenon caused by gravitational lensing, namely, the distribution of matter in the Universe (including galaxies and clusters of galaxies) bending light as it travels through space, which results in the apparent brightness and size of distant objects being magnified or distorted \citep[e.g.][]{SCH92}. The magnification bias effect manifests as a significant cross-correlation function between two source samples with different redshift ranges, indicating that the large-scale structure traced by the foreground sources is amplifying the background ones. This phenomenon has been observed in various situations, such as the correlation between galaxies and quasars \citep[e.g.][]{SCR05, MEN10}, Herschel sources and Lyman-break galaxies \citep[][]{HIL13}, and between the cosmic microwave background (CMB) and other sources \citep[][]{BIA15, BIA16}. In particular, submillimeter galaxies (SMGs) are very suitable as background sources due to their exceptional properties, which include a steep luminosity function, high redshifts, and faint optical band emission \citep[see e.g.][]{GON12, BLA96, Neg07, NEG10, NEG17, GON17, BUS12, BUS13, FU12, WAR13, CAL14, NAY16, Bak20}. 
Previous studies have demonstrated the magnification bias effect on SMGs and have measured it to derive both astrophysical and cosmological information \citep[][]{GON17, BON20, CUE21, GON21}. Additionally, the ability to divide the foreground sample into distinct redshift bins allows for a more detailed tomographic analysis \citep[][]{BON21, CUE22}.

Indeed, similarly to shear studies \citep{BA00,VW00,WI00,RHO01}, cosmological parameters could benefit from a tomographic analysis, especially by allowing for better estimation of the halo occupation distribution (HOD) parameters. Slight variations in these parameters are expected when explored in distinct redshift bins. \cite{GON17} demonstrated this by performing tomographic studies on the HOD parameters, separating the foreground sample into four redshift bins: $0.1-0.2$, $0.2-0.3$, $0.3-0.5$, and $0.5-0.8$. 
It was discovered that the values of $M_{min}$, the mean minimum mass for a halo to host a (central) galaxy, exhibited a clear evolution, increasing with redshift as predicted by theoretical calculations. 
Furthermore, in \cite{GON21}, a simplified tomographic analysis is conducted using a background sample of Herschel Astrophysical Terahertz Large Area Survey (H-ATLAS) galaxies with photometric redshifts of $z > 1.2$, and two independent foreground samples: Galaxy And Mass Assembly (GAMA) survey galaxies with spectroscopic redshifts and Sloan Digital Sky Survey (SDSS) galaxies with photometric redshifts, with $0.2 < z < 0.8$. The study obtained constraints on the cosmological parameters $\Omega_M$ ($0.50^{+0.14}_{-0.20}$ ) and $\sigma_8$ ($0.75^{+0.07}_{-0.10}$) as a 68\% confidence interval (C.I.).

\cite{BON21} investigated the possibility of improving constraints on cosmological parameters $\Omega_M$, $\sigma_8$, and $H_0$ by using unbiased measurements of the H-ATLAS/GAMA cross-correlation function, as concluded by the methodological analysis of \cite{GON21} and a tomographic analysis. These authors found a trend towards higher values of $M_{min}$ at higher redshifts.
They also explored different values of the dark energy equation of state parameter, $\omega$. Their results showed maximum posterior values, at a 68\% C.I., of $0.26^{+0.15}_{-0.09}$ for $\Omega_M$, and $0.87^{+0.13}_{-0.12}$ for $\sigma_8$ in the $\Lambda$CDM model. 
The study also examined a more general $\omega_0$CDM model, allowing for values of $\omega_0 \ne -1$, and found that the constraints on $\Omega_M$ and $\sigma_8$ were similar. However, the maximum posterior value for $\omega_0$ was found to be at $-1.00^{+0.53}_{-0.56}$, with a 68\% C.I. 
In the $\omega_0$$\omega_a$CDM model, where a possible z dependence of the $\omega$ parameter is allowed, the results showed a maximum posterior value of $-1.09^{+0.43}_{-0.63}$ with a 68\% C.I. for $\omega_0$ and a maximum posterior value of $-0.19^{+1.29}_{-1.69}$ with a 68\% C.I. for $\omega_a$.
The tomographic analysis improved the constraints on the $\sigma_8$ - $\Omega_M$ plane and demonstrated that the magnification bias does not to show the degeneracy found with cosmic shear measurements. Finally, the study revealed a trend of higher $\omega_0$ values for lower $H_0$ values concerning dark energy.

This work is the companion paper to \citet{Cue23}, also referred as Paper I, that focuses on deriving cosmological and astrophysical constraints using a single broad foreground redshift bin. Additionally, the theoretical modelling of the signal is reexamined with respect to the work of \cite{BON21} and \cite{GON21} to assess the significance of the logarithmic slope of background galaxy number counts and a specific numerical correction.

The goal of this second work is to assess the performance of this observable in a tomographic setup after the methodological refinement described in Paper I.
Given the numerous parameters to be jointly analysed in tomography and the numerical complexity of the theoretical model, the computational time required for this task is large. The aim is to investigate the possibility to optimise CPU time and strategies, such as selecting different redshift bins, to improve the stability and accuracy of the results. This work suggests various strategies to streamline CPU time without compromising the precision of the measurements. It also delves into the selection of different redshift bins and their effect on the ultimate outcome.

In this manuscript, we have structured the content as follows. Section \ref{sec:Methodology} summarises the methodology of this work (data, theoretical framework, and parameter estimation). The results obtained for different combinations of redshift bins and ranges are summarised in Sect. \ref{sec:results}. Finally, in Sect. \ref{sec:concl}, we draw conclusions based on our findings.

\begin{figure}[ht]
\includegraphics[width=0.5\textwidth]{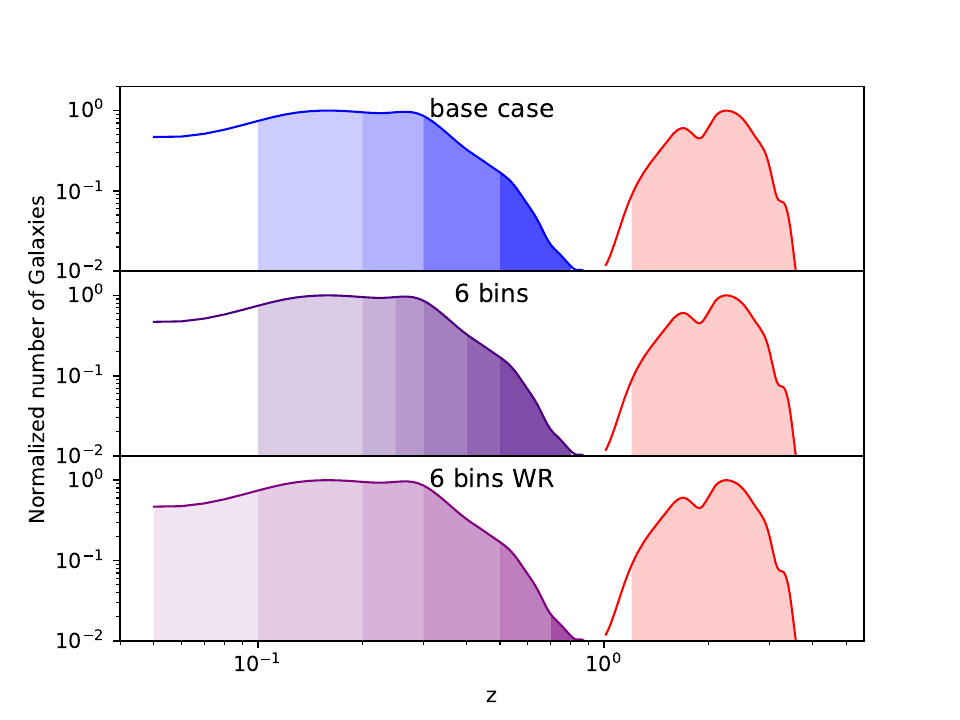}
 \caption{Redshift distributions of the background SMGs (in red) and foreground lenses with the adopted redshift bins indicated by different shades of the same colour. From top to bottom, the "base case" is shown in blue, the "six bins" case in the $0.1<z<0.8$ range in indigo and the "six bins-WR" case in the wider $0.01<z<0.9$ range in purple, respectively. 
 }
 \label{fig:z_distro}
\end{figure}
 
\section{Methodology}
\label{sec:Methodology}
\subsection{Data and redshift subsample}
\label{subsec:data}

The details of the data used for computing the cross-correlation signal are discussed in detail in Paper I \citep{Cue23}. The background sample primarily consists of sources with photometric redshifts, while the foreground sample comprises sources with spectroscopic redshifts. The analysis involves dividing the foreground sample into redshift bins, necessitating the use of spectroscopic measurements for this particular sample in order to avoid overlapping between bins. Photometric errors for the background sample are considered to smooth the background redshift distribution required for the theoretical model but do not present any concerns in this context.

Specifically, the background sample is selected from the Herschel Astrophysical Terahertz Large Area Survey \citep[H-ATLAS;][]{EAL10} sources in the Galaxy And Mass Assembly (GAMA) fields (G09, G12 and G15) which covers approximately $\sim147$ deg$^2$ of common area and part of the south Galactic pole (SGP), totalling around 60 deg$^2$. A photometric redshift selection of $1.2<z<4.0$ was applied to avoid redshift overlap with the lenses. As a result, we obtained a sample of approximately 66,000 sources with an average photometric redshift of 2.20. The redshift distribution of the background sources used in the study is estimated as $p(z|W)$ for galaxies selected by a window function with a redshift range of $1.2<z<4.0$. The distribution takes into account the effect of random errors in photometric redshifts, following the methodology described in \citet{GON17}.

The foreground or lens sample used in this work is referred to as the "z$_{spec}$ sample" in \citet{GON21}, consisting of approximately 150,000 galaxies with spectroscopic redshifts in the range of 0.2 < z$_{spec}$ < 0.8. These redshift values were obtained from the GAMA II spectroscopic survey. The total common area used in this study was approximately 207 deg$^2$. To facilitate our analysis, we divided the foreground sample into different sub-samples based on their redshift. Following the previous results from \citet{BON21}, we adopted the "base case" which involved dividing the entire redshift range into four bins: $0.1<z<0.2$ (bin 1), $0.2<z<0.3$ (bin 2), $0.3<z<0.5$ (bin 3), and $0.5<z<0.8$ (bin 4). Since the spectroscopic redshifts for the foreground sources have negligible errors, there is no overlap in the redshift distribution among bins, and no source is misplaced in the wrong bin, although this detail is not critical for cross-correlation analyses with a non-overlapping background sample.

To investigate the possible dependence on the number of redshift bins but maintaining the same redshfit range, we conducted a joint analysis using six redshift bins, referred to as the "six bins" scenario. In this scenario, the entire range was divided into $0.1<z<0.2$ (bin 1), $0.2<z<0.25$( bin 2a), $0.25<z<0.3$ (bin 2b), $0.3<z<0.4$ (bin 3a), $0.4<z<0.5$ (bin 3b), and $0.5<z<0.8$ (bin 4). Finally, we also conducted a joint analysis using six bins of redshift, but with a wider range, denoted as the "six bins-WR" case. In this case, the range extended from 0.01 to 0.9, with the following bin divisions: $0.01<z<0.1$ (bin 0), $0.1<z<0.2$ (bin 1), $0.2<z<0.3$ (bin 2), $0.3<z<0.5$ (bin 3), $0.5<z<0.7$ (bin 4), and $0.7<z<0.9$ (bin 5). 

The redshift distributions of the foreground and background sources are shown in Fig. \ref{fig:z_distro}. The bins of redshift adopted in the foreground sample are highlighted with varying shades of the same colour. The three different sets of bins, consisting of four bins, six bins and six bins in a wider range, are displayed from top to bottom in blue, indigo, and purple, respectively.

\subsection{Measurements}
\label{subsec:method}
In previous works, the available area is divided in small regular regions, mini-tiles, due to its computational efficiency and accuracy in measuring the angular clustering of galaxies \citep[e.g.][]{GON17, BON19, BON20, BON21, CUE21, GON21,BON22, CUE22}. However, it requires an integral constraint (IC) correction to obtain an unbiased estimate of the true function, particularly crucial for constraining cosmological parameters on the largest angular scales. Moreover, using mini-tiles poses a practical challenge as the vertices of each of them must be manually defined, which is poorly scalable when expanding the study to other wide field surveys observed by Herschel that may include  hundreds of mini-tiles. 

To overcome these challenges, we have adopted a statistically rigorous approach that leverages the full field area and combines the number of different foreground-background pairs from each field into a single estimation. This new approach, described in Paper I \citep[see also][for a more detailed discussion]{GON23}, reduces statistical uncertainty and effectively utilises all available information in the data, as opposed to separately measuring cross-correlation functions for individual fields. The cross-correlation function is then measured using the modified version of the \cite{LAN93} estimator by \cite{HER01},
\begin{equation}
\hat{w}_{\text{fb}}(\theta)=\frac{\rm{D}_f\rm{D}_b(\theta)-\rm{D}_f\rm{R}_b(\theta)-\rm{D}_b\rm{R}_f(\theta)+\rm{R}_f\rm{R}_b(\theta)}{\rm{R}_f\rm{R}_b(\theta)}.
\end{equation}
For the angular separation, $\theta$, the values of $\rm{D}_f\rm{D}_b$, $\rm{D}_f\rm{R}_b$, $\rm{D}_b\rm{R}_f$, and $\rm{R}_f\rm{R}_b$ represent the normalised foreground-background, foreground-random, background-random, and random-random pair counts coming from the whole available area. The background random samples are generated taking into account the surface density variation due to the \textit{Herschel} scanning strategy.

Furthermore, to estimate the covariance matrix, we employed a Bootstrap method by dividing each field into at least five patches, ensuring a sufficient number of measurements. 
Instead of using the already mentioned mini-tiles, we have implemented a more adaptable procedure for defining subregions that retains similar characteristics to the previous approach but can be easily scaled for future studies.
As explained in Paper I, these patches were created using a k-mean clustering algorithm, ensuring that the spatial dependence structure is retained during resampling. 
For defining the subregions, we took inspiration from TreeCorr, a widely used software package for gauging galaxy clustering \citep{TreeCorr}. TreeCorr employs a k-means clustering algorithm to divide the data into subregions referred to as 'patches,' which are similar in size and shape to our mini-tiles. Specifically, we utilise the k-means algorithm available in the SciPy library. The objective of the algorithm is to minimise the sum of squared distances between data points and their assigned centroid. To determine the optimal number of clusters (i.e. patches), we enforced a minimum area requirement for each cluster. Additionally, we introduced an extra step by repeating the procedure ten times with different random initial centroids and selecting the run that demonstrated the most consistent number of data points across the various clusters.

This new methodology has demonstrated enhanced robustness and significantly reduced uncertainties in the obtained measurements. 
In essence, the overall area was divided into 22 subregions and $N_b$=10000 Bootstrap samples were created with an oversampling factor of 3. Therefore, the covariance matrix is given by

\begin{equation}
    \text{Cov}(\theta_i,\theta_j)=\frac{1}{N_b-1}\sum_{k=1}^{N_b}\,\bigg[\hat{w}_k(\theta_i)-\bar{\hat{w}}(\theta_i)\bigg]\bigg[\hat{w}_k(\theta_j)-\bar{\hat{w}}(\theta_j)\bigg]\label{covariance},
\end{equation}

where $\hat{w}_k$ denotes the measured cross-correlation function from the $k^{\text{th}}$ Bootstrap sample and $\bar{\hat{w}}$ is the corresponding average value over all Bootstrap samples. It should be noted that in this work, no attempt of correction of the  super-sample covariance \citep[SSC, e.g.][]{LAC17} is made since our main source of uncertainty at the moment is cosmic variance. It is due to the non-negligible variations that we have among the spatially separated fields (G09, G12, G15, and SGP), which is studied and commented on  in Paper I.

\begin{figure}[ht]
\includegraphics[width=0.5\textwidth]{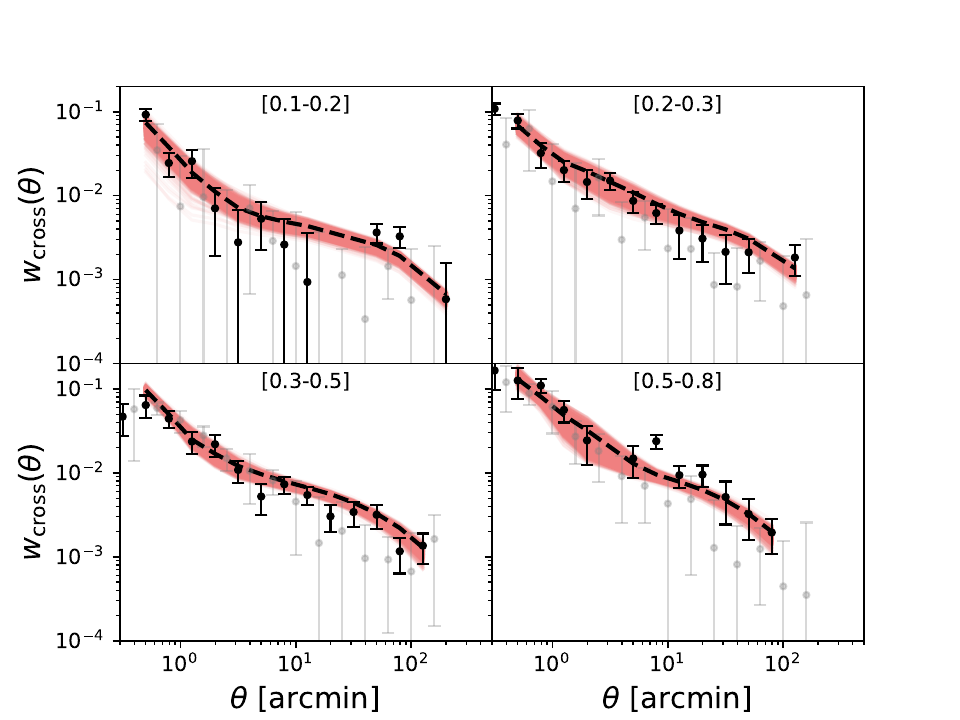}
 \caption{Cross-correlation data (black points) and the results obtained with the tomographic case: the posterior sampling of the cross-correlation function is shown in red and the best fit with the black dashed line. Bins 1 to 4 are shown from left to right and top to bottom. For comparison, the cross-correlation data by \cite{BON21} are shown in grey. 
 }
 \label{fig:sampled_tomo_zmed}
\end{figure}

In this work, we conduct a tomographic analysis by calculating the cross-correlation functions for each redshift bin defined in the previous section. The cross-correlation measurements, along with the outcomes of each case, are presented in Sect. \ref{sec:results}. Specifically, in Fig. \ref{fig:sampled_tomo_zmed}, we compare the cross-correlation measurements obtained by \citet{BON21} (light grey circles) with those derived using our new methodology (black circles), thereby confirming the improved results also in the context of tomography.

\subsection{Theoretical framework}
\label{subsec:framework}
The low redshift galaxy-mass correlation and the cross-correlation between foreground and background sources are connected through weak lensing. The mass density field traced by the foreground galaxy sample causes weak lensing and this affects the number counts of the background galaxy sample through magnification bias, as described in detail in Paper I. To compute the correlation between the foreground and background sources, we adopt the halo model formalism described in \cite{COO02} and the Limber and flat-sky approximations. The correlation can be evaluated using the following equation:
\begin{equation}
\label{eq:w_fb}
    \begin{split}
    w_{\text{fb}}(\theta)=2(\beta -1)\int^{\infty}_0 \frac{dz}{\chi^2(z)}\frac{dN_f}{dz}W^{lens}(z) \\
    \int_{0}^{\infty}\frac{ldl}{2\pi}P_{\text{g-m}}(l/\chi(z),z)J_0(l\theta) 
    \end{split}
,\end{equation}
where $W^{\text{lens}}(z)$ is defined as:
\begin{equation}
W^{\text{lens}}(z)=\frac{3}{2}\frac{H_0^2}{c^2}E^2(z)\int_z^{\infty} dz' \frac{\chi(z)\chi(z'-z)}{\chi(z')}\frac{dN_b}{dz'}
.\end{equation}
Here, $E(z)=\sqrt{\Omega_m(1+z)^3+\Omega_{\Lambda}}$, $dN_b/dz$, and $dN_f/dz$ are the unit-normalised background and foreground redshift distributions, $\chi(z)$ is the comoving distance to redshift, z, and $P_{\text{g-m}}$ is the galaxy-matter cross-power spectrum.  The logarithmic slope of the background sources number counts is denoted by $\beta$, and the details of our handling of such parameter in the analysis are described in Paper I. 

Due to the considerable number of integrals that need to be computed for each evaluation of the model, a mean-redshift approximation has been utilised. 
Such an approximation speeds up the computational time dramatically, by a factor of 10, enabling us to perform a variety of different tests in a reasonable time that would have otherwise been impossible due to our computer resources. Moreover, given the small redshift ranges used in tomography, we anticipate that this approximation will remain useful in the future, where improved parameter constraints are expected due to advancements in data, such as those from new surveys.

This approximation refers to the evaluation of the outermost integrals in Eq. \ref{eq:w_fb} not being carried out directly for the entire redshift distribution, but instead done through computation at the sample's mean redshift:
\begin{equation}
\label{eq:w_fb_approx}
    w_{\text{fb}}(\theta)\approx 2(\beta-1)\frac{W^{\text{lens}}(\bar{z})}{\chi^2(\bar{z})}\int_0^{\infty}\frac{ldl}{2\pi}P_{\text{g-m}}(l/\chi(\bar{z}),\bar{z})J_0(l\theta).
\end{equation}

As detailed in Paper I, the cross-power spectrum of galaxies and matter is expressed as the sum of two terms: the 1-halo term, which describes galaxy-matter correlations within the same halo, and the two-halo term, which refers to the cross-correlation between different halos.

The cross-correlation function depends on the value of beta and the cosmological parameters of the assumed underlying model, but it also needs information about the way galaxies populate dark matter halos. This is described by the halo occupation distribution (HOD), for which we assume the simple three-parameter model by Zehavi et al. (2005). According to it,
when the mass of a halo exceeds a certain threshold, $M_{min}$, a galaxy is located at its centre. Any other galaxies are considered satellites and their distribution is directly proportional to the halo mass profile, as detailed in works such as \cite{ZHE05}. If the mass of a halo exceeds a different threshold, $M_1$, it will host satellites, and the number of satellites present is described by a power-law function with a coefficient of $\alpha$.

Consequently, it is possible to represent the probability of a central galaxy being present as a step function:
\begin{equation}
    \label{eq:ncen01}
    N_\text{c}(M) =
    \begin{cases}
    0 \quad \text{if}\ M < M_\text{min},\\
    1 \quad \text{otherwise}.
    \end{cases}
\end{equation}
The satellite galaxies occupation can be described as:
\begin{equation}
    \label{eq:nsat01}
    N_\text{s}(M) = N_\text{c}(M) \cdot \biggl(\dfrac{M}{M_1}\biggr)^{\alpha},
\end{equation}
with $M_\text{min}$, $M_1$, and $\alpha$ the free-parameters of the model.

It is important to emphasise that the analysis only considers cross-correlation data in the weak lensing regime, namely, $\theta \ge 0.5$ arcmin since we are working within such approximation in our theoretical model \citep[for a more comprehensive discussion, refer to][]{GON17,BON19}. Moreover, as discussed in Paper I, the validity of the standard halo model at small scales should initially be taken with caution. Physical effects such as baryonic feedback could affect the 1-halo cross-correlation function via the modification of the galaxy-matter power spectrum \citep[][]{DAA14, REN20, DAA20, AMO23}. An approach that has been used in the literature to address this issue implies adopting a different normalisation for dark matter halo profiles and for the satellite galaxy distribution \citep{CACC13,VIOLA15,VANUITERT16,DVORNIK18,DVO23}. We checked that introducing this modification affects only the sub-arcmin data points. However, these variations can be regarded as negligible when compared to the current uncertainties in the observational data. Therefore, in this work we do not discuss the effect of the baryonic feedback, which should be taken into account in future works, when more constraining results will be achieved.

\subsection{Parameter estimation}
\label{subsec:paramest}

In order to estimate the parameters, we use the open source software package called {\texttt{emcee}}, which is licensed under MIT, and utilises a Markov chain Monte Carlo (MCMC) algorithm to estimate the parameters. It is based on the Affine Invariant MCMC Ensemble sampler by \cite{GOO10} and is implemented purely in Python.
The MCMC runs in this work are performed with a number of walkers set to three times the number of parameters to be estimated, $N$, and a fixed number of iterations set to 5000. This generates $15000 N$ posterior samples per run, which are then reduced by discarding the initial iterations and introducing a burn-in when flattening the chain. 
We examined each parameter's walkers separately to identify the burn-in value. We selected the iterations where all walkers had moved to explore a specific area rather than wandering randomly. The number of iterations we kept depended on the specific case. Finally, we confirmed the chains' convergence using the common tests provided by \texttt{GETDIST} \citep{GETDIST}.

In our analysis, we aim to estimate both astrophysical and cosmological parameters. The astrophysical parameters of interest are $M_{min}$, $M_1$, and $\alpha$, with different values for each redshift bin. With respect to \cite{BON21}, the parameter $\beta$ is also added to the analysis, since, as studied in Paper I, fixing it to the previously used value of 3 could affect the results.
On the other hand, the cosmological parameters include $\Omega_M$, $\sigma_8$, and $h$, which is defined as $H_0=100\times h\,\text{km}\,\text{s}^{-1}\text{Mpc}^{-1}$. 
We assume a flat universe where $\Omega_{\text{DE}}=1-\Omega_M$, and we hold the values of $\Omega_b$ and $n_s$ fixed at their best-fit values from \citet{PLA18_VI}, which are $\Omega_b=0.0486$ and $n_s=0.9667$, respectively.
In the weak lensing approximation, only the cross-correlation function data with angular scales greater than or equal to $0.5$ arcmin are considered, as discussed in detail in \citet{BON19}.

When using the Bayesian method for parameter estimation, it is necessary to define both a prior and a likelihood distribution. For the likelihood distribution, a conventional Gaussian function is employed.
In terms of selecting prior distributions for the astrophysical parameters, we used uniform priors. In particular, in the scenarios using the four bins, that is "base case", "w/o G15", "bins 1+4," and "bins 2+3", we adopted the $\log{M_{min}}$ and $\log{M_{1}}$
priors listed in Table \ref{tab:prior_4bins}. In the "six-bin" case we use the priors specified in Table \ref{tab:prior_6bins}. For the "six bins-WR" case, we give those in Table \ref{tab:prior_6bins_WR}.
For all bins and all cases, $\alpha\sim\mathcal{U}[0.5,1.5]$ and $\beta$ priors are Gaussian with $\beta\sim\mathcal{G}[2.9,0.04]$, following the discussion in Paper I.

\begin{table*}[t]
\caption{Parameter prior distributions adopted in the scenarios using the four bins, that is "base case," "w/o G15," "bins 1+4," and "bins 2+3"} 
\label{tab:prior_4bins} 
\centering 
\begin{tabular}{c c c c c} 
\hline 
\hline \\[-1.2ex]
Parameter&bin 1&bin 2&bin 3&bin 4\\ 
\hline 
\\[-1ex]
$\log M_{\text{min}}$ & $\mathcal{U}[10.0,13.0]$ & $\mathcal{U}[11.0,13.0]$ & $\mathcal{U}[11.5,13.5]$ & $\mathcal{U}[12.0,14.5]$\\ [0.3ex]  
$\log M_1$ & $\mathcal{U}[11.0,15.5]$ & $\mathcal{U}[12.0,15.5]$ & $\mathcal{U}[12.5,15.5]$ & $\mathcal{U}[13.0,15.5]$\\   
\\[-1ex]
\hline 
\hline 
\end{tabular} 
\end{table*} 

\begin{table*}[t]
\caption{Parameter prior distributions adopted in the "six bins"} 
\label{tab:prior_6bins} 
\centering 
\begin{tabular}{c c c c c c c} 
\hline 
\hline \\[-1.2ex]
Parameter&bin 1&bin 2a&bin 2b&bin 3a&bin 3b&bin 4\\ 
\hline 
\\[-1ex]
$\log M_{\text{min}}$ & $\mathcal{U}[10.0,13.0]$ & $\mathcal{U}[11.0,13.0]$ & $\mathcal{U}[11.0,13.0]$ & $\mathcal{U}[11.5,13.5]$ & $\mathcal{U}[11.5,13.5]$ & $\mathcal{U}[12.0,14.5]$\\ [0.3ex]  
$\log M_1$ & $\mathcal{U}[11.0,15.5]$ & $\mathcal{U}[12.0,15.5]$ & $\mathcal{U}[12.0,15.5]$ & $\mathcal{U}[12.5,15.5]$ & $\mathcal{U}[12.5,15.5]$ & $\mathcal{U}[13.0,15.5]$\\   
\\[-1ex]
\hline 
\hline 
\end{tabular} 
\label{table1}
\end{table*} 

\begin{table*}[t]
\caption{Parameter prior distributions adopted in the "six bins-WR"} 
\label{tab:prior_6bins_WR} 
\centering 
\begin{tabular}{c c c c c c c} 
\hline 
\hline \\[-1.2ex]
Parameter&bin 0&bin 1&bin 2&bin 3&bin 4&bin 5\\ 
\hline 
\\[-1ex]
$\log M_{\text{min}}$ & $\mathcal{U}[10.0,13.0]$ & $\mathcal{U}[10.0,13.0]$ & $\mathcal{U}[11.0,13.0]$ & $\mathcal{U}[11.5,13.5]$ & $\mathcal{U}[12.0,14.5]$ & $\mathcal{U}[12.0,14.5]$\\ [0.3ex]  
$\log M_1$ & $\mathcal{U}[11.0,15.5]$ & $\mathcal{U}[11.0,15.5]$ & $\mathcal{U}[12.0,15.5]$ & $\mathcal{U}[12.5,15.5]$ & $\mathcal{U}[13.0,15.5]$ & $\mathcal{U}[13.0,15.5]$\\   
\\[-1ex]
\hline 
\hline 
\end{tabular} 
\end{table*} 

The prior distributions for the cosmological parameters that we aim to estimate are the same as those employed in \cite{BON20, BON21}. More specifically, $\Omega_M\sim\mathcal{U}[0.1-0.8]$, $\sigma_8\sim\mathcal{U}[0.6-1.2]$, and $h\sim\mathcal{U}[0.5-1.0]$.

\section{Results}
\label{sec:results}

In this section we describe and discuss the results obtained by jointly analysing different combination of bins of redshift for the foreground sample, the lenses. In particular, Sect.~\ref{subsec:results_4bins} describes the "base case", while Sect.~\ref{subsec:results_noG15} presents a study of the potential effect in the results due to the anomalous measurements in the G15 field, pointed out in Paper I.
Finally, in Sect.~\ref{subsec:results_bins_changes} the effect on the constraints of using different number and ranges of bins for the tomographic analysis is discussed.
On one hand, we address the possibility of reducing the joint analysis to only a pair of bins ("bins 1+4" and "bins 2+3"), which would diminish the computational time of the parameters estimation. On the other hand, we increase the number of bins to six to explore both the effect of a thinner slicing ("six bin" case) of the foreground sample in the same redshift range of the "base case" and the effect of introducing two more bins, one at each end of the four bin cases ("six bins-WR" case).

\subsection{Base case: Four redshift bins}
\label{subsec:results_4bins}

The first step of our study consists in the comparison between the results obtained with the "full model" and the mean-redshift approximation, as described in Paper I. This approximation reduces the computational time by about a factor of 10 and it is of crucial importance when performing tomographic analyses, given the large number of parameters to be estimated in such a configuration. The same improved data and methodology described in Paper I are used in both cases. The comparative plots are shown in Appendix \ref{app:comp_full_MzA}. The HOD parameters prior distributions used for both cases are summarised in Table \ref{tab:prior_4bins}.

Figures \ref{fig:sampled_tomo} and \ref{fig:sampled_tomo_zmed} present the posterior sampling of the cross-correlation function (in red), the best fit (dashed black line), and the data (black dots) in the four redshift bins for  the "full model" and the mean-redshift approximation, respectively. The panels represent the results for the different bins of redshift: bins 1 to 4 from left to right, top to bottom, respectively. The redshift range is indicated in every panel. 
The data have comparable errorbars in the four bins -- except in bin 1, which present larger errorbars at intermediate and large angular scales.
Overall, the best fit and sampled area area in good agreement with the data in both cases.

Moreover, Fig. \ref{fig:corner_tomo_MzA} illustrates the marginalised posterior distribution and probability contours\footnote{The probability contours in all the corner plots are set to 0.393 and 0.865.} for all the estimated parameters. Given the large number of involved parameters, in order to clearly show the results, we divide the corner plot according to the number of bins. In this way we depict the posterior distribution of the HOD parameters of each bin together with the cosmological parameters (that are the same for every bin). With the same aim of making the presentation of the results more intelligible, we decide not to include in the plots the $\beta$ parameter, since its posterior distribution barely deviates from its (Gaussian) prior. 
As expected, given the small bin ranges used for tomography, the mean-redshift approximation ("base case") results are in good agreement with the "full model" case.

\begin{figure}[ht]
\includegraphics[width=0.5\textwidth]{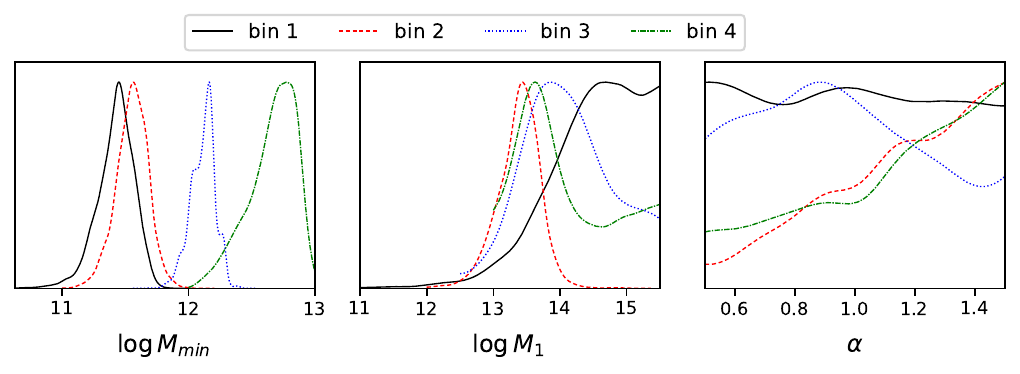}\\
\includegraphics[width=0.5\textwidth]{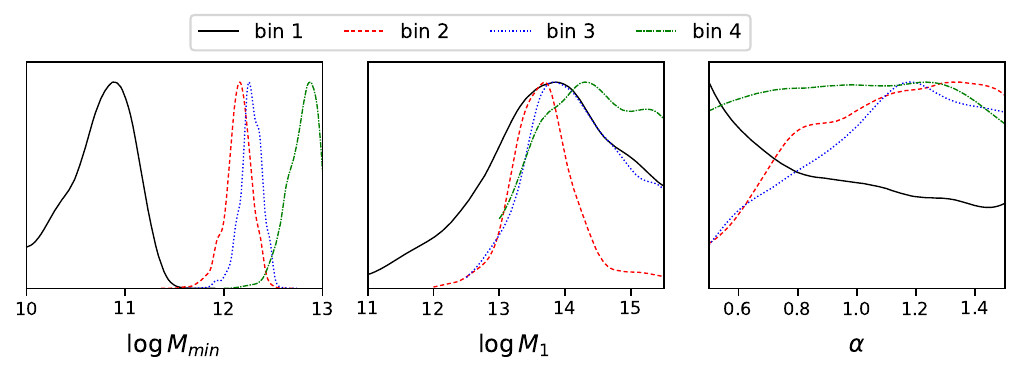}
 \caption{HOD parameters for the tomographic case (top). HOD parameters for the tomographic case excluding the data from the G15 field (bottom). The solid black, dashed red, dotted blue and dash-dotted green lines are the bin 1, bin 2, bin 3, and bin 4, respectively. The mass values are expressed in $M_{\odot}/h$ units. 
 }
 \label{fig:astro_tomo_zmed}
\end{figure}

The top panel of Figure \ref{fig:astro_tomo_zmed} displays the marginalised posterior distributions of the HOD parameters for the "base case." The solid black, dashed red, dotted blue, and dash-dotted green lines correspond to bin 1, bin 2, bin 3, and bin 4, respectively. A notable trend can be observed, with the value of $\log{M_{min}}$ increasing as a function of redshift from bin 1 to bin 4. The distribution peaks are located at 11.35, 11.45, 11.87, and 12.71, respectively.\footnote{Throughout the manuscript, all halo masses are expressed in $M_{\odot}/h$.} This behaviour is consistent with previous tomographic analyses \citep{GON17, BON21, CUE22}.

Regarding the $\log{M_1}$ parameter, the marginalised posteriors appear relatively homogeneous across the four bins, with peaks ranging from 13.22 to 13.58; with the exception of bin 1, which exhibits slightly higher values peaking at 14.08 along with a broader distribution. On the other hand, the $\alpha$ parameter remains poorly constrained in all bins.

Comparing our results to those of \cite{BON21}, we generally find narrower distributions and consistent findings. The main discrepancy lies in the $\log{M_{min}}$ distributions for bin 4, where the values obtained in the "base case" are lower. Additionally, some subtle distinctions can be observed in the opposite trends of the $\alpha$ distributions for bin 2, although it should be noted that $\alpha$ is not constrained in either case.

\begin{figure}[ht]
\includegraphics[width=0.5\textwidth]{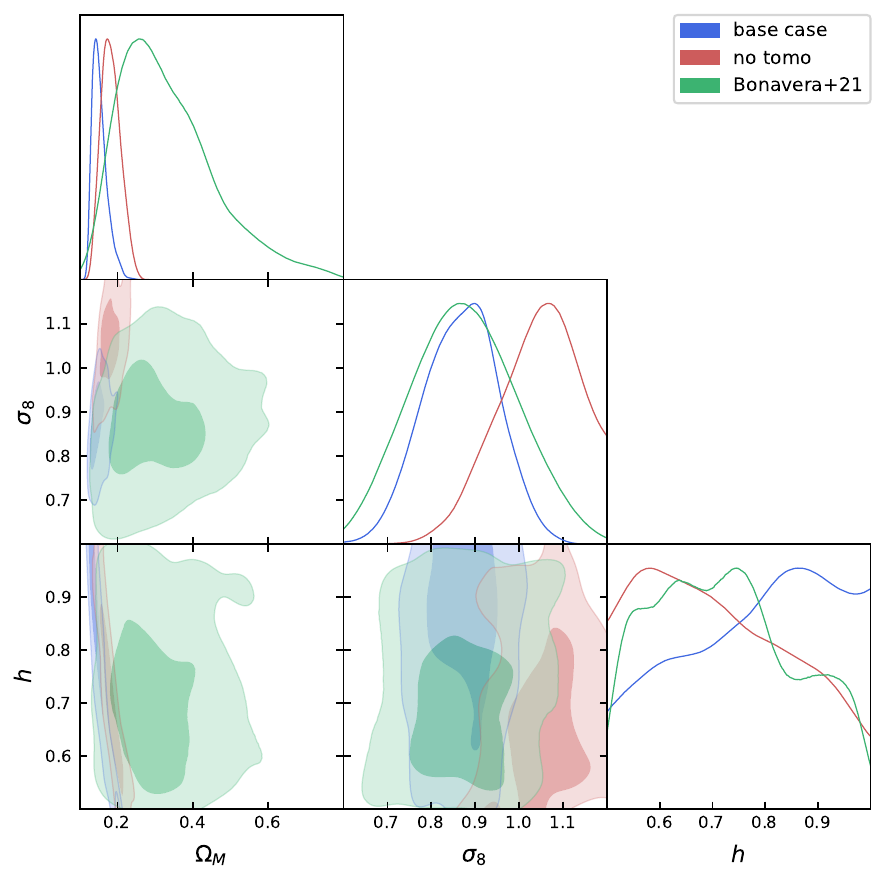}
 \caption{Comparison of the marginalised posterior distribution and probability contours (set to 0.393 and 0.865) for the cosmological parameters obtained in the "base case" in blue, with single bin from Paper I in red and by \cite{BON21} in green.}
 \label{fig:cosmo_comp_tomo_notomo}
\end{figure}

Figure \ref{fig:cosmo_comp_tomo_notomo} presents the estimated cosmological parameters for the "base case" represented by the blue curves. The marginalised distributions exhibit peaks at 0.16 for $\Omega_M$ and 0.70 for $\sigma_8$. The results from the non-tomographic analysis in Paper I and the tomographic analysis by \citet{BON21} are also displayed in red and green, respectively. The impact of the new methodology, particularly the reduction in measurement uncertainties, is evident for $\Omega_M$, as indicated by the significantly narrower posterior distributions compared to the results obtained by \citet{BON21}. However, it is noteworthy that there is a preference for lower values of $\Omega_M$, with the posterior distributions deviating considerably from the standard model. This issue will be further explored in the subsequent subsection.

Regarding $\sigma_8$, there is also a preference for lower values, with the newer analysis exhibiting a slightly narrower posterior distribution. The higher value observed in the non-tomographic case is consistent with the comparison between the non-tomographic results from \cite{GON21} and the tomographic analysis by \cite{BON21}. Notably, the parameter $h$ remains unconstrained in all three cases.

\begin{figure}[ht]
\includegraphics[width=0.5\textwidth]{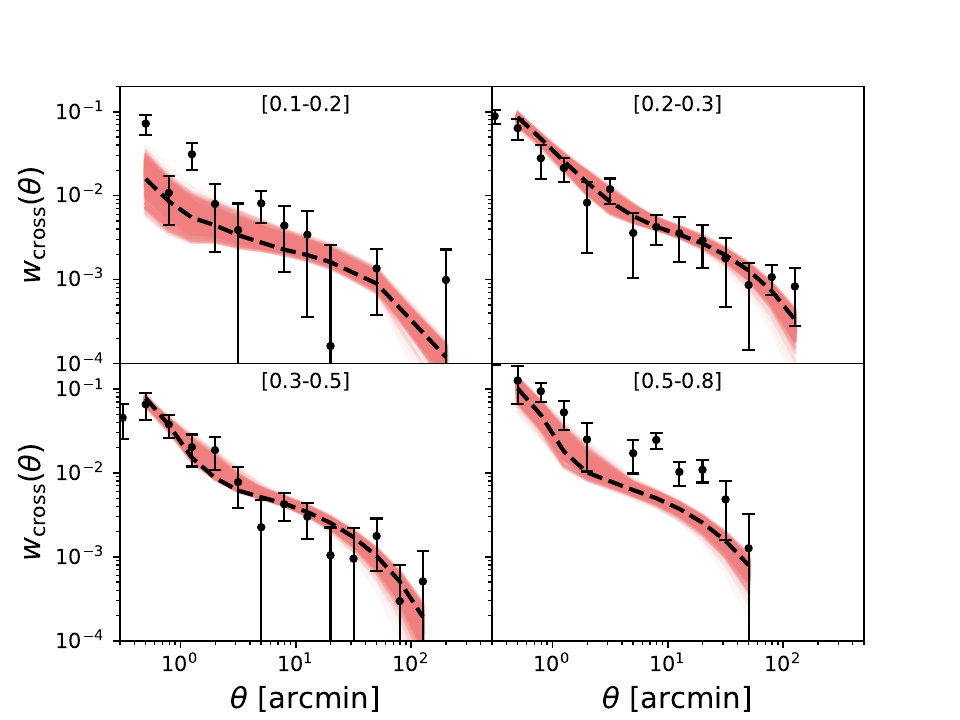}
 \caption{Cross-correlation data (black points) and the results obtained with the tomographic case excluding G15: the posterior sampling of the cross-correlation function is shown in red and the best fit with the black dashed line. Bins 1 to 4 are shown from left to right, top to bottom, and the redshift ranges are indicated in each panel.
 }
 \label{fig:sampled_tomo_zmed_noG15}
\end{figure}

\subsection{Without G15: Four redshift bins without the G15 field}
\label{subsec:results_noG15}

The inclusion or exclusion of the G15 field has a significant impact on the measured cross-correlation function at larger angular scales, as previously discussed in Paper I. Since the cosmological constraints heavily rely on these angular scales, the effect of the G15 field on the results becomes crucial, as demonstrated in the same work for a broad single redshift bin. In light of this, we explore the potential effects of excluding the G15 field in the tomographic analysis in this study.

The posterior sampling of the cross-correlation function (shown in red in Figure \ref{fig:sampled_tomo_zmed_noG15}) confirms the effect of removing the G15 field on the largest angular scales, resulting in a steeper decrement of the signal. Due to the reduced statistics compared to the case where all four fields are included, the uncertainties of the data are larger. Comparing both sets of measurements, it is evident that the removal of the G15 field leads to a worsening of the best fit and posterior sampling for bin 1 and, especially, for the highest redshift bin. This can be attributed to the statistical weight of the central bins in the overall best fit due to the relatively better uncertainties in the measurements.

\begin{figure}[ht]
\includegraphics[width=0.5\textwidth]{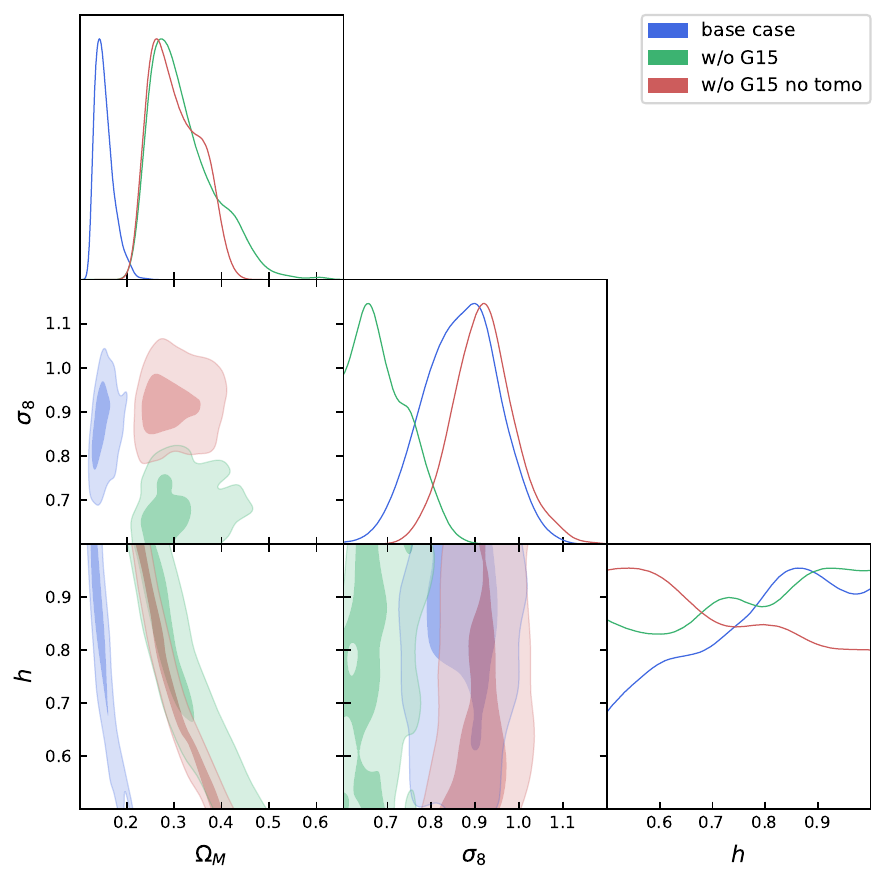}
 \caption{Comparison of the marginalised posterior distribution and probability contours (set to 0.393 and 0.865) of the cosmological parameters obtained in the "w/o G15" case (in green) and in the corresponding case with single bin from Paper I (in red). The "base case" results are also shown in blue. 
 }
 \label{fig:cosmo_comp_tomo_notomo_noG15}
\end{figure}

Figure \ref{fig:corner_tomo_zmed_noG15} displays the posterior distribution of the estimated parameters, organised per bin, similar to Fig. \ref{fig:corner_tomo_MzA}. The marginalised posterior distributions of the estimated HOD parameters, presented in the bottom panel of Fig. \ref{fig:astro_tomo_zmed}, reveal that (similarly to the "base case") the $\log{M_{min}}$ parameters exhibit increasing peak values with the bin range. Specifically, the distributions peak at 10.70, 11.94, 12.05, and 12.74 for bins 1 to 4, respectively. Comparing with the "base case" in the upper panel, the main difference lies in the $\log{M_{min}}$ distribution of bin 2, which shifts to higher values and of bin 1, which shifts to lower values, when the G15 field is excluded. This shift may be related to the substantial difference in the signal levels observed in bin 1 and bin 2 between these two cases. The parameter $\log{M_1}$ displays a worse but similar behaviour to that of the "base case," while the $\alpha$ distributions are mostly unconstrained and with slightly opposite trend than in the "base case."

Figure \ref{fig:cosmo_comp_tomo_notomo_noG15} presents the marginalised posterior distributions and probability contours of the cosmological parameters estimated for the "w/o G15" case (in green). The marginalised posterior distribution for $\Omega_M$ peaks at a higher value of 0.32 compared to the "base case" (in blue), while it peaks at a lower value of 0.60 for $\sigma_8$. The parameter $h$ remains unconstrained. In the same plot, we also compare the results with the corresponding non-tomographic analysis from Paper I (in red). The main difference lies in the higher value of $\sigma_8$, which peaks at approximately 0.73 in the non-tomographic case, but the qualitative behaviour is the same: the removal of the G15 region produces lower values of $\sigma_8$ and higher values of $\Omega_m$
both in the non-tomographic and tomographic cases. 

In summary, we confirm in this study that the exclusion of the G15 field leads to a decrease in the cross-correlation signal at the largest angular separation distances, as observed in Paper I. While this effect is noticeable in all redshift bins, it is particularly evident across all angular scales in bins 1 and 2. Consequently, the main difference with respect to the "base case" is observed in the $\log{M_{min}}$ distribution for those bins, which shifts towards significantly lower and higher values, respectively. These findings suggest that the anomaly associated with the G15 field may be related to a sample variance problem within the large-scale structure in the line of sight, but also within a specific range of redshifts.

\subsection{Tomographic analysis with different bins of redshift}
\label{subsec:results_bins_changes}

In this section, we assess the impact of considering different numbers of redshift bins and exploring different redshift ranges. One interesting aspect to investigate from a computational perspective is the possibility of using only two bins of redshift, which would effectively halve the number of HOD parameters to be estimated compared to the "base case" (see Sect. \ref{sec:2bins}). Additionally, we aim to discern the major contributor to cosmological parameter estimation: whether it is more influenced by the extreme redshift bins ("bins 1+4" case), where the redshift evolution of the measurements plays a prominent role, or the central bins ("bins 2+3" case), which offer better statistical robustness.

In Section \ref{sec:6bins}, despite not reducing computational time, we further explore the possibilities with a finer slicing of the default redshift range. Specifically, we investigate the potential benefits of jointly analysing six bins of redshift ("six bin" case). Lastly, we extend the total redshift range by incorporating two additional bins at the endpoints of the original "base case" in the foreground sample. Through these analyses, we aim to gain insights into the implications of different redshift binning strategies and explore the trade-offs between computational efficiency and precision in cosmological parameter estimation.

\subsubsection{Two-bin analysis}
\label{sec:2bins}

\begin{figure*}[ht]
\includegraphics[width=0.5\textwidth]{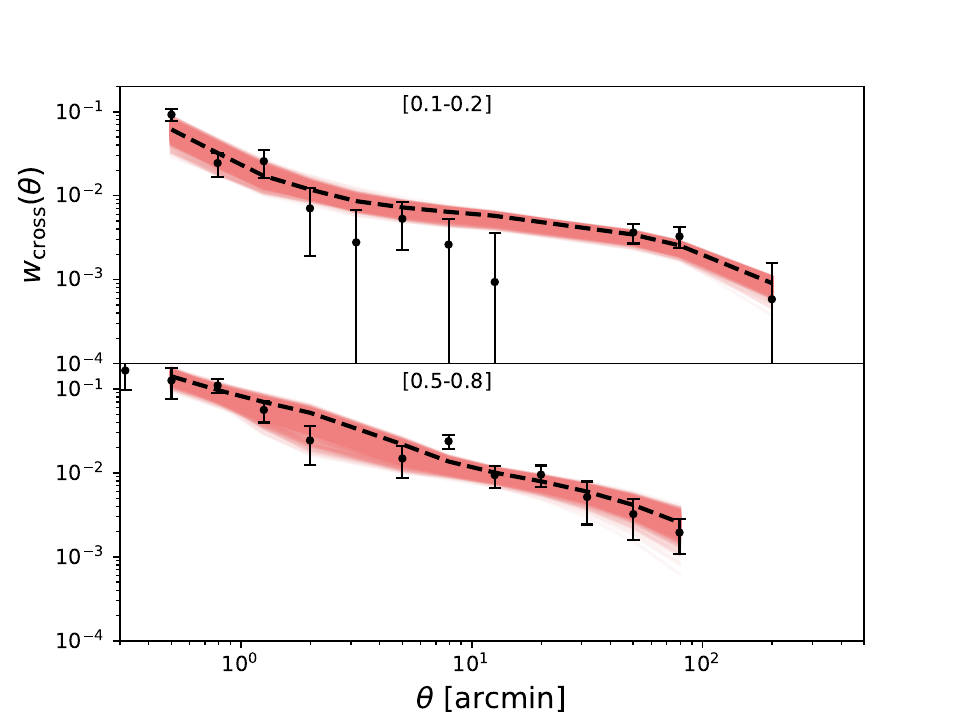}
\includegraphics[width=0.5\textwidth]{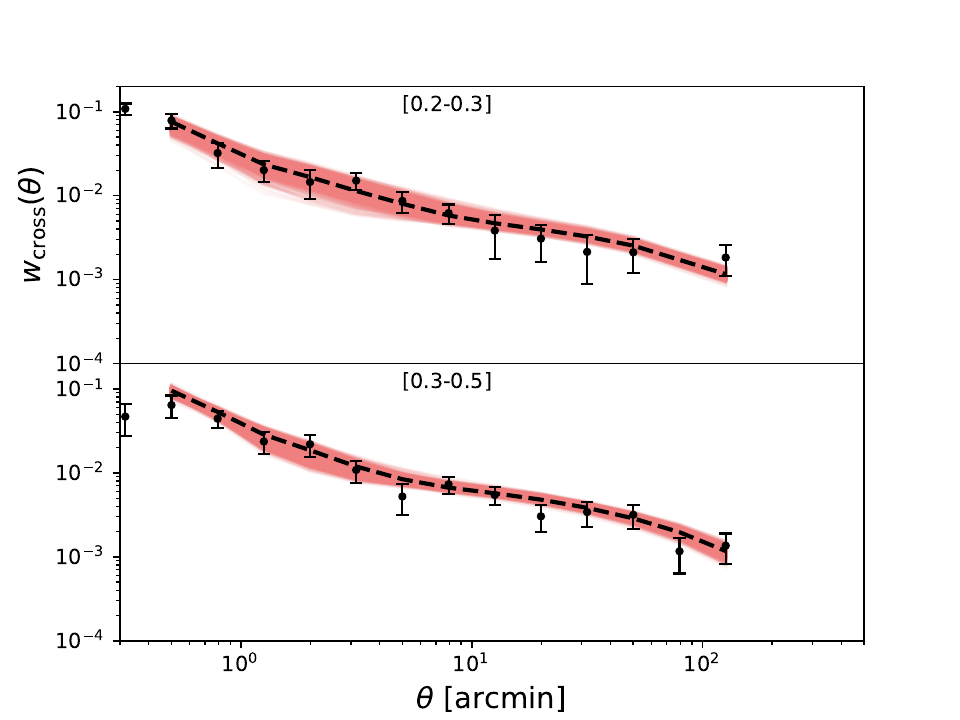}
 \caption{Cross-correlation data (black points), best fit (black dashed line) and the posterior sampling (in red) in the "bins 1+4" scenario (left panel) and in the "bins 2+3" case (right panel).
 }
 \label{fig:sampled_tomo_zmed_bin14}
\end{figure*}
\begin{figure}[ht]
\includegraphics[width=0.5\textwidth]{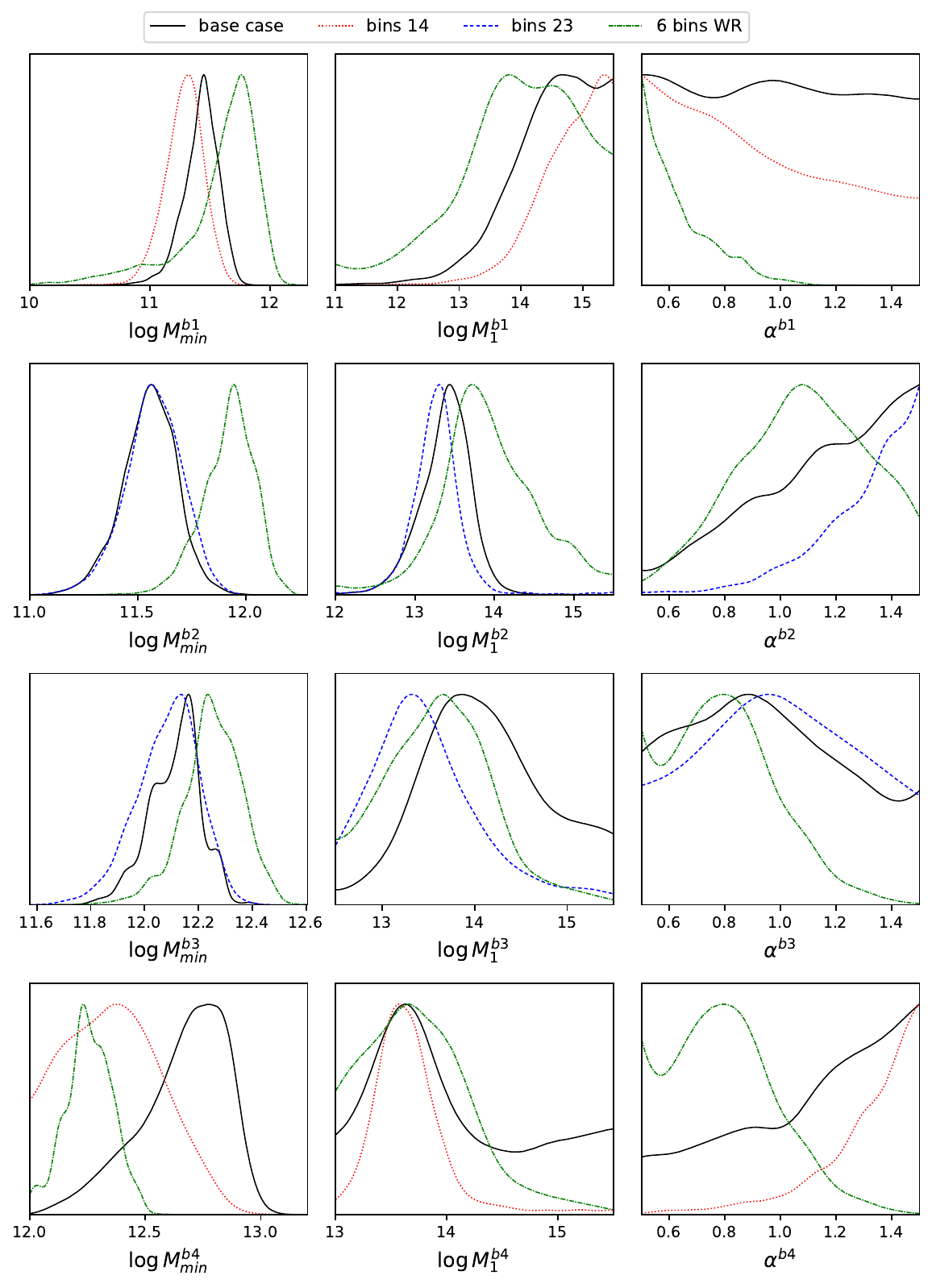}
 \caption{Comparison of the marginalised posterior distributions of the HOD parameters ($\log{M_{min}}$, $\log{M_1}$ and $\alpha$ from left to right, respectively) derived in the different analysed cases, for bin 1, bin 2, bin 3, and bin 4, respectively (from top to bottom). The "base case" is shown with the black solid line, "bins 1+4" with the red dotted line, "bins 2+3" with the blue dashed line and the "six bins-WR" with the green dash-dotted line. The mass values are expressed in $M_{\odot}/h$ units.
 }
 \label{fig:comp_astro_all_cases}
\end{figure}

\begin{figure}[ht]
\includegraphics[width=0.5\textwidth]{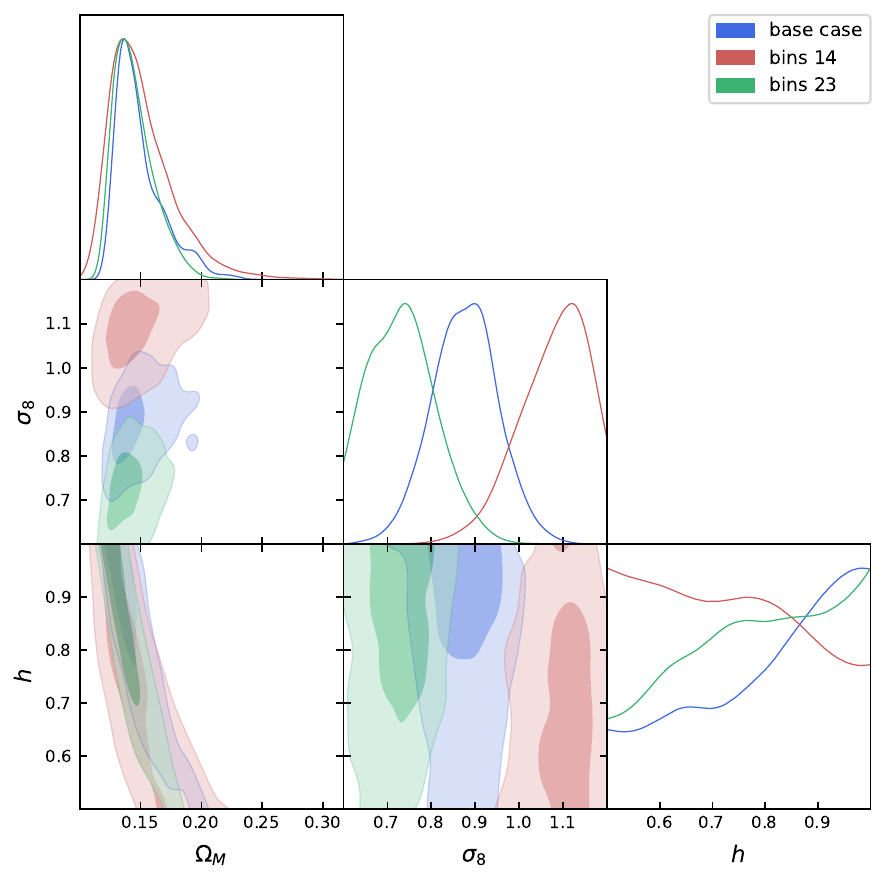}
 \caption{ Comparison of the marginalised posterior distribution and probability contours (set to 0.393 and 0.865) of the cosmological parameters obtained in the "base case" (in blue) and in the two bins cases, "bins 1+4" (in red) and "bins 2+3" (in green).
 }
 \label{fig:corner_comp_cosmo_2bins}
\end{figure}

The panels in Fig. \ref{fig:sampled_tomo_zmed_bin14} depict the data (black dots), the posterior sampling of the cross-correlation function (in red), and the best fit (dashed black line) obtained in the joint analysis of bin 1 and bin 4 ("bins 1+4," left panel) as well as in the joint analysis of bin 2 and bin 3 ("bins 2+3," right panel). In both cases, the upper panel refers to the lowest bin in the pair and the bottom to the highest. The redshift range is indicated in every panel. In most cases the best fit and sampled area are in agreement with the data.
The marginalised and probability distribution are shown in Figs. \ref{fig:corner_tomo_bin14} (for the joint bin 1 and bin 4 case) and \ref{fig:corner_tomo_bin23} (for the joint bin 2 and bin 3 case). As with the previous plots, the results are divided according to the number of jointly analysed bins --  two in this case -- for an improved visualisation. 

The HOD marginalised posterior distributions for the two-bins cases are shown in Fig. \ref{fig:comp_astro_all_cases}: the red dotted line indicates the results of "bins 1+4" and the blue dashed line those of "bins 2+3". These cases are compared with the "base case" (black solid line) and the common bins in the "six bins-WR" case. 
This figure shows the three analysed HOD parameters $\log{M_{min}}$, $\log{M_1}$, and $\alpha$ in the columns from left to right, respectively. The four rows represents the four bins (bins 1 to 4) from top to bottom.

For the "bins 1+4" case, the ${M_{min}}$ parameter 
is slightly shifted towards lower values with that of the "base case", peaking at 11.16 for bin 1 and 12.46 for bin 4. The $\log{M_1}$ parameter aligns with the "base case" peaking at 13.38 in bin 4 but it is slightly higher, 14.91, in bin 1. The $\alpha$ parameter remains unconstrained in bin 1 and provides a lower limit in bin 4. In the "bins 2+3" case, $\log{M_{min}}$ agrees with the "base case," with peaks at 11.44 in bin 2 and 11.86 in bin 3. The $\log{M_1}$ results are also consistent, peaking at 13.10 in bin 2 and in bin 3. The parameter $\alpha$ exhibits a lower limit in bin 2 and shows a tentative peak at 1.23 in bin 3.

The posterior distributions of cosmological parameters are shown in Fig. \ref{fig:corner_comp_cosmo_2bins}, with the "bins 1+4" case displayed in red and the "bins 2+3" case in green. For comparison, the "base case" is plotted in blue, with uncertainties only slightly improved compared to the other two cases. The marginalised posterior distribution of $\Omega_M$ shows good agreement among the three cases, peaking at 0.16. The $\sigma_8$ posteriors reveal an intriguing displacement: the distribution for "bins 2+3" is shifted to lower values (peaking at 0.61) compared to "bins 1+4" (peaking at 0.82), while the "base case" distribution falls in between (peaking at 0.70). The parameter $h$ remains unconstrained as expected. Tomographic analysis achieves substantial constraining power even with only two redshift bins. However, the discrepancy in $\sigma_8$ across the different cases suggests a sample variance issue that can be mitigated by increasing the number of redshift bins. Thus, the sample variance problem identified in Paper I, which affected the spatial distribution, is also manifested as sample variance in redshift, as hinted at in the analysis of the G15 effect in Section \ref{subsec:results_noG15}.

\subsubsection{Six-bin analysis}
\label{sec:6bins}

\begin{figure*}[ht]
\includegraphics[width=0.5\textwidth]{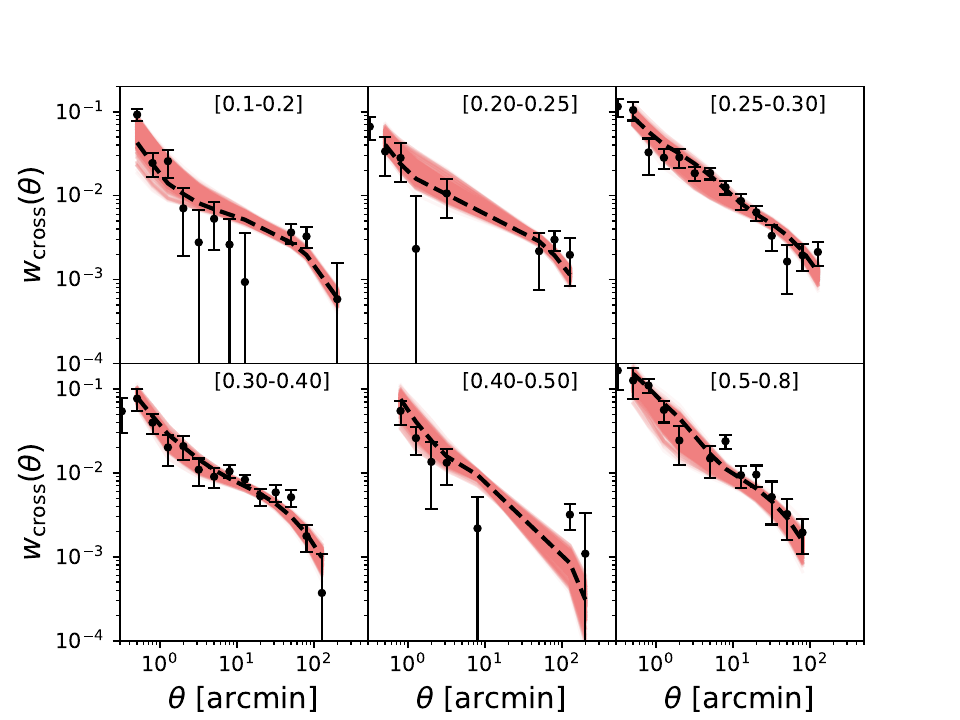}
\includegraphics[width=0.5\textwidth]{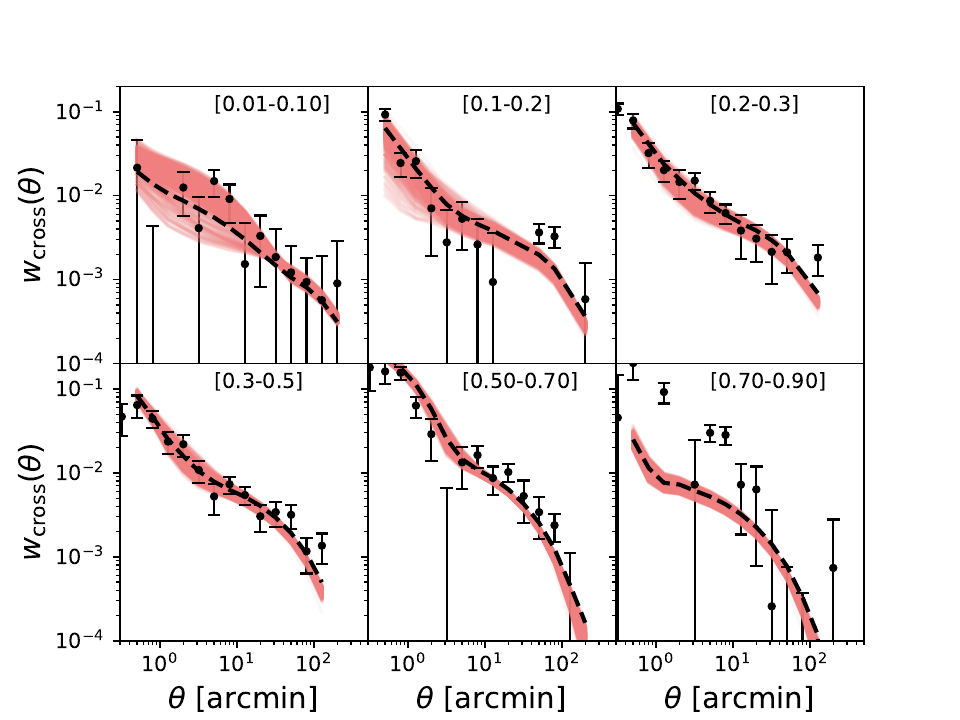}
 \caption{Cross-correlation data (black points), best fit (black dashed line) and sampling posterior (in red) obtained in the "six bin" case (left) and the "six bins-WR" case (right).
 }
 \label{fig:sampled_tomo_zmed_6bins}
\end{figure*}

\begin{figure}[ht]
\includegraphics[width=0.5\textwidth]{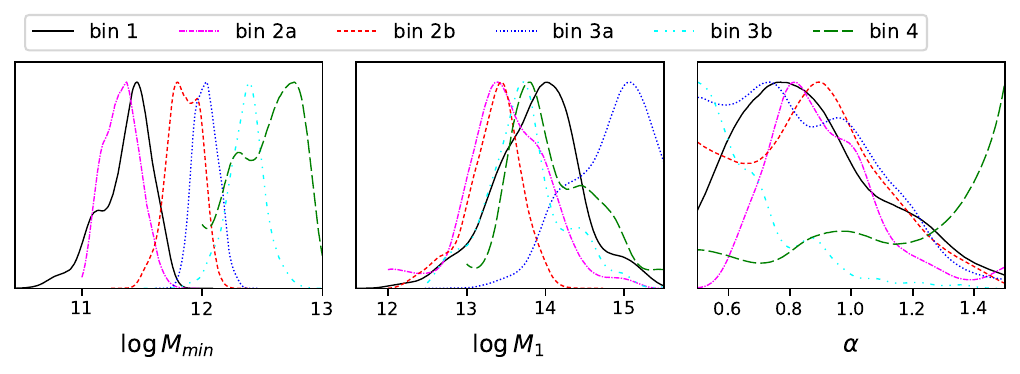}\\
\includegraphics[width=0.5\textwidth]{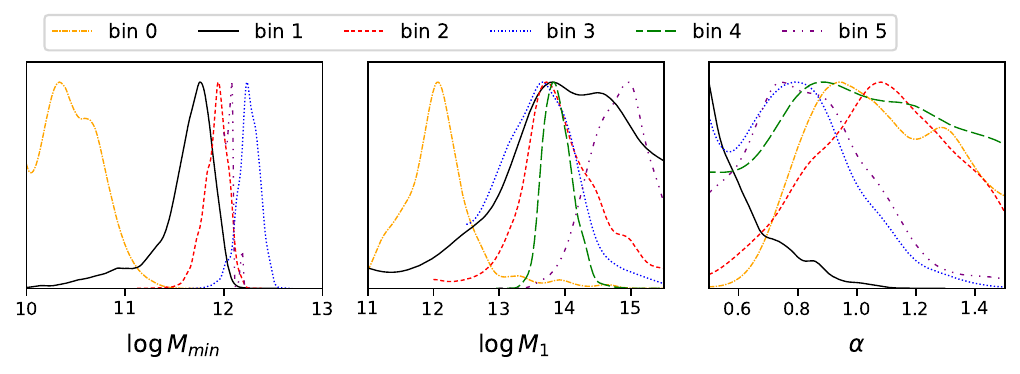}
 \caption{HOD parameters marginalised posterior distributions: $\log{M_{min}}$, $\log{M_1}$ and $\alpha$ in the first, second, and third column, respectively. The top panel shows the results for the "six bin" case and the bottom panel those for the "six bins-WR" case. The mass values are expressed in $M_{\odot}/h$ units.
 }
 \label{fig:astro_tomo_zmed_6bins}
\end{figure}

\begin{figure}[ht]
\includegraphics[width=0.5\textwidth]{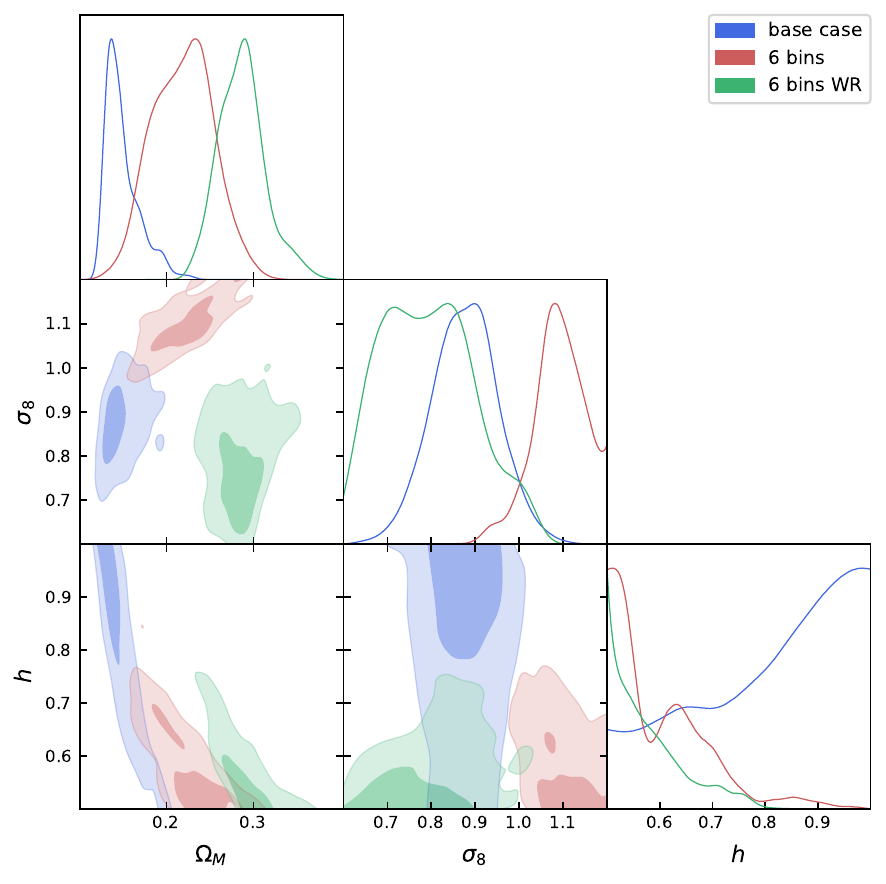}
 \caption{Comparison of the marginalised posterior distribution and probability contours for the cosmological parameters obtained in the "base case" (in blue) and in the six bins cases: "six bins" (in red) and "six bins-WR" (in green).
 }
 \label{fig:corner_comp_cosmo_6bins}
\end{figure}

Considering the conclusions of the previous section, we investigate the effect of increasing the number of redshift bins in two specific cases: maintaining the same redshift range ("six bins") or extending it by adding additional redshift bins ("six bins-WR"). The definition of the redshift bins for both cases was already discussed in Sect. \ref{subsec:data} and the prior distributions used are summarised in Table \ref{tab:prior_6bins} and \ref{tab:prior_6bins_WR}, respectively.

The panels in Fig. \ref{fig:sampled_tomo_zmed_6bins} display the posterior sampling of the cross-correlation function (in red), the best fit (dashed black line), and the data (black dots) for the joint analysis of six bins in the original redshift range ("six bins" case, left plot) and the wider redshift range from 0.01 to 0.9 ("six bins-WR" case, right plot). Each panel represents the results for a specific redshift bin, and the range of each bin is indicated.
In the "six bins" case, the data are visibly affected by poorer statistics compared to the "base case" due to the same total number of sources being spread over more bins. In the "six bins-WR" case, the central bins show similar behaviour to the "base case," while the newly added first and last bins have large error bars due to lower lensing probability \citep[see][]{LAPI12} in the first bin and significantly poorer statistics in the last bin. In most cases, the best fit and sampled area align well with the data, except for the last bin in the "six bins-WR" case, which exhibits poor agreement.

The marginalised and probability distribution of the six-bins analysis are illustrated in Figs. \ref{fig:corner_tomo_6bins} ("six bins" case) and \ref{fig:corner_tomo_6bins_09} ("six bins-WR" case). As in the previous cases, the corner plot is divided according to the number of jointly analysed bins (six in this case)\ for a better visualisation.

Figure \ref{fig:astro_tomo_zmed_6bins} displays the marginalised posterior distributions of the HOD parameters for the "six bins" cases: the "six bins" case in the upper panel and the "six bins-WR" case in the lower panel. The parameters $\log{M_{min}}$, $\log{M_1}$, and $\alpha$ are shown from left to right, and different redshift bins are indicated with various line styles. In the "six bins-WR" case, the most similar bins to the original ones are represented with the same line style. The additional bins added at the extremes of the original range are depicted with the yellow dash-dotted line (bin 0) and the purple dash-dot-dotted line (bin 5) in the bottom panel.

The ${M_{min}}$ parameter generally increases with redshift, except for bin 2a in the "six bins" case, which shows a distribution similar to bin 1, as well as bin 5 in the "six bins-WR" case, positioned between bins 3 and 4. This might simply be due to the much larger errobars and less significant signal in such bins, especially bin 5 in the "six bins-WR" case, which is the one with the worst fit.
All the distributions are well constrained showing peaks (from lower to higher bin) at 11.23, 11.07, 11.66, 11.77, 12.31 and 12.66 for the "six bins" case and at 10.32, 11.53, 11.75, 11.95, 13.36 and 12.09 for the "six bins-WR" case. Notably, in the "six bins-WR" case, the values tend to converge around 12.0 for all redshift bins except bin 0 and bin 4. Fig. \ref{fig:comp_astro_all_cases} compares, for the redshift bins in common, the HOD marginalised posterior distributions of the "six bins-WR" case (green dash-dotted line) with the other three cases studied in this work. The figure shows again how the $\log{M_{min}}$ values tend to converge toward a value near 12.0.

The marginalised posterior distributions of $\log{M_1}$ for the "six bins" show very similar peaks that range from 13.85 to 14.55, except for the last bin.
In the "six bins-WR" case, the distributions do not converge and they do not exactly follow the same behaviour with redshift as in the $\log{M_{min}}$ ones. Their peaks range from 12.37 for bin 0 to 14.96 for bin 5.
The parameter $\alpha$ does not show a clear trend, as the posterior distributions peak at different values within the available range for both cases.

The marginalised posterior distributions of the cosmological parameters are presented in Fig. \ref{fig:corner_comp_cosmo_6bins} in red for the "six bins" case and in green for the "six bins-WR" case. The "base case" results are plotted in blue for comparison. The marginalised posterior distribution of $\Omega_M$ clearly shifts towards higher values, with the "six bins" case peaking at 0.20 and the "six bins-WR" case peaking at 0.32, compared to the "base case" peak at 0.16. The posterior distributions in both cases are wider than in the "base case".

Increasing the number of redshift bins has a similar effect to removing the G15 field, especially when additional bins are added instead of dividing the existing ones. This suggests that the evolution with redshift, i.e. the cosmological model, starts to dominate over the sample variance effect. However, the overall fit is not perfect for each individual bin, resulting in poorer fits for the most biased samples. Extending the redshift range seems to work better than simply increasing the number of redshift bins, as it improves statistics per bin and produces tighter constraints compared to the "w/o G15" case.

The $\sigma_8$ posterior distribution for the "six bins-WR" case is compatible with the "base case" results but shows a wider distribution towards lower values, with a peak (mean) value at 0.66 (0.69).
In contrast, the "six bins" case exhibits a clear peak at 0.99.
This difference can be attributed to the splitting of the two central bins, which reduces their statistics compared to bin 1 and bin 4, resulting in a lower weight in the parameter estimation and producing estimates similar to the "bins 1+4" case.
The secondary peak observed between 0.7-0.8 in the "six bins-WR" case is of particular interest. We have confirmed the stability of the full distribution, ruling out any convergence issues. Instead, we interpret the secondary peak as a local valid minimum within the constraints of our current data.

Finally, the parameter $h$ remains unconstrained, but for the first time in this work, the posterior distributions show tight upper limits of 0.69 at 95\% C.I., in the "six bins-WR". These results align with the findings in Paper I, where a joint analysis of the cross-correlation function and galaxy clustering measurements was performed. 
Although these results should be interpreted with caution, they indicate that with improved data and better-constrained parameters, there is hope for obtaining constraints on $h$.
\section{Conclusions}
\label{sec:concl}
This study and its companion paper examine various aspects associated with the study of the submillimeter galaxy magnification bias, with a specific focus on the tomographic scenario.
It is akin to the work carried out by \cite{BON21}, but using the updated data from Paper I. 
The objective is to refine the methodology employed in constraining the free parameters of the HOD model, that is, $M_{min}$, $M_1$, and $\alpha$, and the cosmological parameters $\Omega_M$, $\sigma_8$, and $h$ of a flat $\Lambda$CDM model on a tomographic scenario. 

Due to the large number of parameters that must be jointly analysed in tomography, a significant amount of computational time is required to accomplish this kind of task. Therefore, one of the aims of this work is to explore ways to optimise CPU time and strategies for analysing different redshift bins without compromising the precision of the measurements.
Various data sets have been explored to this end, which involve the combination of different redshift bins and the inclusion or exclusion of the G15 field suspected to have an anomalous behaviour. Each data set consists in the measurement of the cross-correlation function between background H-ATLAS submillimeter galaxies and foreground GAMA galaxies within a tomographic setup.

As in \citet{BON21} and \citet{CUE22}, we divided the entire redshift range into four bins: bin 1 ($0.1<z<0.2$), bin 2 ($0.2<z<0.3$), bin 3 ($0.3<z<0.5$), and bin 4 ($0.5<z<0.8$). In the "base case" and "w/o G15" scenarios, we examine all four bins together, including SGP and all the GAMA fields and excluding the G15 field, respectively.

We first compare the "base case" which makes use of the mean-redshift approximation with the "full model" case. This approximation reduces the computational time by about a factor of 10 and it is of crucial importance when performing tomographic analyses, given the large number of parameters to be estimated in such a configuration. The comparative plots confirm that given the thinness of the redshift bins used for tomography, the "base case" results are in good agreement with the "full model" case. 

The marginalised posterior distributions exhibit a systematic increase in $\log{M_{min}}$ from bin 1 to bin 4 as redshift increases, while the distributions of $\log{M_1}$ show relatively consistent values across the four bins. In addition to the HOD parameters, we compare the inferred cosmological parameters with the results obtained from the non-tomographic runs in Paper I and the tomographic analysis conducted by \cite{BON21}. The impact of our new methodology, particularly the reduction in measurement uncertainties, becomes evident when examining $\Omega_M$, as indicated by the considerably narrower posterior distributions compared to the findings of \citet{BON21}. However, it is worth noting that our results exhibit a preference for lower values of $\Omega_M$, with the posterior distributions deviating noticeably from the predictions of the standard model.

In relation to the "w/o G15" scenario, our study confirms the findings of Paper I, which demonstrate that excluding the G15 field results in a reduction of the cross-correlation signal at larger angular separations. While this effect is observed across all redshift bins, it is particularly pronounced in bins 1 and 2, affecting all angular scales. As a result, the most significant deviation from the "base case" is observed in the $\log{M_{min}}$ distribution for these bins, which shifts towards lower and higher values, respectively. These observations suggest that the peculiar behavior associated with the G15 field might be attributed to sample variance within the large-scale structure along the line of sight, particularly within a specific range of redshifts.

For the sake of computational efficiency, we narrow our focus to the joint analysis of two bin pairs: "bins 1+4" (the first and fourth bin) and "bins 2+3" (the second and third bin). Remarkably, the HOD parameters obtained from these analyses are consistent with those of the "base case". The marginalised posterior distribution of $\Omega_M$ exhibits strong agreement across all three cases, peaking at 0.16. However, the $\sigma_8$ posteriors reveal an intriguing disparity: the distribution for "bins 2+3" is shifted towards lower values (peaking at 0.61), whereas "bins 1+4" yields higher values (peaking at 0.82), and the "base case" falls in between (peaking at 0.70). This discrepancy suggests the presence of sample variance in redshift, which can be mitigated by increasing the number of redshift bins. Thus, the sample variance issue identified in Paper I, which affected the spatial distribution, also manifests as sample variance in the redshift domain, as implied by the analysis of the G15 effect.

We investigate the impact of increasing the number of redshift bins in two cases: "six bins" (maintaining the same redshift range) and "six bins-WR" (extending the range with additional bins). In the "six bins" case, the data are affected by poorer statistics due to the same number of sources being spread over more bins. In the "six bins-WR" case, the central bins resemble the "base case," while the newly added first and last bins have larger error bars and significantly poorer statistics.

The $\log{M_{min}}$ distributions are well-constrained, with peaks increasing with redshift, as expected. In the "six bins-WR" case, the values show a convergence around the value 12.0 for most bins. 
The marginalised posterior distributions of $\log{M_1}$ tend to peak around the value 14 for all the bins in both cases, although there is more dispersion in the "six bins-WR" case.
The $\Omega_M$ distribution shifts towards higher values, peaking at 0.20 in the "six bins" case and 0.32 in the "six bins-WR" case, compared to the "base case" peak at 0.16. Increasing the number of redshift bins has a similar effect as removing the G15 field, suggesting the dominance of the cosmological model over sample variance. However, individual bin fits are not perfect, resulting in poorer fits for the most biased samples. Extending the redshift range improves statistics and produces tighter constraints compared to the "w/o G15" case.

The $\sigma_8$ posterior distribution for the "six bins-WR" case is compatible with the "base case" results but shows a wider distribution towards lower values, with a peak (mean) value of 0.66 (0.69). In contrast, the "six bins" case exhibits a clear peak at 0.99. This difference is due to reduced statistics in the two central bins compared to bins 1 and 4, producing estimates more similar to the "bins 1+4" case. The parameter $h$ remains unconstrained, but the posterior distributions show tight upper limit for the "six bins-WR" case, consistent with previous findings in Paper I.

As a concluding remark, this study helps to refine the methodology for tomographic analyses and highlights the importance of considering sample variance and redshift binning in obtaining precise measurements of HOD and cosmological parameters. The spatial sample variance as well as (most likely) the redshift one can be minimised by increasing the number of independent fields, for instance, by incorporating additional Herschel wide area surveys such us the \textit{Herschel} Multi-tiered Extragalactic Survey \citep[HerMES;][]{Oli12} or the Great Observatories Origins Deep Survey-\textit{Herschel} \citep[GOODS-H;][]{Elb11}. 
However, our results also indicate that enhancing both the number of redshift bins and the redshift range yields more accurate and robust outcomes, mitigating the potential impact of both types of sample variance. 
To implement this second approach effectively, it is important to adopt updated foreground catalogues with improved statistics for our purposes, such as the Dark Energy Survey \citep[DES;][]{DES05} or the future Euclid mission \citep{EuclidI_22}.

\begin{acknowledgements}
LB, JGN, JMC and DC acknowledge the PID2021-125630NB-I00 project funded by MCIN/AEI/10.13039/501100011033/FEDER, UE. LB also acknowledges the CNS2022-135748 project funded by MCIN/AEI/10.13039/501100011033 and by the EU “NextGenerationEU/PRTR”. JMC also acknowledges financial support from the SV-PA-21-AYUD/2021/51301 project.\\
We deeply acknowledge the CINECA award under the ISCRA initiative, for the availability of high performance computing resources and support. In particular the projects `SIS22\_lapi', `SIS23\_lapi' in the framework `Convenzione triennale SISSA-CINECA'.\\
The Herschel-ATLAS is a project with Herschel, which is an ESA space observatory with science instruments provided by European-led Principal Investigator consortia and with important participation from NASA. The H-ATLAS web- site is http://www.h-atlas.org. GAMA is a joint European- Australasian project based around a spectroscopic campaign using the Anglo- Australian Telescope. The GAMA input catalogue is based on data taken from the Sloan Digital Sky Survey and the UKIRT Infrared Deep Sky Survey. Complementary imaging of the GAMA regions is being obtained by a number of independent survey programs including GALEX MIS, VST KIDS, VISTA VIKING, WISE, Herschel-ATLAS, GMRT and ASKAP providing UV to radio coverage. GAMA is funded by the STFC (UK), the ARC (Australia), the AAO, and the participating institutions. The GAMA web- site is: http://www.gama-survey.org/.\\
This research has made use of the python packages \texttt{ipython} \citep{ipython}, \texttt{matplotlib} \citep{matplotlib} and \texttt{Scipy} \citep{scipy}.
\end{acknowledgements}

\bibliographystyle{aa} 
\bibliography{paperIII} 

\begin{thebibliography}{46}
\expandafter\ifx\csname natexlab\endcsname\relax\def\natexlab#1{#1}\fi

\bibitem[{{Amon} {et~al.}(2023){Amon}, {Robertson}, {Miyatake}, {Heymans}, {White},
  {DeRose}, {Yuan}, {Wechsler}, {Varga}, {Bocquet}, {Dvornik}, {More}, {Ross},
  {Hoekstra}, {Alarcon}, {Asgari}, {Blazek}, {Campos}, {Chen}, {Choi}, {Crocce},
  {Diehl}, {Doux}, {Eckert}, {Elvin-Poole}, {Everett}, {Fert{\'e}}, {Gatti},
  {Giannini}, {Gruen}, {Gruendl}, {Hartley}, {Herner}, {Hildebrandt}, {Huang}, {Huff},
  {Joachimi}, {Lee}, {MacCrann}, {Myles}, {Navarro-Alsina}, {Nishimichi}, {Prat}, {Secco},
  {Sevilla-Noarbe}, {Sheldon}, {Shin}, {Tr{\"o}ster}, {Troxel}, {Tutusaus}, {Wright},
  {Yin}, {Aguena}, {Allam}, {Annis}, {Bacon}, {Bilicki}, {Brooks}, {Burke}, {Carnero Rosell},
  {Carretero}, {Castander}, {Cawthon}, {Costanzi}, {da Costa}, {Pereira}, {de Jong},
  {De Vicente}, {Desai}, {Dietrich}, {Doel}, {Ferrero}, {Frieman}, {Garc{\'\i}a-Bellido},
  {Gerdes}, {Gschwend}, {Gutierrez}, {Hinton}, {Hollowood}, {Honscheid}, {Huterer},
  {Kannawadi}, {Kuehn}, {Kuropatkin}, {Lahav}, {Lima}, {Maia}, {Marshall}, {Menanteau},
  {Miquel}, {Mohr}, {Morgan}, {Muir}, {Paz-Chinch{\'o}n}, {Pieres}, {Plazas Malag{\'o}n},
  {Porredon}, {Rodriguez-Monroy}, {Roodman}, {Sanchez}, {Serrano}, {Shan}, {Suchyta},
  {Swanson}, {Tarle}, {Thomas}, {To}, \& {Zhang}}]{AMO23}
{Amon}, A., {Robertson}, N.~C., {Miyatake}, H., {et~al.} 2023, \mnras, 518, 1       

\bibitem[{Bacon {et~al.}(2000)Bacon, Refregier, \& Ellis}]{BA00}
Bacon, D.~J., Refregier, A.~R., \& Ellis, R.~S. 2000, Monthly Notices of the
  Royal Astronomical Society, 318, 625

\bibitem[{{Bakx} {et~al.}(2020){Bakx}, {Eales}, \& {Amvrosiadis}}]{Bak20}
{Bakx}, T. J.~L.~C., {Eales}, S., \& {Amvrosiadis}, A. 2020, \mnras, 493, 4276

\bibitem[{{Bianchini} {et~al.}(2015){Bianchini}, {Bielewicz}, {Lapi},
  {Gonzalez-Nuevo}, {Baccigalupi}, {de Zotti}, {Danese}, {Bourne}, {Cooray},
  {Dunne}, {Dye}, {Eales}, {Ivison}, {Maddox}, {Negrello}, {Scott}, {Smith}, \&
  {Valiante}}]{BIA15}
{Bianchini}, F., {Bielewicz}, P., {Lapi}, A., {et~al.} 2015, \apj, 802, 64

\bibitem[{{Bianchini} {et~al.}(2016){Bianchini}, {Lapi}, {Calabrese},
  {Bielewicz}, {Gonzalez-Nuevo}, {Baccigalupi}, {Danese}, {de Zotti}, {Bourne},
  {Cooray}, {Dunne}, {Eales}, \& {Valiante}}]{BIA16}
{Bianchini}, F., {Lapi}, A., {Calabrese}, M., {et~al.} 2016, \apj, 825, 24

\bibitem[{{Blain}(1996)}]{BLA96}
{Blain}, A.~W. 1996, \mnras, 283, 1340

\bibitem[{{Bonavera} {et~al.}(2022){Bonavera}, {Cueli}, \&
  {Gonzalez-Nuevo}}]{BON22}
{Bonavera}, L., {Cueli}, M.~M., \& {Gonzalez-Nuevo}, J. 2022, Proceedings of
  the MG16 Meeting on General Relativity, R. Ruffini \& G. Vereshchagin eds.,
  World Scientific., arXiv:2112.02959

\bibitem[{{Bonavera} {et~al.}(2021){Bonavera}, {Cueli}, {Gonz{\'a}lez-Nuevo},
  {Ronconi}, {Migliaccio}, {Lapi}, {Casas}, \& {Crespo}}]{BON21}
{Bonavera}, L., {Cueli}, M.~M., {Gonz{\'a}lez-Nuevo}, J., {et~al.} 2021, \aap,
  656, A99

\bibitem[{{Bonavera} {et~al.}(2020){Bonavera}, {Gonz{\'a}lez-Nuevo}, {Cueli},
  {Ronconi}, {Migliaccio}, {Dunne}, {Lapi}, {Maddox}, \& {Negrello}}]{BON20}
{Bonavera}, L., {Gonz{\'a}lez-Nuevo}, J., {Cueli}, M.~M., {et~al.} 2020, \aap,
  639, A128

\bibitem[{{Bonavera} {et~al.}(2019){Bonavera}, {Gonz{\'a}lez-Nuevo},
  {Su{\'a}rez G{\'o}mez}, {Lapi}, {Bianchini}, {Negrello}, {D{\'\i}ez Alonso},
  {Santos}, \& {de Cos Juez}}]{BON19}
{Bonavera}, L., {Gonz{\'a}lez-Nuevo}, J., {Su{\'a}rez G{\'o}mez}, S.~L.,
  {et~al.} 2019, \jcap, 2019, 021

\bibitem[{{Bussmann} {et~al.}(2012){Bussmann}, {Gurwell}, {Fu}, {Smith}, {Dye},
  {Auld}, {Baes}, {Baker}, {Bonfield}, {Cava}, {Clements}, {Cooray}, {Coppin},
  {Dannerbauer}, {Dariush}, {De Zotti}, {Dunne}, {Eales}, {Fritz}, {Hopwood},
  {Ibar}, {Ivison}, {Jarvis}, {Kim}, {Leeuw}, {Maddox}, {Micha{\l}owski},
  {Negrello}, {Pascale}, {Pohlen}, {Riechers}, {Rigby}, {Scott}, {Temi}, {Van
  der Werf}, {Wardlow}, {Wilner}, \& {Verma}}]{BUS12}
{Bussmann}, R.~S., {Gurwell}, M.~A., {Fu}, H., {et~al.} 2012, \apj, 756, 134

\bibitem[{{Bussmann} {et~al.}(2013){Bussmann}, {P{\'e}rez-Fournon}, {Amber},
  {Calanog}, {Gurwell}, {Dannerbauer}, {De Bernardis}, {Fu}, {Harris}, {Krips},
  {Lapi}, {Maiolino}, {Omont}, {Riechers}, {Wardlow}, {Baker}, {Birkinshaw},
  {Bock}, {Bourne}, {Clements}, {Cooray}, {De Zotti}, {Dunne}, {Dye}, {Eales},
  {Farrah}, {Gavazzi}, {Gonz{\'a}lez Nuevo}, {Hopwood}, {Ibar}, {Ivison},
  {Laporte}, {Maddox}, {Mart{\'\i}nez-Navajas}, {Michalowski}, {Negrello},
  {Oliver}, {Roseboom}, {Scott}, {Serjeant}, {Smith}, {Smith}, {Streblyanska},
  {Valiante}, {van der Werf}, {Verma}, {Vieira}, {Wang}, \& {Wilner}}]{BUS13}
{Bussmann}, R.~S., {P{\'e}rez-Fournon}, I., {Amber}, S., {et~al.} 2013, \apj,
  779, 25 

\bibitem[{{Cacciato} {et~al.}(2013), {Cacciato}, {van den Bosch}, {More}, {Mo}, \&  {Yang}}]{CACC13}
{Cacciato}, M., {van den Bosch}, F.~C., {More}, S., {et~al.} 2013, \mnras, 430, 2
  
\bibitem[{{Calanog} {et~al.}(2014){Calanog}, {Fu}, {Cooray}, {Wardlow}, {Ma},
  {Amber}, {Baker}, {Baes}, {Bock}, {Bourne}, {Bussmann}, {Casey}, {Chapman},
  {Clements}, {Conley}, {Dannerbauer}, {De Zotti}, {Dunne}, {Dye}, {Eales},
  {Farrah}, {Furlanetto}, {Harris}, {Ivison}, {Kim}, {Maddox}, {Magdis},
  {Messias}, {Micha{\l}owski}, {Negrello}, {Nightingale}, {O'Bryan}, {Oliver},
  {Riechers}, {Scott}, {Serjeant}, {Simpson}, {Smith}, {Timmons}, {Thacker},
  {Valiante}, \& {Vieira}}]{CAL14}
{Calanog}, J.~A., {Fu}, H., {Cooray}, A., {et~al.} 2014, \apj, 797, 138

\bibitem[{{Cooray} \& {Sheth}(2002)}]{COO02}
  {Cooray}, A. \& {Sheth}, R. 2002, \physrep, 372, 1

\bibitem[{{Cueli} {et~al.}(2023){Cueli}, {Gonz{\'a}lez-Nuevo}, {Bonavera}, {Lapi},
  {Crespo}, \& {Casas}}]{Cue23}
  {Cueli}, M.~M., {Gonz{\'a}lez-Nuevo}, J., {Bonavera}, L., {et~al.} 2024, arXiv:2305.13835

\bibitem[{{Cueli} {et~al.}(2022){Cueli}, {Bonavera}, {Gonz{\'a}lez-Nuevo},
  {Crespo}, {Casas}, \& {Lapi}}]{CUE22}
{Cueli}, M.~M., {Bonavera}, L., {Gonz{\'a}lez-Nuevo}, J., {et~al.} 2022, \aap,
  662, A44

\bibitem[{{Cueli} {et~al.}(2021){Cueli}, {Bonavera}, {Gonz{\'a}lez-Nuevo}, \&
  {Lapi}}]{CUE21}
{Cueli}, M.~M., {Bonavera}, L., {Gonz{\'a}lez-Nuevo}, J., \& {Lapi}, A. 2021,
  \aap, 645, A126

\bibitem[{{Dvornik} {et~al.}(2023){Dvornik}, {Heymans}, {Asgari}, {Mahony},
  {Joachimi}, {Bilicki}, {Chisari}, {Hildebrandt}, {Hoekstra}, {Johnston},
  {Kuijken}, {Mead}, {Miyatake}, {Nishimichi}, {Reischke}, {Unruh}, \& {Wright}}]{DVO23}
{Dvornik}, A., {Heymans}, C., {Asgari}, M., {et~al.} 2023, \aap, 675, A189

\bibitem[{{Dvornik} {et~al.}(2018){Dvornik}, {Hoekstra}, {Kuijken},
  {Schneider}, {Amon}, {Nakajima}, {Viola}, {Choi}, {Erben}, {Farrow},
  {Heymans}, {Hildebrandt}, {Sif{\'o}n}, \& {Wang}}]{DVORNIK18}
{Dvornik}, A., {Hoekstra}, H., {Kuijken}, K., {et~al.} 2018, \mnras, 479, 1
  
\bibitem[{{Eales} {et~al.}(2010){Eales}, {Dunne}, {Clements}, {Cooray}, {De
  Zotti}, {Dye}, {Ivison}, {Jarvis}, {Lagache}, {Maddox}, {Negrello},
  {Serjeant}, {Thompson}, {Van Kampen}, {Amblard}, {Andreani}, {Baes},
  {Beelen}, {Bendo}, {Benford}, {Bertoldi}, {Bock}, {Bonfield}, {Boselli},
  {Bridge}, {Buat}, {Burgarella}, {Carlberg}, {Cava}, {Chanial}, {Charlot},
  {Christopher}, {Coles}, {Cortese}, {Dariush}, {da Cunha}, {Dalton}, {Danese},
  {Dannerbauer}, {Driver}, {Dunlop}, {Fan}, {Farrah}, {Frayer}, {Frenk},
  {Geach}, {Gardner}, {Gomez}, {Gonz{\'a}lez-Nuevo}, {Gonz{\'a}lez-Solares},
  {Griffin}, {Hardcastle}, {Hatziminaoglou}, {Herranz}, {Hughes}, {Ibar},
  {Jeong}, {Lacey}, {Lapi}, {Lawrence}, {Lee}, {Leeuw}, {Liske},
  {L{\'o}pez-Caniego}, {M{\"u}ller}, {Nandra}, {Panuzzo}, {Papageorgiou},
  {Patanchon}, {Peacock}, {Pearson}, {Phillipps}, {Pohlen}, {Popescu},
  {Rawlings}, {Rigby}, {Rigopoulou}, {Robotham}, {Rodighiero}, {Sansom},
  {Schulz}, {Scott}, {Smith}, {Sibthorpe}, {Smail}, {Stevens}, {Sutherland},
  {Takeuchi}, {Tedds}, {Temi}, {Tuffs}, {Trichas}, {Vaccari}, {Valtchanov},
  {van der Werf}, {Verma}, {Vieria}, {Vlahakis}, \& {White}}]{EAL10}
{Eales}, S., {Dunne}, L., {Clements}, D., {et~al.} 2010, \pasp, 122, 499

\bibitem[{{Elbaz} {et~al.}(2011){Elbaz}, {Dickinson}, {Hwang},
  {D{\'\i}az-Santos}, {Magdis}, {Magnelli}, {Le Borgne}, {Galliano},
  {Pannella}, {Chanial}, {Armus}, {Charmandaris}, {Daddi}, {Aussel}, {Popesso},
  {Kartaltepe}, {Altieri}, {Valtchanov}, {Coia}, {Dannerbauer}, {Dasyra},
  {Leiton}, {Mazzarella}, {Alexander}, {Buat}, {Burgarella}, {Chary}, {Gilli},
  {Ivison}, {Juneau}, {Le Floc'h}, {Lutz}, {Morrison}, {Mullaney}, {Murphy},
  {Pope}, {Scott}, {Brodwin}, {Calzetti}, {Cesarsky}, {Charlot}, {Dole},
  {Eisenhardt}, {Ferguson}, {F{\"o}rster Schreiber}, {Frayer}, {Giavalisco},
  {Huynh}, {Koekemoer}, {Papovich}, {Reddy}, {Surace}, {Teplitz}, {Yun}, \&
  {Wilson}}]{Elb11}
{Elbaz}, D., {Dickinson}, M., {Hwang}, H.~S., {et~al.} 2011, \aap, 533, A119

\bibitem[{{Euclid Collaboration} {et~al.}(2022){Euclid Collaboration},
  {Scaramella}, {Amiaux}, {Mellier}, {Burigana}, {Carvalho}, {Cuillandre}, {Da
  Silva}, {Derosa}, {Dinis}, {Maiorano}, {Maris}, {Tereno}, {Laureijs},
  {Boenke}, {Buenadicha}, {Dupac}, {Gaspar Venancio}, {G{\'o}mez-{\'A}lvarez},
  {Hoar}, {Lorenzo Alvarez}, {Racca}, {Saavedra-Criado}, {Schwartz}, {Vavrek},
  {Schirmer}, {Aussel}, {Azzollini}, {Cardone}, {Cropper}, {Ealet}, {Garilli},
  {Gillard}, {Granett}, {Guzzo}, {Hoekstra}, {Jahnke}, {Kitching}, {Maciaszek},
  {Meneghetti}, {Miller}, {Nakajima}, {Niemi}, {Pasian}, {Percival},
  {Pottinger}, {Sauvage}, {Scodeggio}, {Wachter}, {Zacchei}, {Aghanim},
  {Amara}, {Auphan}, {Auricchio}, {Awan}, {Balestra}, {Bender}, {Bodendorf},
  {Bonino}, {Branchini}, {Brau-Nogue}, {Brescia}, {Candini}, {Capobianco},
  {Carbone}, {Carlberg}, {Carretero}, {Casas}, {Castander}, {Castellano},
  {Cavuoti}, {Cimatti}, {Cledassou}, {Congedo}, {Conselice}, {Conversi},
  {Copin}, {Corcione}, {Costille}, {Courbin}, {Degaudenzi}, {Douspis},
  {Dubath}, {Duncan}, {Dusini}, {Farrens}, {Ferriol}, {Fosalba}, {Fourmanoit},
  {Frailis}, {Franceschi}, {Franzetti}, {Fumana}, {Gillis}, {Giocoli},
  {Grazian}, {Grupp}, {Haugan}, {Holmes}, {Hormuth}, {Hudelot}, {Kermiche},
  {Kiessling}, {Kilbinger}, {Kohley}, {Kubik}, {K{\"u}mmel}, {Kunz},
  {Kurki-Suonio}, {Lahav}, {Ligori}, {Lilje}, {Lloro}, {Mansutti}, {Marggraf},
  {Markovic}, {Marulli}, {Massey}, {Maurogordato}, {Melchior}, {Merlin},
  {Meylan}, {Mohr}, {Moresco}, {Morin}, {Moscardini}, {Munari}, {Nichol},
  {Padilla}, {Paltani}, {Peacock}, {Pedersen}, {Pettorino}, {Pires}, {Poncet},
  {Popa}, {Pozzetti}, {Raison}, {Rebolo}, {Rhodes}, {Rix}, {Roncarelli},
  {Rossetti}, {Saglia}, {Schneider}, {Schrabback}, {Secroun}, {Seidel},
  {Serrano}, {Sirignano}, {Sirri}, {Skottfelt}, {Stanco}, {Starck},
  {Tallada-Cresp{\'\i}}, {Tavagnacco}, {Taylor}, {Teplitz}, {Toledo-Moreo},
  {Torradeflot}, {Trifoglio}, {Valentijn}, {Valenziano}, {Verdoes Kleijn},
  {Wang}, {Welikala}, {Weller}, {Wetzstein}, {Zamorani}, {Zoubian}, {Andreon},
  {Baldi}, {Bardelli}, {Boucaud}, {Camera}, {Di Ferdinando}, {Fabbian},
  {Farinelli}, {Galeotta}, {Graci{\'a}-Carpio}, {Maino}, {Medinaceli}, {Mei},
  {Neissner}, {Polenta}, {Renzi}, {Romelli}, {Rosset}, {Sureau}, {Tenti},
  {Vassallo}, {Zucca}, {Baccigalupi}, {Balaguera-Antol{\'\i}nez}, {Battaglia},
  {Biviano}, {Borgani}, {Bozzo}, {Cabanac}, {Cappi}, {Casas}, {Castignani},
  {Colodro-Conde}, {Coupon}, {Courtois}, {Cuby}, {de la Torre}, {Desai},
  {Dole}, {Fabricius}, {Farina}, {Ferreira}, {Finelli}, {Flose-Reimberg},
  {Fotopoulou}, {Ganga}, {Gozaliasl}, {Hook}, {Keihanen}, {Kirkpatrick},
  {Liebing}, {Lindholm}, {Mainetti}, {Martinelli}, {Martinet}, {Maturi},
  {McCracken}, {Metcalf}, {Morgante}, {Nightingale}, {Nucita}, {Patrizii},
  {Potter}, {Riccio}, {S{\'a}nchez}, {Sapone}, {Schewtschenko}, {Schultheis},
  {Scottez}, {Teyssier}, {Tutusaus}, {Valiviita}, {Viel}, {Vriend}, \&
  {Whittaker}}]{EuclidI_22}
{Euclid Collaboration}, {Scaramella}, R., {Amiaux}, J., {et~al.} 2022, \aap,
  662, A112

\bibitem[{{Fu} {et~al.}(2012){Fu}, {Jullo}, {Cooray}, {Bussmann}, {Ivison},
  {P{\'e}rez-Fournon}, {Djorgovski}, {Scoville}, {Yan}, {Riechers}, {Aguirre},
  {Auld}, {Baes}, {Baker}, {Bradford}, {Cava}, {Clements}, {Dannerbauer},
  {Dariush}, {De Zotti}, {Dole}, {Dunne}, {Dye}, {Eales}, {Frayer}, {Gavazzi},
  {Gurwell}, {Harris}, {Herranz}, {Hopwood}, {Hoyos}, {Ibar}, {Jarvis}, {Kim},
  {Leeuw}, {Lupu}, {Maddox}, {Mart{\'\i}nez-Navajas}, {Micha{\l}owski},
  {Negrello}, {Omont}, {Rosenman}, {Scott}, {Serjeant}, {Smail}, {Swinbank},
  {Valiante}, {Verma}, {Vieira}, {Wardlow}, \& {van der Werf}}]{FU12}
{Fu}, H., {Jullo}, E., {Cooray}, A., {et~al.} 2012, \apj, 753, 134

\bibitem[{{Gonz{\'a}lez-Nuevo} {et~al.}(2023){Gonz{\'a}lez-Nuevo}, {Bonavera}, {Cueli},
   {Crespo},  \& {Casas}}]{GON23}
{Gonz{\'a}lez-Nuevo}, J., {Bonavera}, L., {Cueli}, M.~M., {et~al.} 2023, arXiv:2305.13834

\bibitem[{{Gonz{\'a}lez-Nuevo} {et~al.}(2021){Gonz{\'a}lez-Nuevo}, {Cueli},
  {Bonavera}, {Lapi}, {Migliaccio}, {Arg{\"u}eso}, \& {Toffolatti}}]{GON21}
{Gonz{\'a}lez-Nuevo}, J., {Cueli}, M.~M., {Bonavera}, L., {et~al.} 2021, \aap,
  646, A152

\bibitem[{{Gonz{\'a}lez-Nuevo} {et~al.}(2017){Gonz{\'a}lez-Nuevo}, {Lapi},
  {Bonavera}, {Danese}, {de Zotti}, {Negrello}, {Bourne}, {Cooray}, {Dunne},
  {Dye}, {Eales}, {Furlanetto}, {Ivison}, {Loveday}, {Maddox}, {Smith}, \&
  {Valiante}}]{GON17}
{Gonz{\'a}lez-Nuevo}, J., {Lapi}, A., {Bonavera}, L., {et~al.} 2017, \jcap,
  2017, 024

\bibitem[{{Gonz{\'a}lez-Nuevo} {et~al.}(2012){Gonz{\'a}lez-Nuevo}, {Lapi},
  {Fleuren}, {Bressan}, {Danese}, {De Zotti}, {Negrello}, {Cai}, {Fan},
  {Sutherland}, {Baes}, {Baker}, {Clements}, {Cooray}, {Dannerbauer}, {Dunne},
  {Dye}, {Eales}, {Frayer}, {Harris}, {Ivison}, {Jarvis}, {Micha{\l}owski},
  {L{\'o}pez-Caniego}, {Rodighiero}, {Rowlands}, {Serjeant}, {Scott}, {van der
  Werf}, {Auld}, {Buttiglione}, {Cava}, {Dariush}, {Fritz}, {Hopwood}, {Ibar},
  {Maddox}, {Pascale}, {Pohlen}, {Rigby}, {Smith}, \& {Temi}}]{GON12}
{Gonz{\'a}lez-Nuevo}, J., {Lapi}, A., {Fleuren}, S., {et~al.} 2012, \apj, 749,
  65

\bibitem[{{Goodman} \& {Weare}(2010)}]{GOO10}
{Goodman}, J. \& {Weare}, J. 2010, Communications in Applied Mathematics and
  Computational Science, 5, 65

\bibitem[{{Herranz}(2001)}]{HER01}
{Herranz}, D. 2001, in Cosmological Physics with Gravitational Lensing, ed.
  J.~{Tran Thanh Van}, Y.~{Mellier}, \& M.~{Moniez}, 197

\bibitem[{{Hildebrandt} {et~al.}(2013){Hildebrandt}, {van Waerbeke}, {Scott},
  {B{\'e}thermin}, {Bock}, {Clements}, {Conley}, {Cooray}, {Dunlop}, {Eales},
  {Erben}, {Farrah}, {Franceschini}, {Glenn}, {Halpern}, {Heinis}, {Ivison},
  {Marsden}, {Oliver}, {Page}, {P{\'e}rez-Fournon}, {Smith}, {Rowan-Robinson},
  {Valtchanov}, {van der Burg}, {Vieira}, {Viero}, \& {Wang}}]{HIL13}
{Hildebrandt}, H., {van Waerbeke}, L., {Scott}, D., {et~al.} 2013, \mnras, 429,
  3230

\bibitem[{Hunter (2007)}]{matplotlib}
Hunter, J.~D. 2007, Computing in Science and Engineering, 9, 3

\bibitem[{Jarvis (2015)Javis}]{TreeCorr}
Jarvis, M. 2015, {TreeCorr}: Two-point correlation functions

\bibitem[{Jones {et~al.}(2001)Jones, Oliphant, Peterson, {et~al.}}]{scipy}
Jones, E., Oliphant, T., Peterson, P., {et~al.} 2001, {SciPy}: Open source
  scientific tools for {Python}

\bibitem[{{Lacasa} \& {Kunz}(2017)}]{LAC17}
{Lacasa}, F. \& {Kunz}, M. 2017, 
  {Landy}, S.~D. \& {Szalay}, A.~S. 1993, \aap, 604, A104
  
\bibitem[{{Landy} \& {Szalay}(1993)}]{LAN93}
{Landy}, S.~D. \& {Szalay}, A.~S. 1993, \apj, 412, 64

\bibitem[{{Lapi} {et~al.}(2012){Lapi}, {Negrello}, {Gonz{\'a}lez-Nuevo},
  {et~al.}}]{LAPI12}
{Lapi}, A., {Negrello}, M., {Gonz{\'a}lez-Nuevo}, J., {et~al.} 2012, The
  Astrophysical Journal, 755, 46

\bibitem[{Lewis(2019)}]{GETDIST}
Lewis, A. 2019 [\eprint[arXiv]{1910.13970}]

\bibitem[{{M{\'e}nard} {et~al.}(2010){M{\'e}nard}, {Scranton}, {Fukugita}, \&
  {Richards}}]{MEN10}
{M{\'e}nard}, B., {Scranton}, R., {Fukugita}, M., \& {Richards}, G. 2010,
  \mnras, 405, 1025

\bibitem[{{Nayyeri} {et~al.}(2016){Nayyeri}, {Keele}, {Cooray}, {Riechers},
  {Ivison}, {Harris}, {Frayer}, {Baker}, {Chapman}, {Eales}, {Farrah}, {Fu},
  {Marchetti}, {Marques-Chaves}, {Martinez-Navajas}, {Oliver}, {Omont},
  {Perez-Fournon}, {Scott}, {Vaccari}, {Vieira}, {Viero}, {Wang}, \&
  {Wardlow}}]{NAY16}
{Nayyeri}, H., {Keele}, M., {Cooray}, A., {et~al.} 2016, \apj, 823, 17

\bibitem[{{Negrello} {et~al.}(2017){Negrello}, {Amber}, {Amvrosiadis}, {Cai},
  {Lapi}, {Gonzalez-Nuevo}, {De Zotti}, {Furlanetto}, {Maddox}, {Allen},
  {Bakx}, {Bussmann}, {Cooray}, {Covone}, {Danese}, {Dannerbauer}, {Fu},
  {Greenslade}, {Gurwell}, {Hopwood}, {Koopmans}, {Napolitano}, {Nayyeri},
  {Omont}, {Petrillo}, {Riechers}, {Serjeant}, {Tortora}, {Valiante}, {Verdoes
  Kleijn}, {Vernardos}, {Wardlow}, {Baes}, {Baker}, {Bourne}, {Clements},
  {Crawford}, {Dye}, {Dunne}, {Eales}, {Ivison}, {Marchetti}, {Micha{\l}owski},
  {Smith}, {Vaccari}, \& {van der Werf}}]{NEG17}
{Negrello}, M., {Amber}, S., {Amvrosiadis}, A., {et~al.} 2017, \mnras, 465,
  3558

\bibitem[{{Negrello} {et~al.}(2010){Negrello}, {Hopwood}, {De Zotti}, {Cooray},
  {Verma}, {Bock}, {Frayer}, {Gurwell}, {Omont}, {Neri}, {Dannerbauer},
  {Leeuw}, {Barton}, {Cooke}, {Kim}, {da Cunha}, {Rodighiero}, {Cox},
  {Bonfield}, {Jarvis}, {Serjeant}, {Ivison}, {Dye}, {Aretxaga}, {Hughes},
  {Ibar}, {Bertoldi}, {Valtchanov}, {Eales}, {Dunne}, {Driver}, {Auld},
  {Buttiglione}, {Cava}, {Grady}, {Clements}, {Dariush}, {Fritz}, {Hill},
  {Hornbeck}, {Kelvin}, {Lagache}, {Lopez-Caniego}, {Gonzalez-Nuevo}, {Maddox},
  {Pascale}, {Pohlen}, {Rigby}, {Robotham}, {Simpson}, {Smith}, {Temi},
  {Thompson}, {Woodgate}, {York}, {Aguirre}, {Beelen}, {Blain}, {Baker},
  {Birkinshaw}, {Blundell}, {Bradford}, {Burgarella}, {Danese}, {Dunlop},
  {Fleuren}, {Glenn}, {Harris}, {Kamenetzky}, {Lupu}, {Maddalena}, {Madore},
  {Maloney}, {Matsuhara}, {Micha{\l}owski}, {Murphy}, {Naylor}, {Nguyen},
  {Popescu}, {Rawlings}, {Rigopoulou}, {Scott}, {Scott}, {Seibert}, {Smail},
  {Tuffs}, {Vieira}, {van der Werf}, \& {Zmuidzinas}}]{NEG10}
{Negrello}, M., {Hopwood}, R., {De Zotti}, G., {et~al.} 2010, Science, 330, 800

\bibitem[{{Negrello} {et~al.}(2007){Negrello}, {Perrotta},
  {Gonz{\'a}lez-Nuevo}, {Silva}, {de Zotti}, {Granato}, {Baccigalupi}, \&
  {Danese}}]{Neg07}
{Negrello}, M., {Perrotta}, F., {Gonz{\'a}lez-Nuevo}, J., {et~al.} 2007,
  \mnras, 377, 1557

\bibitem[{{Oliver} {et~al.}(2012){Oliver}, {Bock}, {Altieri}, {Amblard},
  {Arumugam}, {Aussel}, {Babbedge}, {Beelen}, {B{\'e}thermin}, {Blain},
  {Boselli}, {Bridge}, {Brisbin}, {Buat}, {Burgarella},
  {Castro-Rodr{\'\i}guez}, {Cava}, {Chanial}, {Cirasuolo}, {Clements},
  {Conley}, {Conversi}, {Cooray}, {Dowell}, {Dubois}, {Dwek}, {Dye}, {Eales},
  {Elbaz}, {Farrah}, {Feltre}, {Ferrero}, {Fiolet}, {Fox}, {Franceschini},
  {Gear}, {Giovannoli}, {Glenn}, {Gong}, {Gonz{\'a}lez Solares}, {Griffin},
  {Halpern}, {Harwit}, {Hatziminaoglou}, {Heinis}, {Hurley}, {Hwang}, {Hyde},
  {Ibar}, {Ilbert}, {Isaak}, {Ivison}, {Lagache}, {Le Floc'h}, {Levenson},
  {Faro}, {Lu}, {Madden}, {Maffei}, {Magdis}, {Mainetti}, {Marchetti},
  {Marsden}, {Marshall}, {Mortier}, {Nguyen}, {O'Halloran}, {Omont}, {Page},
  {Panuzzo}, {Papageorgiou}, {Patel}, {Pearson}, {P{\'e}rez-Fournon}, {Pohlen},
  {Rawlings}, {Raymond}, {Rigopoulou}, {Riguccini}, {Rizzo}, {Rodighiero},
  {Roseboom}, {Rowan-Robinson}, {S{\'a}nchez Portal}, {Schulz}, {Scott},
  {Seymour}, {Shupe}, {Smith}, {Stevens}, {Symeonidis}, {Trichas}, {Tugwell},
  {Vaccari}, {Valtchanov}, {Vieira}, {Viero}, {Vigroux}, {Wang}, {Ward},
  {Wardlow}, {Wright}, {Xu}, \& {Zemcov}}]{Oli12}
{Oliver}, S.~J., {Bock}, J., {Altieri}, B., {et~al.} 2012, \mnras, 424, 1614

\bibitem[{P\'erez \& Granger(2007)}]{ipython}
P\'erez, F. \& Granger, B.~E. 2007, Computing in Science and Engineering, 9, 21

\bibitem[{{Planck Collaboration} {et~al.}(2021){Planck Collaboration},
  {Aghanim}, {Akrami}, {Ashdown}, {Aumont}, {Baccigalupi}, {Ballardini},
  {Banday}, {Barreiro}, {Bartolo}, {Basak}, {Battye}, {Benabed}, {Bernard},
  {Bersanelli}, {Bielewicz}, {Bock}, {Bond}, {Borrill}, {Bouchet}, {Boulanger},
  {Bucher}, {Burigana}, {Butler}, {Calabrese}, {Cardoso}, {Carron},
  {Challinor}, {Chiang}, {Chluba}, {Colombo}, {Combet}, {Contreras}, {Crill},
  {Cuttaia}, {de Bernardis}, {de Zotti}, {Delabrouille}, {Delouis}, {Di
  Valentino}, {Diego}, {Dor{\'e}}, {Douspis}, {Ducout}, {Dupac}, {Dusini},
  {Efstathiou}, {Elsner}, {En{\ss}lin}, {Eriksen}, {Fantaye}, {Farhang},
  {Fergusson}, {Fernandez-Cobos}, {Finelli}, {Forastieri}, {Frailis},
  {Fraisse}, {Franceschi}, {Frolov}, {Galeotta}, {Galli}, {Ganga},
  {G{\'e}nova-Santos}, {Gerbino}, {Ghosh}, {Gonz{\'a}lez-Nuevo}, {G{\'o}rski},
  {Gratton}, {Gruppuso}, {Gudmundsson}, {Hamann}, {Handley}, {Hansen},
  {Herranz}, {Hildebrandt}, {Hivon}, {Huang}, {Jaffe}, {Jones}, {Karakci},
  {Keih{\"a}nen}, {Keskitalo}, {Kiiveri}, {Kim}, {Kisner}, {Knox},
  {Krachmalnicoff}, {Kunz}, {Kurki-Suonio}, {Lagache}, {Lamarre}, {Lasenby},
  {Lattanzi}, {Lawrence}, {Le Jeune}, {Lemos}, {Lesgourgues}, {Levrier},
  {Lewis}, {Liguori}, {Lilje}, {Lilley}, {Lindholm}, {L{\'o}pez-Caniego},
  {Lubin}, {Ma}, {Mac{\'\i}as-P{\'e}rez}, {Maggio}, {Maino}, {Mandolesi},
  {Mangilli}, {Marcos-Caballero}, {Maris}, {Martin}, {Martinelli},
  {Mart{\'\i}nez-Gonz{\'a}lez}, {Matarrese}, {Mauri}, {McEwen}, {Meinhold},
  {Melchiorri}, {Mennella}, {Migliaccio}, {Millea}, {Mitra},
  {Miville-Desch{\^e}nes}, {Molinari}, {Montier}, {Morgante}, {Moss}, {Natoli},
  {N{\o}rgaard-Nielsen}, {Pagano}, {Paoletti}, {Partridge}, {Patanchon},
  {Peiris}, {Perrotta}, {Pettorino}, {Piacentini}, {Polastri}, {Polenta},
  {Puget}, {Rachen}, {Reinecke}, {Remazeilles}, {Renzi}, {Rocha}, {Rosset},
  {Roudier}, {Rubi{\~n}o-Mart{\'\i}n}, {Ruiz-Granados}, {Salvati}, {Sandri},
  {Savelainen}, {Scott}, {Shellard}, {Sirignano}, {Sirri}, {Spencer},
  {Sunyaev}, {Suur-Uski}, {Tauber}, {Tavagnacco}, {Tenti}, {Toffolatti},
  {Tomasi}, {Trombetti}, {Valenziano}, {Valiviita}, {Van Tent}, {Vibert},
  {Vielva}, {Villa}, {Vittorio}, {Wandelt}, {Wehus}, {White}, {White},
  {Zacchei}, \& {Zonca}}]{PLA18_VI}
{Planck Collaboration}, {Aghanim}, N., {Akrami}, Y., {et~al.} 2021, \aap, 652,
  C4

\bibitem[{{Renneby} {et~al.}(2020){Renneby}, {Henriques}, {Hilbert}, {Nelson},
    {Vogelsberger}, {Angulo}, {Springel}, \&{Hernquist}}]{REN20}
  {Renneby}, M., {Henriques}, B. M.~B., {Hilbert}, S., {et~al.} 2020, \mnras, 498, 4

  
\bibitem[{Rhodes {et~al.}(2001)Rhodes, Refregier, \& Groth}]{RHO01}
Rhodes, J., Refregier, A., \& Groth, E.~J. 2001, The Astrophysical Journal,
  552, L85

\bibitem[{{Schneider} {et~al.}(1992){Schneider}, {Ehlers}, \& {Falco}}]{SCH92}
{Schneider}, P., {Ehlers}, J., \& {Falco}, E.~E. 1992, {Gravitational Lenses}

\bibitem[{{Scranton} {et~al.}(2005){Scranton}, {M{\'e}nard}, {Richards},
  {Nichol}, {Myers}, {Jain}, {Gray}, {Bartelmann}, {Brunner}, {Connolly},
  {Gunn}, {Sheth}, {Bahcall}, {Brinkman}, {Loveday}, {Schneider}, {Thakar}, \&
  {York}}]{SCR05}
{Scranton}, R., {M{\'e}nard}, B., {Richards}, G.~T., {et~al.} 2005, \apj, 633,
  589

\bibitem[{{The Dark Energy Survey Collaboration}(2005)}]{DES05}
  {The Dark Energy Survey Collaboration}. 2005, arXiv e-prints, astro

\bibitem[{{van Daalen} {et~al.}(2020){van Daalen},  {McCarthy}, \& {Schaye}}]{DAA20}
{van Daalen}, M.~P., {McCarthy}, I.~G., \& {Schaye}, J. 2020,\mnras, 491, 2

\bibitem[{{van Daalen} {et~al.}(2014){van Daalen}, {Schaye}, {McCarthy}, {Booth},
    \& {Dalla Vecchia}}]{DAA14}
{van Daalen}, M.~P., {Schaye}, J., {McCarthy}, I.~G. {et~al.} 2014,\mnras, 440, 4

\bibitem[{{van Uitert} {et~al.}(2016){van Uitert}, {Cacciato}, {Hoekstra},
    {Brouwer}, {Sif{\'o}n}, {Viola}, {Baldry}, {Bland-Hawthorn}, {Brough}, {Brown},
    {Choi}, {Driver}, {Erben}, {Heymans}, {Hildebrandt}, {Joachimi}, {Kuijken},
    {Liske}, {Loveday}, {McFarland}, {Miller}, {Nakajima}, {Peacock}, {Radovich},
    {Robotham}, {Schneider}, {Sikkema}, {Taylor}, \& {Verdoes Kleijn}}]{VANUITERT16}
{van Uitert}, E., {Cacciato}, M., {Hoekstra}, H. {et~al.} 2016, \mnras, 459, 3

\bibitem[{{Van Waerbeke} {et~al.}(2000){Van Waerbeke}, {Mellier}, {Erben},
  {Cuillandre}, {Bernardeau}, {Maoli}, {Bertin}, {McCracken}, {Le F{\`e}vre},
  {Fort}, {Dantel-Fort}, {Jain}, \& {Schneider}}]{VW00}
{Van Waerbeke}, L., {Mellier}, Y., {Erben}, T., {et~al.} 2000, \aap, 358, 30

\bibitem[{{Viola} {et~al.}(2015){Viola}, {Cacciato}, {Brouwer}, {Kuijken}, {Hoekstra},
  {Norberg}, {Robotham}, {van Uitert}, {Alpaslan}, {Baldry}, {Choi}, {de Jong},
  {Driver}, {Erben}, {Grado}, {Graham}, {Heymans}, {Hildebrandt}, {Hopkins},
  {Irisarri}, {Joachimi}, {Loveday}, {Miller}, {Nakajima}, {Schneider},
  {Sif{\'o}n}, \& {Verdoes Kleijn}}]{VIOLA15}
{Viola}, M., {Cacciato}, M., {Brouwer}, M.,  {et~al.} 2015, \mnras, 452, 4


\bibitem[{{Wardlow} {et~al.}(2013){Wardlow}, {Cooray}, {De Bernardis},
  {Amblard}, {Arumugam}, {Aussel}, {Baker}, {B{\'e}thermin}, {Blundell},
  {Bock}, {Boselli}, {Bridge}, {Buat}, {Burgarella}, {Bussmann},
  {Cabrera-Lavers}, {Calanog}, {Carpenter}, {Casey}, {Castro-Rodr{\'\i}guez},
  {Cava}, {Chanial}, {Chapin}, {Chapman}, {Clements}, {Conley}, {Cox},
  {Dowell}, {Dye}, {Eales}, {Farrah}, {Ferrero}, {Franceschini}, {Frayer},
  {Frazer}, {Fu}, {Gavazzi}, {Glenn}, {Gonz{\'a}lez Solares}, {Griffin},
  {Gurwell}, {Harris}, {Hatziminaoglou}, {Hopwood}, {Hyde}, {Ibar}, {Ivison},
  {Kim}, {Lagache}, {Levenson}, {Marchetti}, {Marsden}, {Martinez-Navajas},
  {Negrello}, {Neri}, {Nguyen}, {O'Halloran}, {Oliver}, {Omont}, {Page},
  {Panuzzo}, {Papageorgiou}, {Pearson}, {P{\'e}rez-Fournon}, {Pohlen},
  {Riechers}, {Rigopoulou}, {Roseboom}, {Rowan-Robinson}, {Schulz}, {Scott},
  {Scoville}, {Seymour}, {Shupe}, {Smith}, {Streblyanska}, {Strom},
  {Symeonidis}, {Trichas}, {Vaccari}, {Vieira}, {Viero}, {Wang}, {Xu}, {Yan},
  \& {Zemcov}}]{WAR13}
{Wardlow}, J.~L., {Cooray}, A., {De Bernardis}, F., {et~al.} 2013, \apj, 762,
  59

\bibitem[{{Wittman} {et~al.}(2000){Wittman}, {Tyson}, {Kirkman},
  {Dell'Antonio}, \& {Bernstein}}]{WI00}
{Wittman}, D.~M., {Tyson}, J.~A., {Kirkman}, D., {Dell'Antonio}, I., \&
  {Bernstein}, G. 2000, \nat, 405, 143

\bibitem[{{Zheng} {et~al.}(2005){Zheng}, {Berlind}, {Weinberg}, {Benson},
  {Baugh}, {Cole}, {Dav{\'e}}, {Frenk}, {Katz}, \& {Lacey}}]{ZHE05}
{Zheng}, Z., {Berlind}, A.~A., {Weinberg}, D.~H., {et~al.} 2005, \apj, 633, 791

\end{thebibliography}

\begin{appendix}

\section{Comparison "full model" and "base case"}
\label{app:comp_full_MzA}

\begin{figure}[ht]
\includegraphics[width=0.45\textwidth]{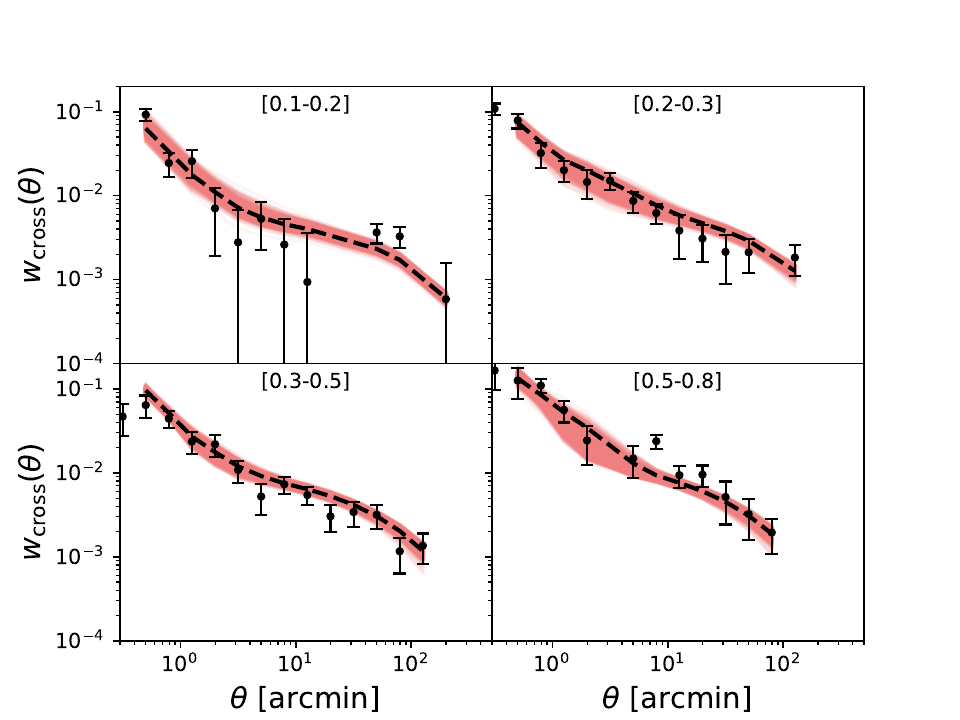}
 \caption{Cross-correlation data (black points) and the results obtained with the full model with no approximations: the black dashed line shows the best fit and the red lines are the sampling posteriors. 
 }
 \label{fig:sampled_tomo}
\end{figure}

This appendix presents the results obtained in the "base case" and the original model, "full model" which is used in previous works \citep[e.g.][]{GON17, BON20, BON21, BON22, CUE21, CUE22}. 
The jointly analysed bins of redshift and the priors are the same in both runs.
Figure~\ref{fig:sampled_tomo} shows the cross-correlation data (black points), the best fit (black dashed line), and the sampled area (in red) obtained with the MCMC. The agreement with the data is good as well as with the findings in Fig. \ref{fig:sampled_tomo_zmed}.

\begin{figure*}
\includegraphics[width=0.45\textwidth]{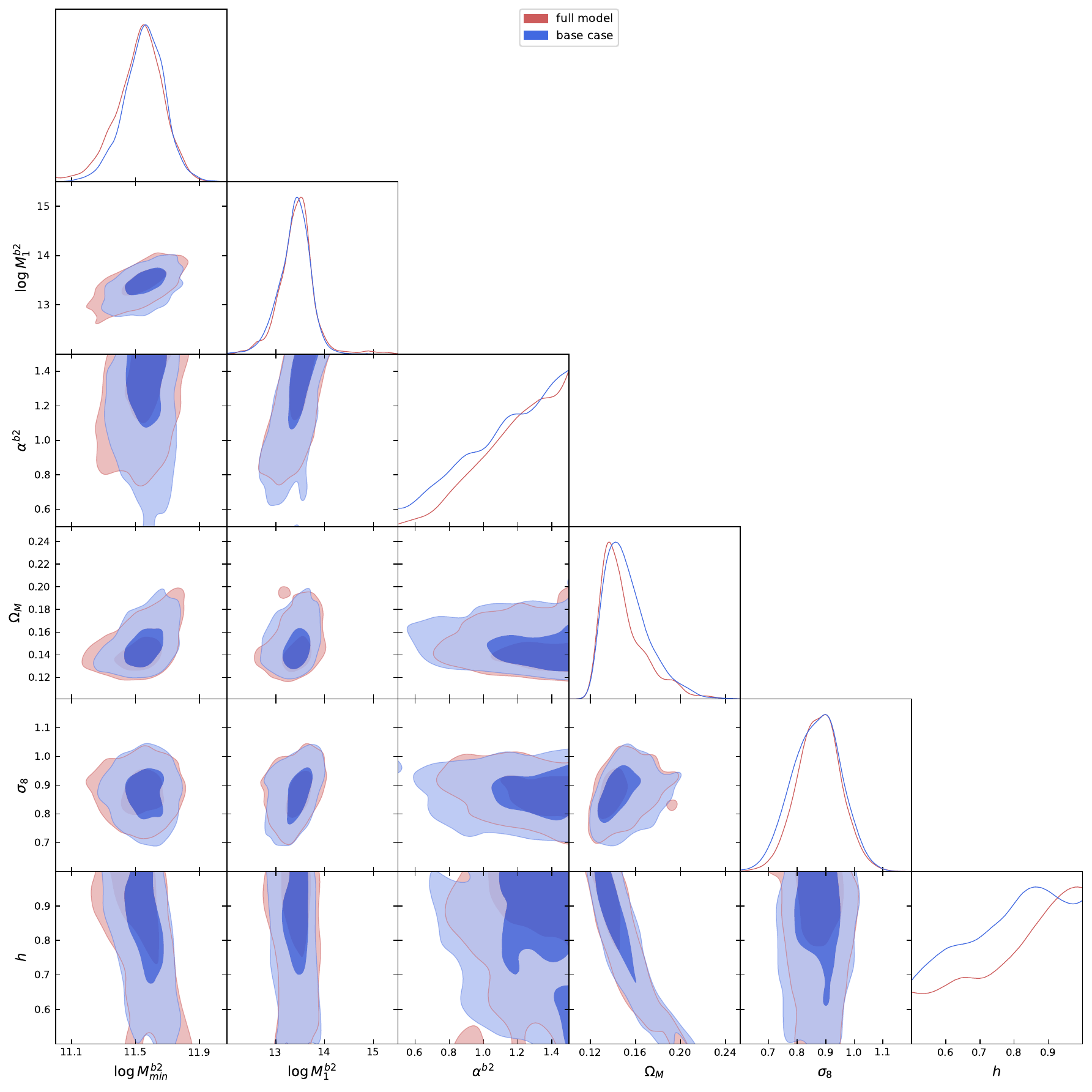}\\
\includegraphics[width=0.45\textwidth]{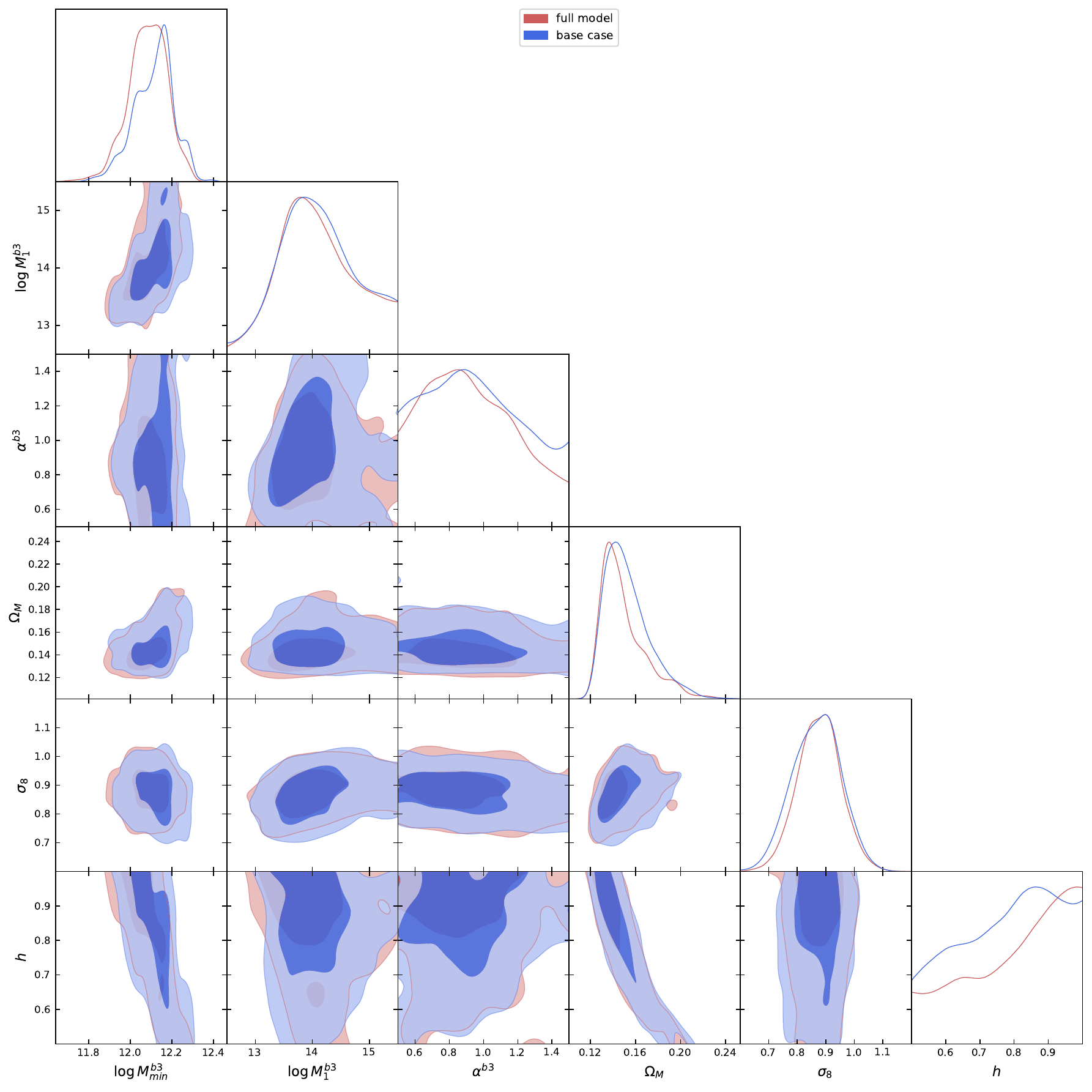}
\includegraphics[width=0.45\textwidth]{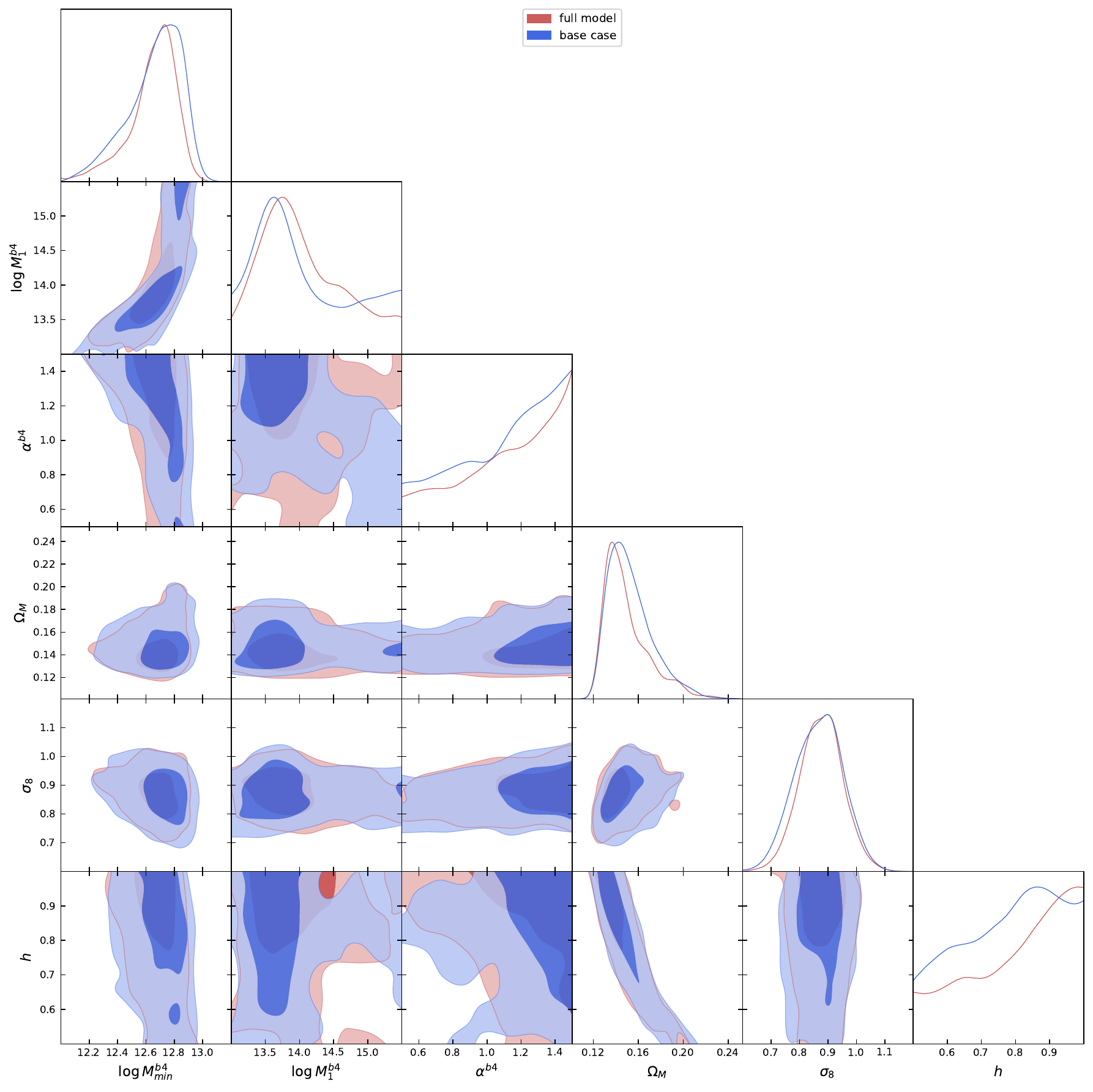}\\
 \caption{Comparison of the marginalised posterior distribution and probability contours (set to 0.393 and 0.865) for the HOD and cosmological parameters obtained in the "base case" (in blue) and in the "full model" case (in red).
 }
 \label{fig:corner_tomo_MzA}
\end{figure*}

Moreover, we show in red the marginalised posterior distributions and probability contours (set to 0.393 and 0.865) in Fig.~\ref{fig:corner_tomo_MzA}. Even if the analysis is jointly performed among the four bins, due to the large amount of parameters to be shown in the figure, we split such results into four panels. Each panel shows the HOD estimated parameters for one bin together with the cosmological parameters, that are the same for the four panels. In particular, bins 1 to 4 are shown from left to right, top to bottom, respectively. 
In this figure we also show in blue the results of the "base case" where we adopt the mean-redshift approximation (detailed in Paper I), which consists of computing the outer integrals in Eq. \ref{eq:w_fb} at the mean redshift of the sample (see Eq. \ref{eq:w_fb_approx}). In the tomographic analysis performed in this work, we can safely adopt such approximation, given the small range of redshift in each in bin. 

The results we find based on these two cases are in very good agreement and the subsequent analyses in this work can be safely performed with it. It provides great advantages from the computational point of view, especially for tomographic analyses, which involve a large number of parameters. 

\FloatBarrier

\section{Posterior distributions of the individual cases}
\label{App:corner_plots}

\begin{figure*}[ht]
\includegraphics[width=0.5\textwidth]{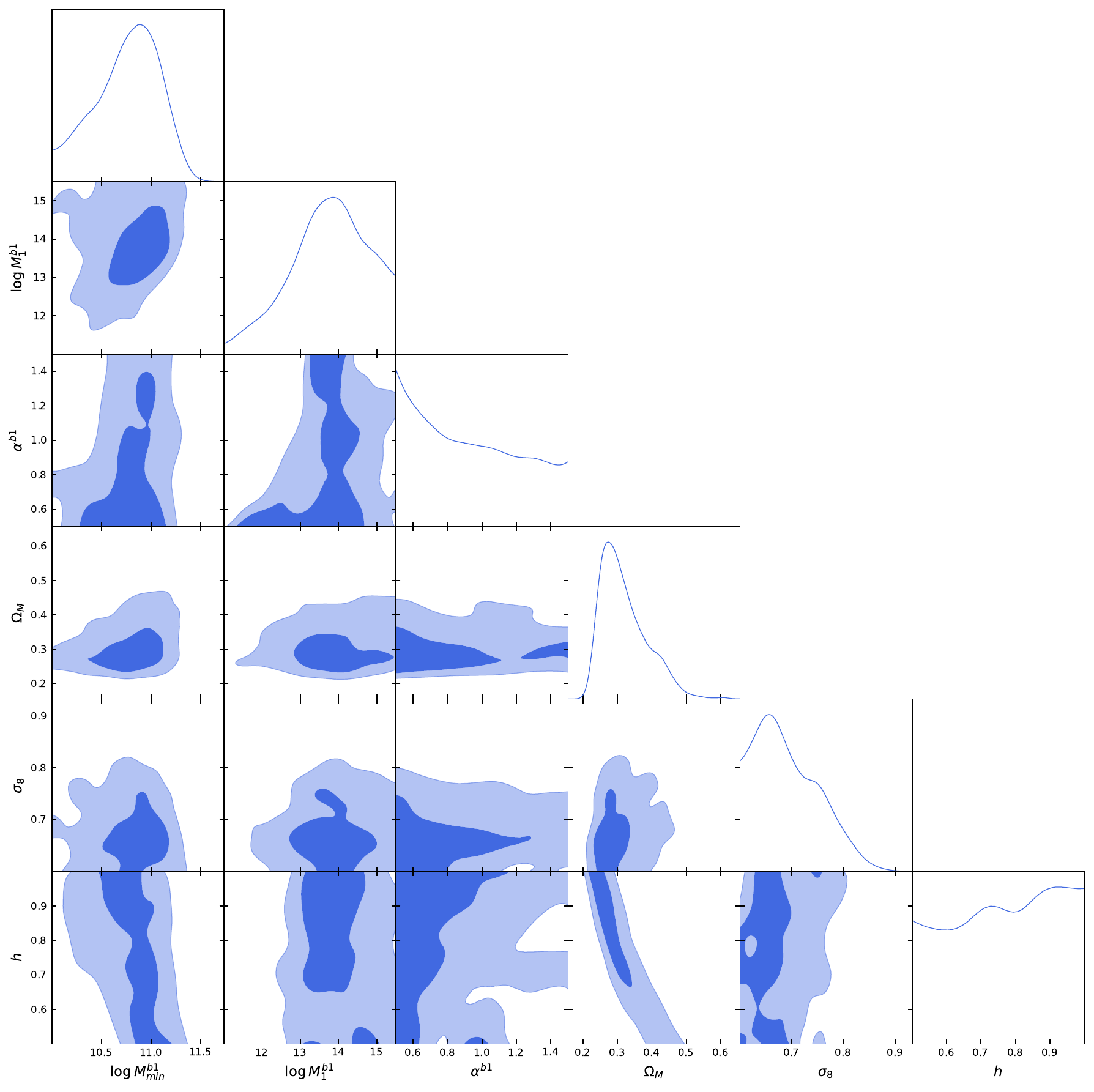}
\includegraphics[width=0.5\textwidth]{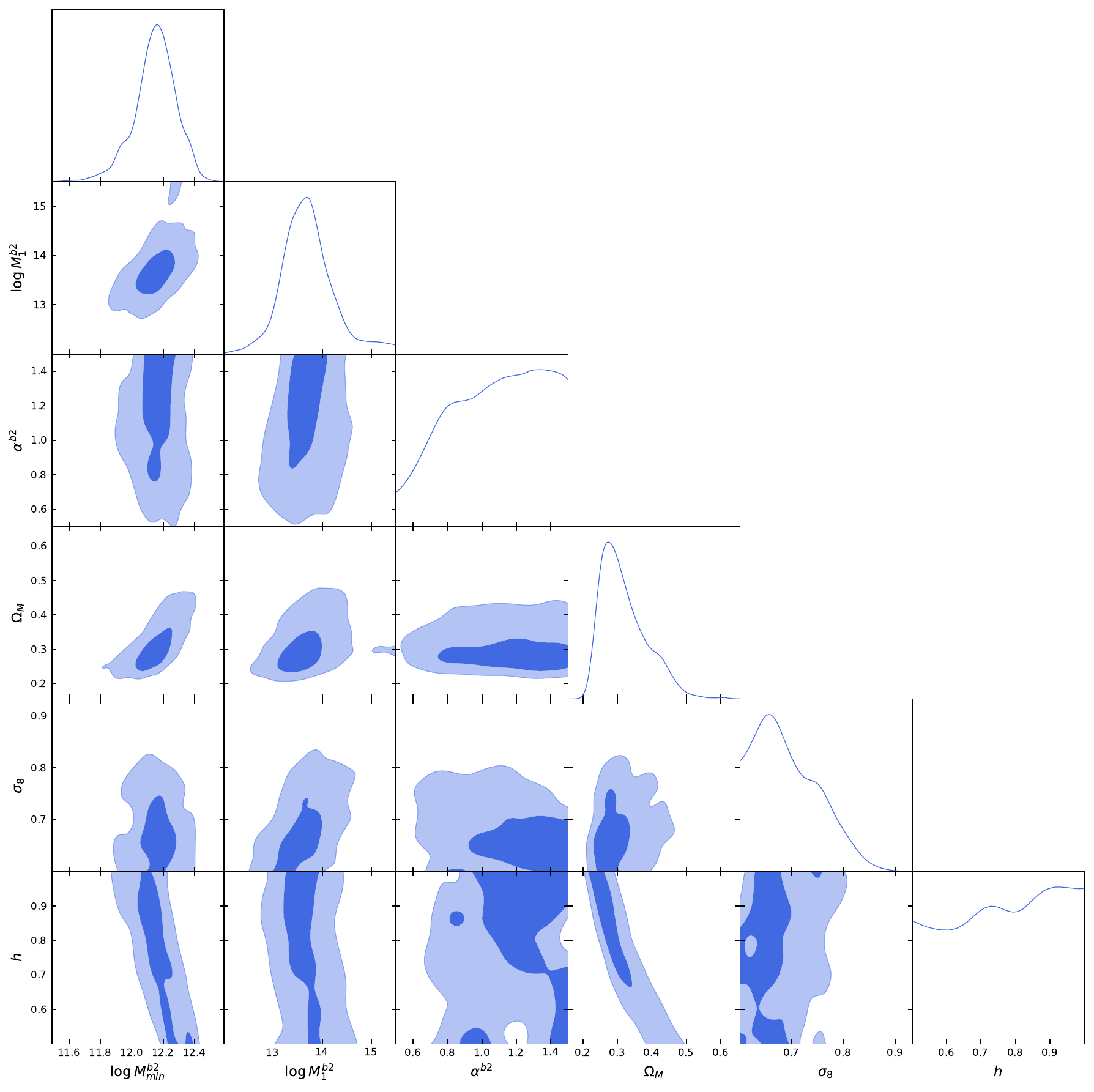}\\
\includegraphics[width=0.5\textwidth]{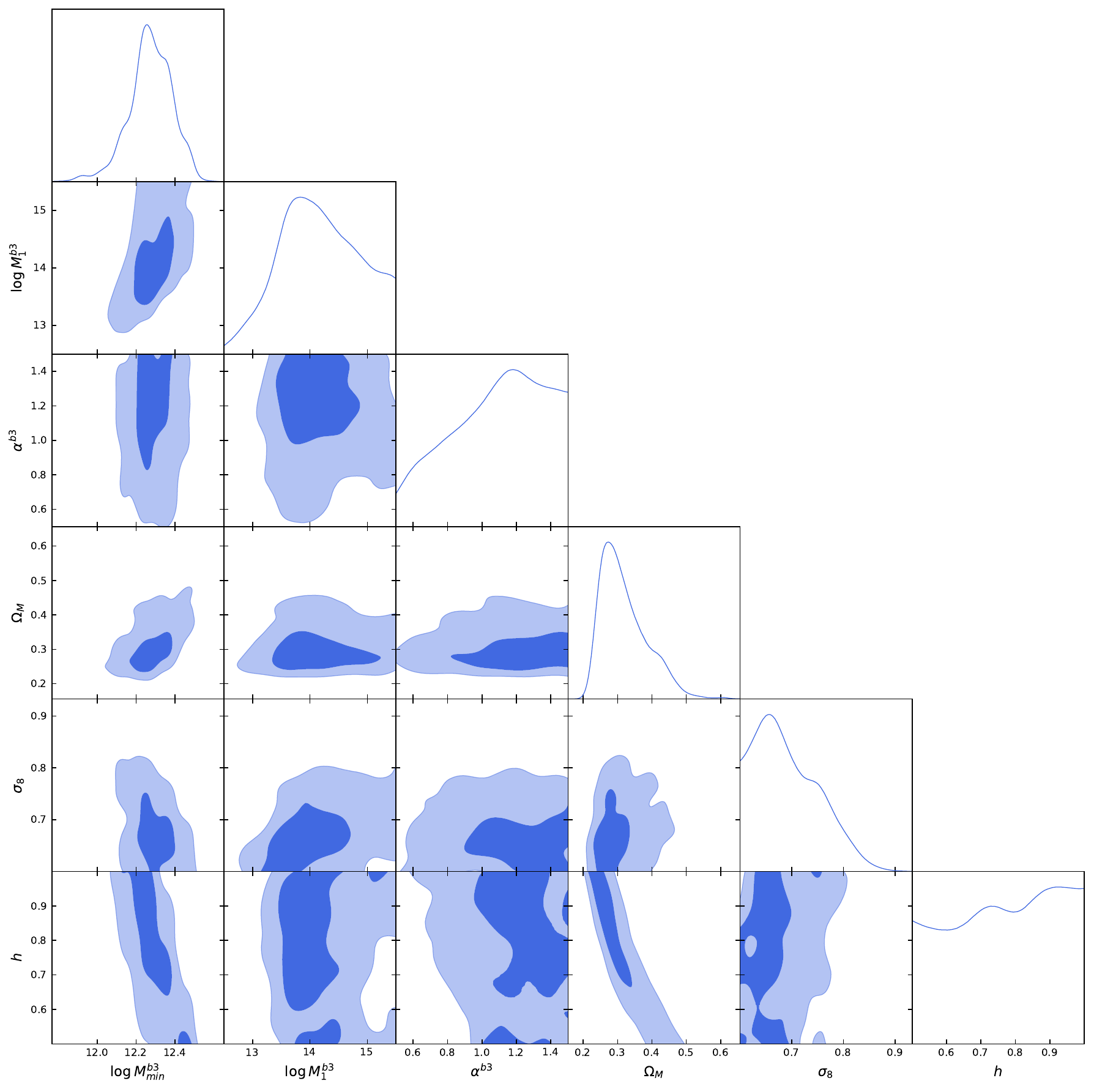}
\includegraphics[width=0.5\textwidth]{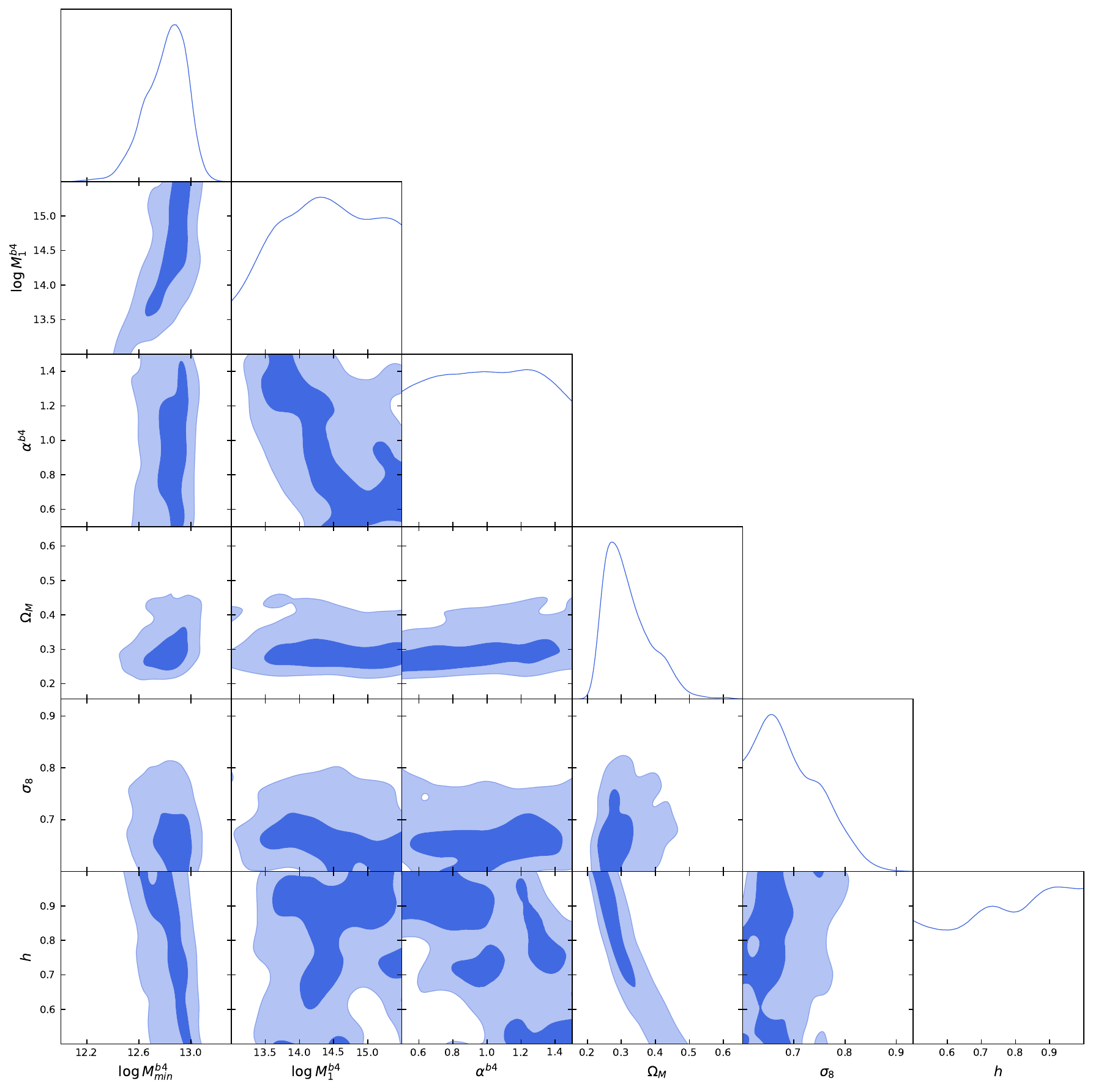}\\
 \caption{Marginalised posterior distribution and probability contours (set to 0.393 and 0.865) for the HOD and cosmological parameters obtained in the "w/o G15" case.}
 \label{fig:corner_tomo_zmed_noG15}
\end{figure*}

This section presents the marginalised posterior distributions and probability contours (set to 0.393 and 0.865) of all the cases addressed in this work and compared to the "base case" presented in section \ref{app:comp_full_MzA}. In particular, Fig. \ref{fig:corner_tomo_zmed_noG15} show the case corresponding to Sect.~\ref{subsec:results_noG15} in which the joint analysis is performed on the four bins bin 1, bin 2, bin 3, and bin 4 without taking into account the G15 field. 

\begin{figure*}[ht]
\includegraphics[width=0.5\textwidth]{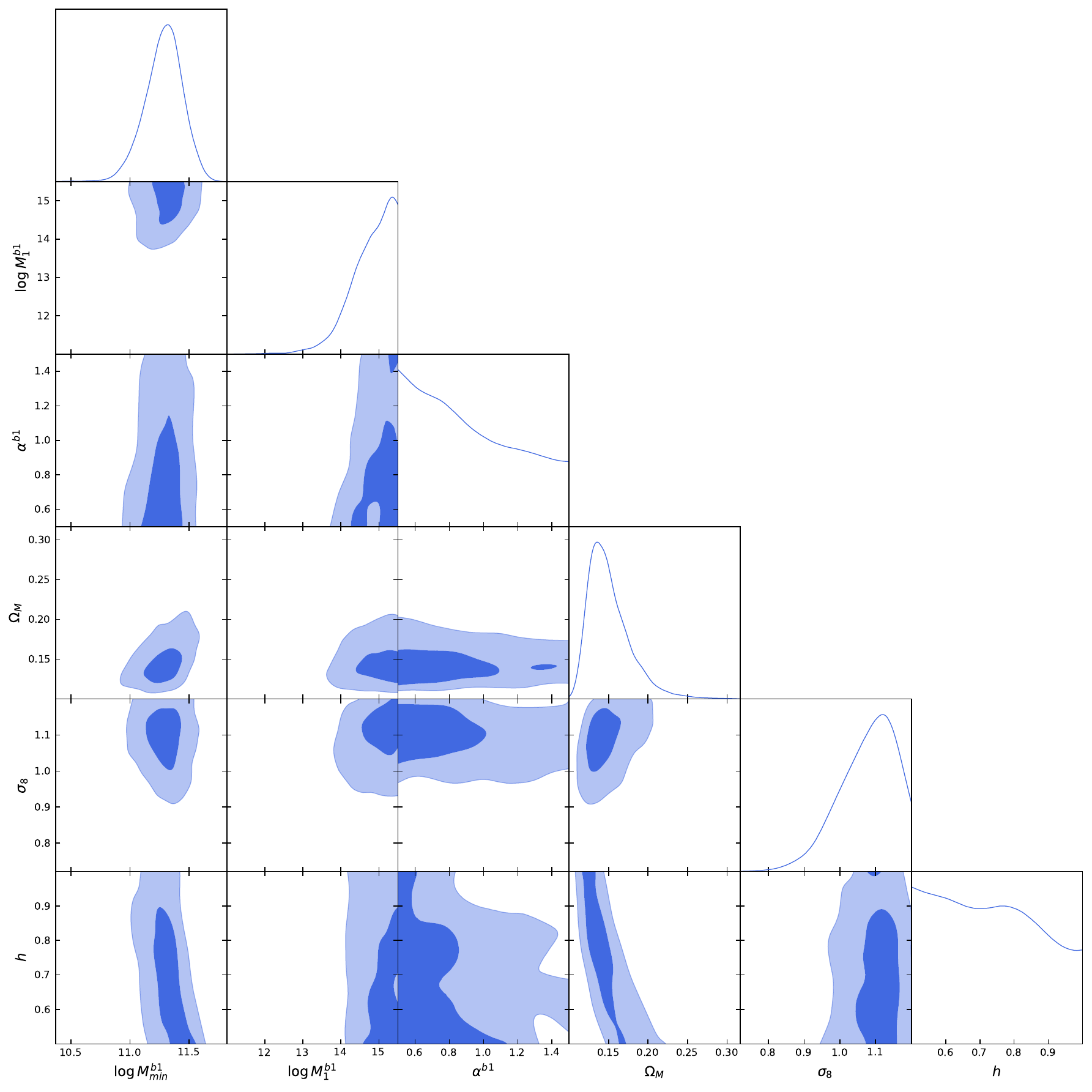}
\includegraphics[width=0.5\textwidth]{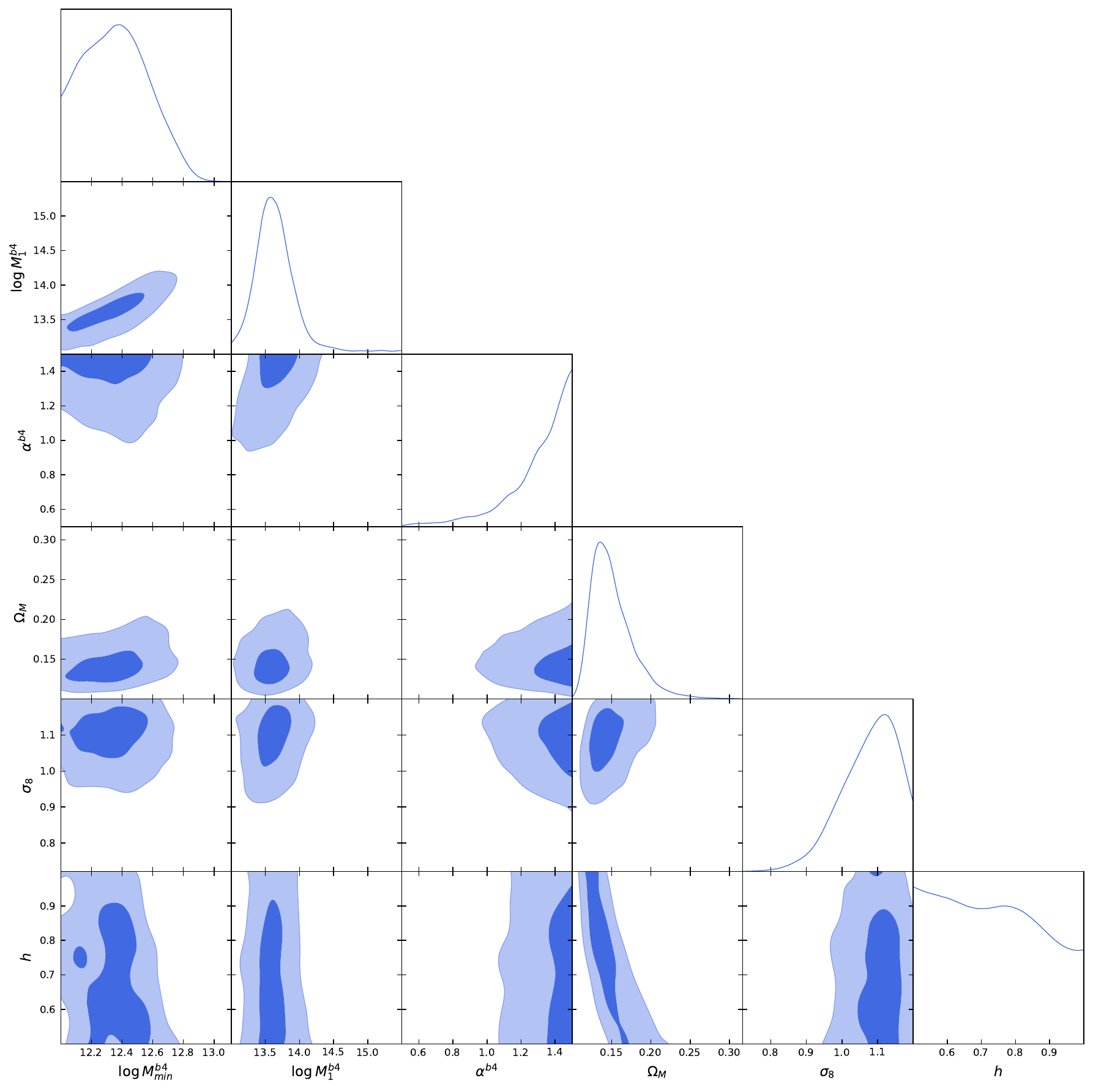}
 \caption{Marginalised posterior distribution and probability contours (set to 0.393 and 0.865) for the HOD and cosmological parameters obtained in the "bins 1+4" case.
 }
 \label{fig:corner_tomo_bin14}
\end{figure*}

\begin{figure*}[ht]
\includegraphics[width=0.5\textwidth]{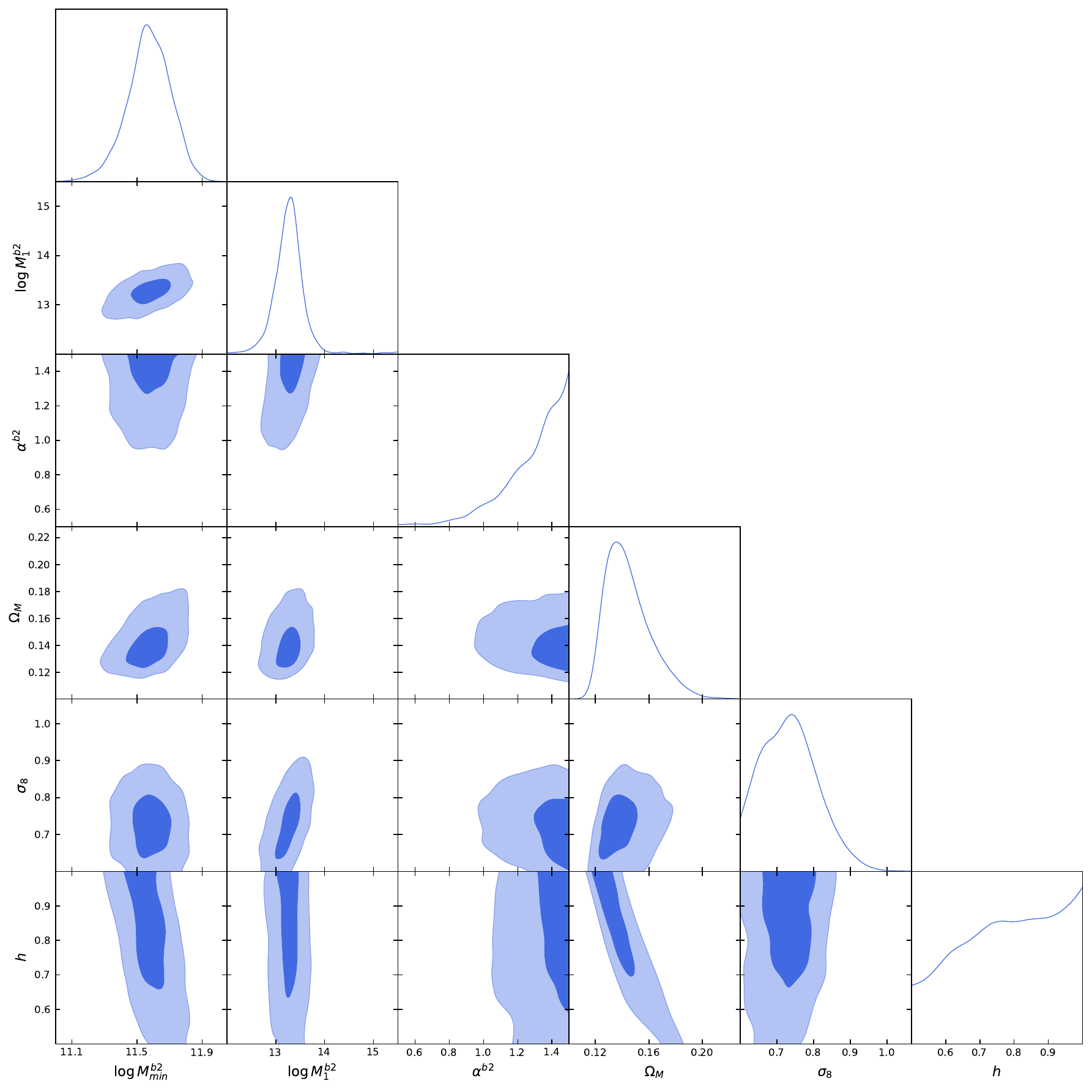}
\includegraphics[width=0.5\textwidth]{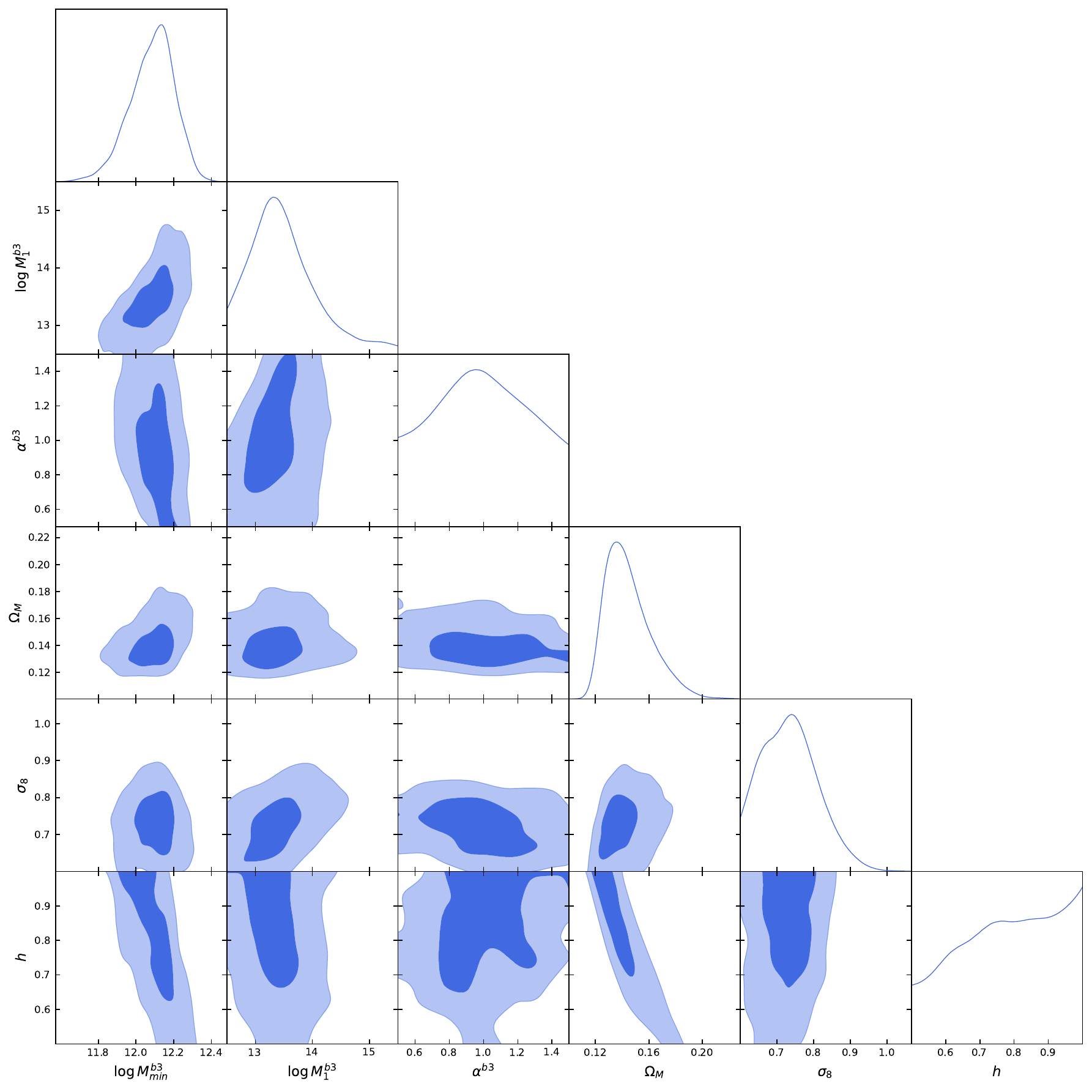}
 \caption{Marginalised posterior distribution and probability contours (set to 0.393 and 0.865) for the HOD and cosmological parameters obtained in the "bins 2+3" case.
 }
 \label{fig:corner_tomo_bin23}
\end{figure*}

The "two bins" case described in Sect.~\ref{sec:2bins}, where only two bins are jointly analysed, are shown in Figs.~\ref{fig:corner_tomo_bin14}  and \ref{fig:corner_tomo_bin23}. In particular, Fig.~\ref{fig:corner_tomo_bin14} corresponds to the "bins 1+4" case, where just bin 1 and bin 4 are jointly analysed and Fig.~\ref{fig:corner_tomo_bin23} shows the results obtained in the "bins 2+3" case, that is, considering only bin 2 and bin 3.

\begin{figure*}[ht]
\includegraphics[width=0.43\textwidth]{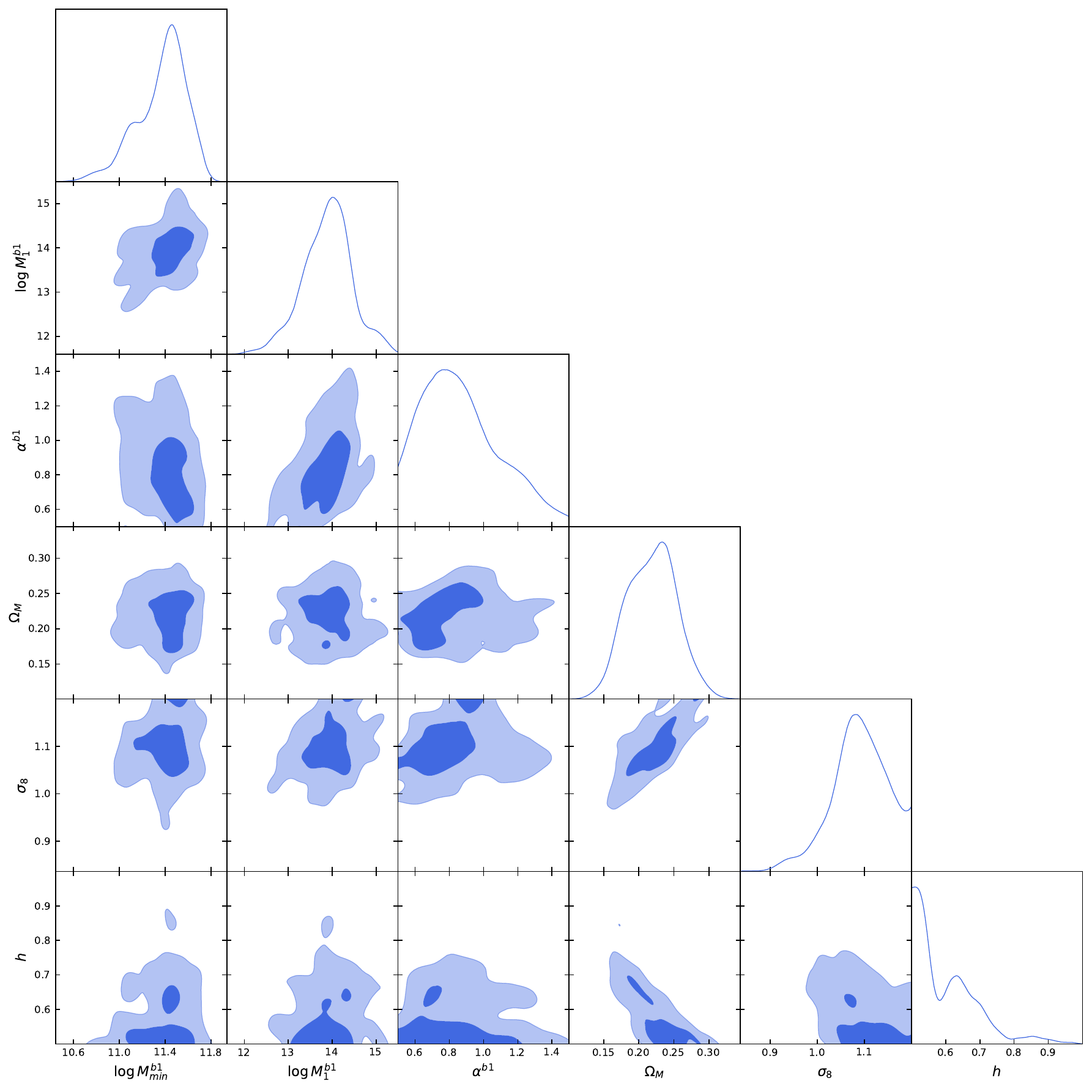}
\includegraphics[width=0.43\textwidth]{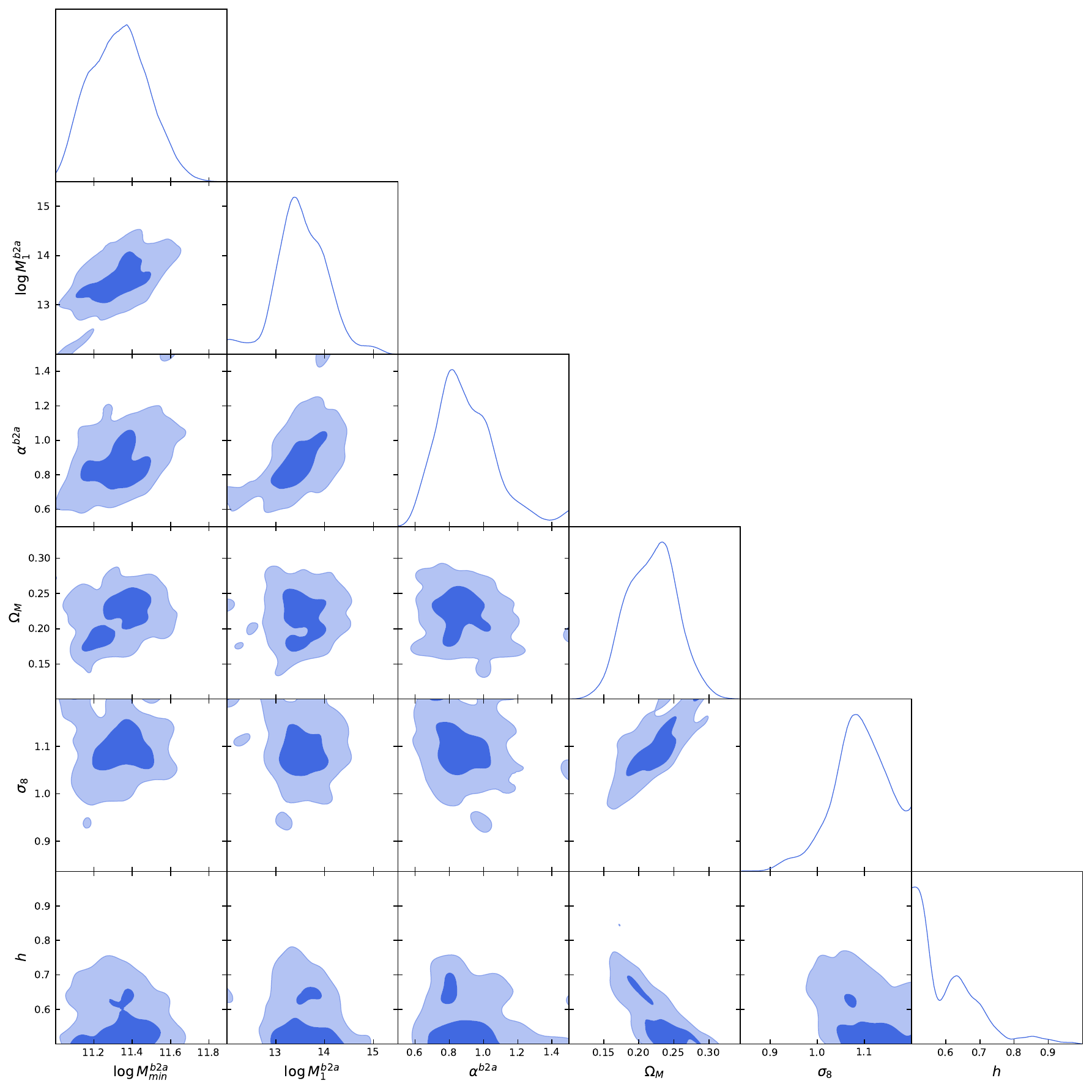}\\
\includegraphics[width=0.43\textwidth]{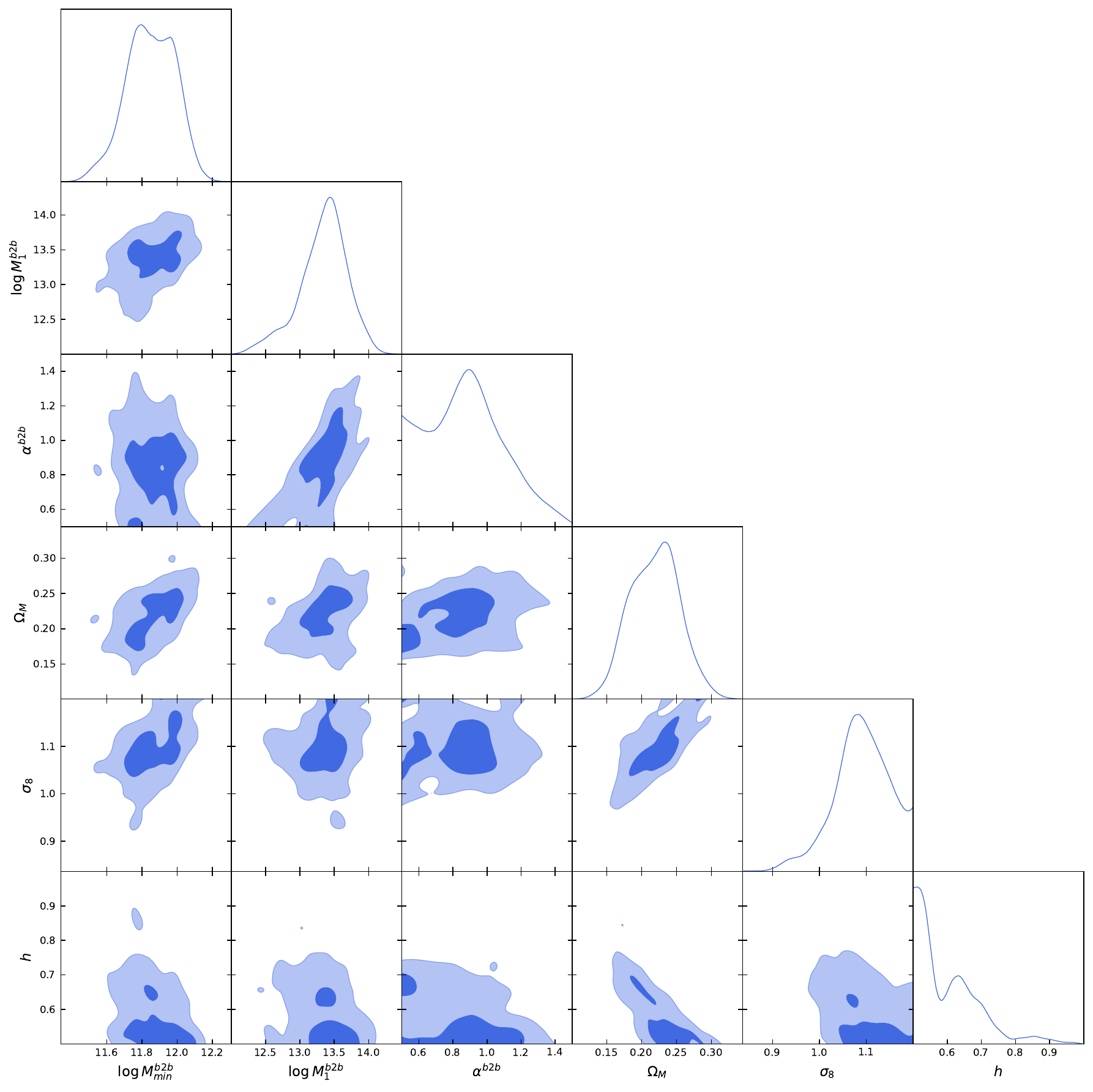}
\includegraphics[width=0.43\textwidth]{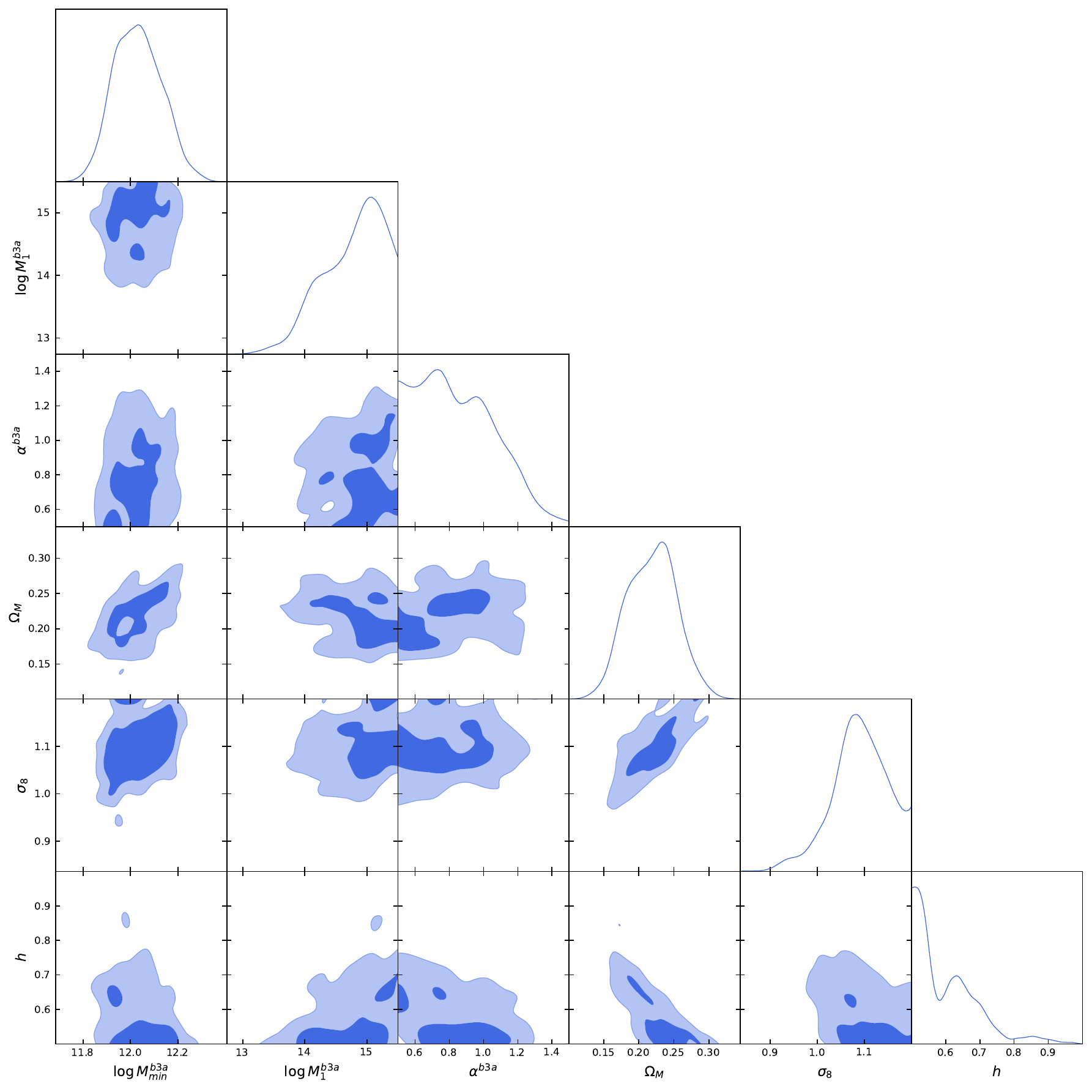}\\
\includegraphics[width=0.43\textwidth]{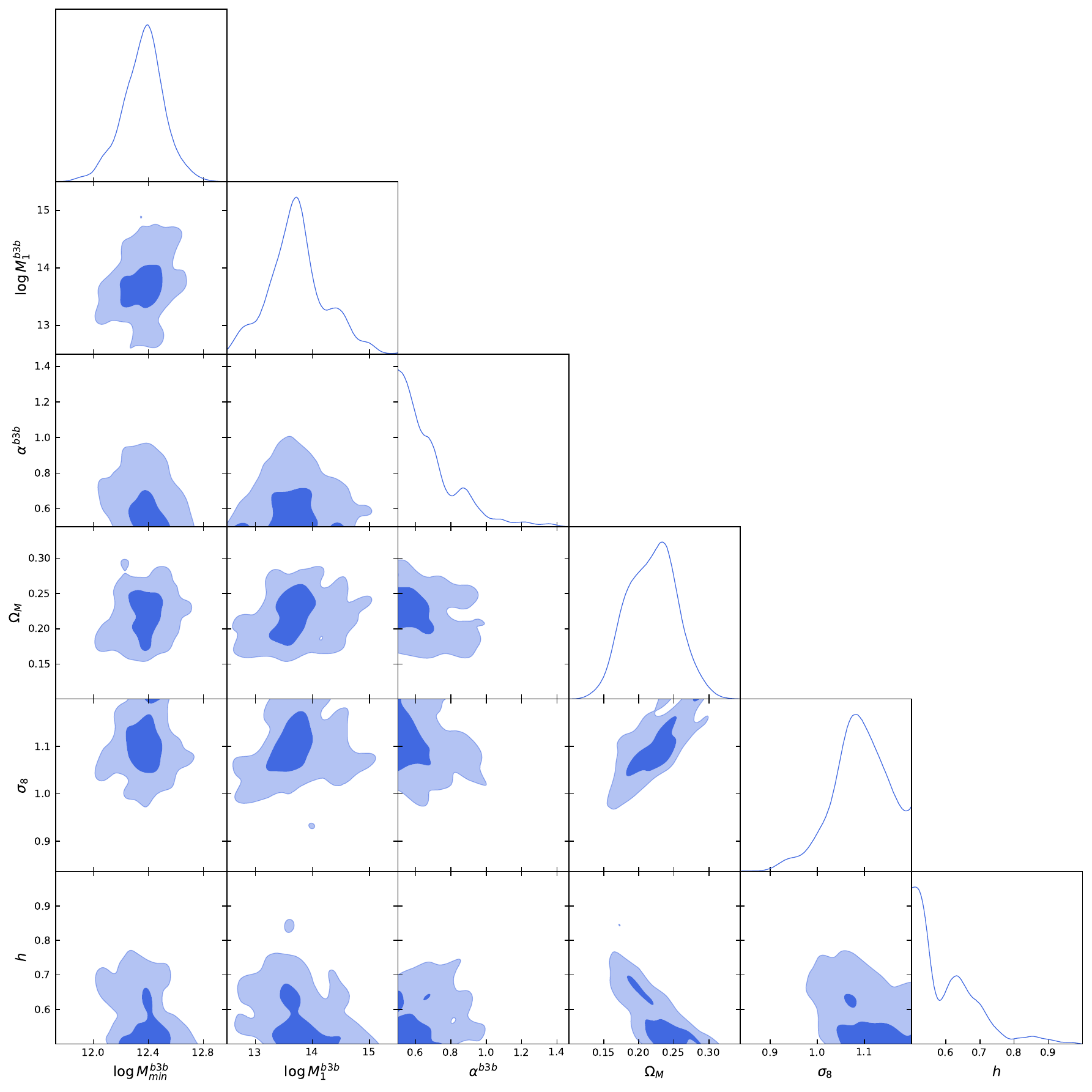}
\includegraphics[width=0.43\textwidth]{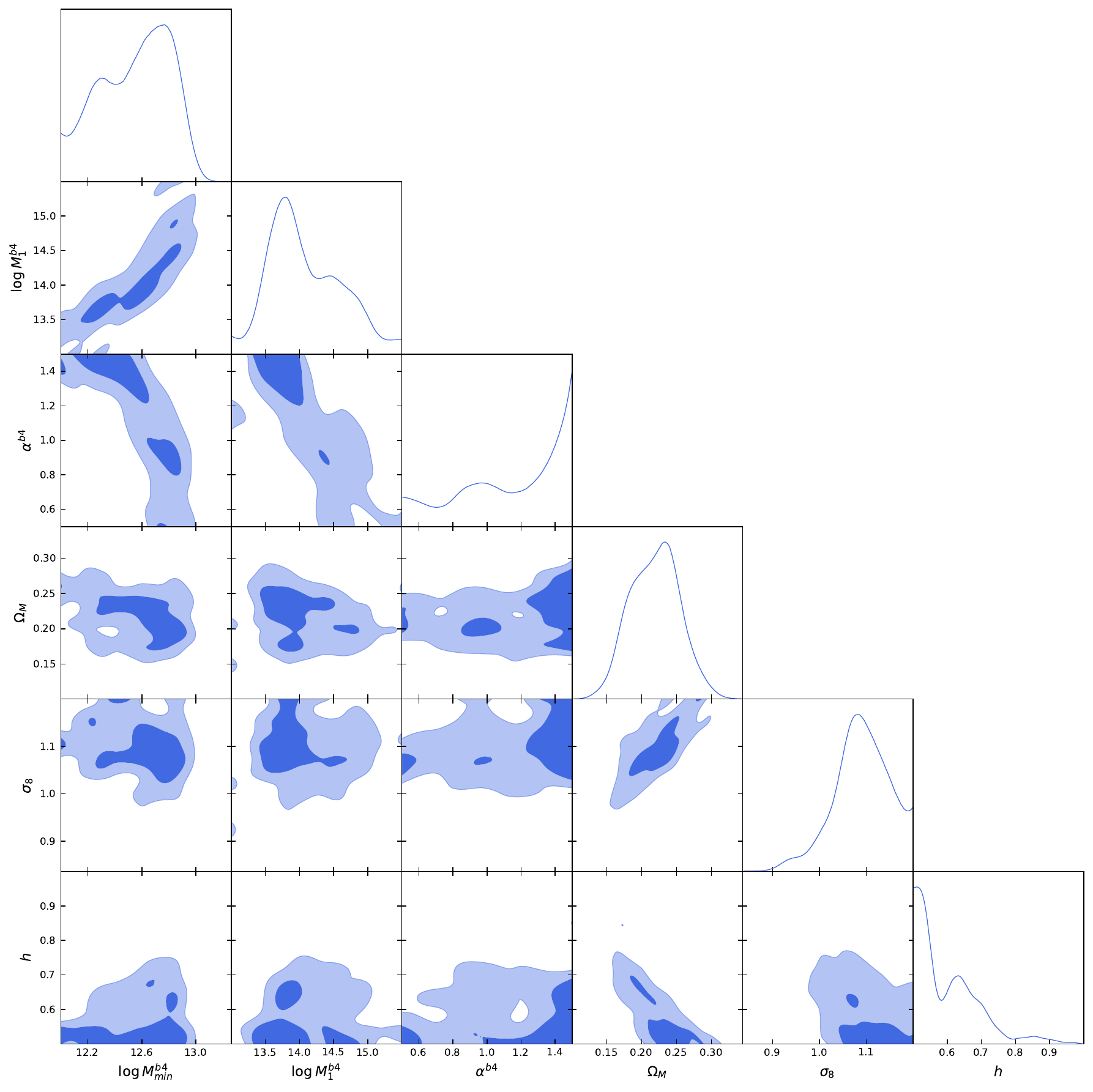}
 \caption{Marginalised posterior distribution and probability contours (set to 0.393 and 0.865) for the HOD and cosmological parameters obtained in the " six bins" (i.e. $0.1<z<0.8$) case.
 }
 \label{fig:corner_tomo_6bins}
\end{figure*}

\begin{figure*}[ht]
\includegraphics[width=0.43\textwidth]{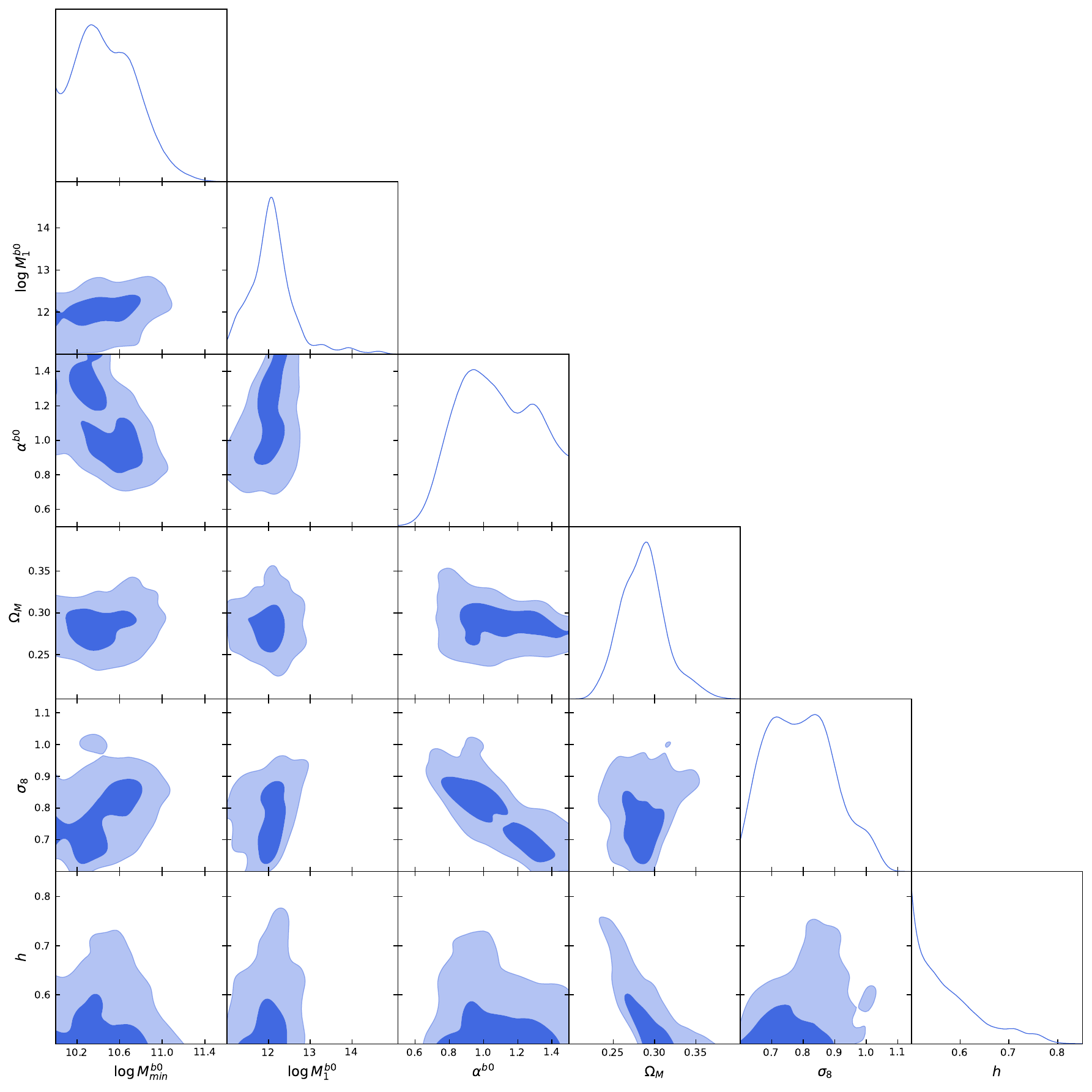}
\includegraphics[width=0.43\textwidth]{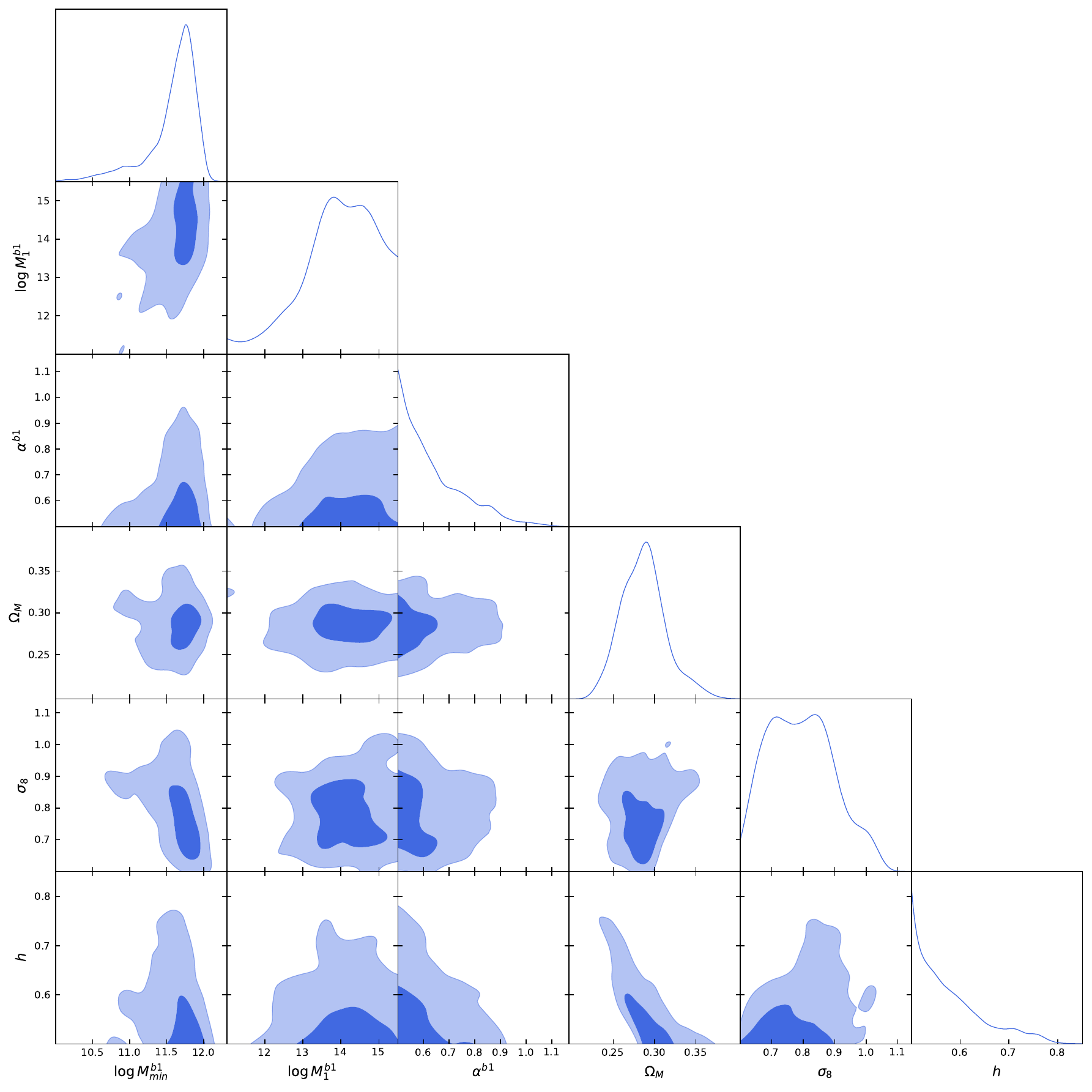}\\
\includegraphics[width=0.43\textwidth]{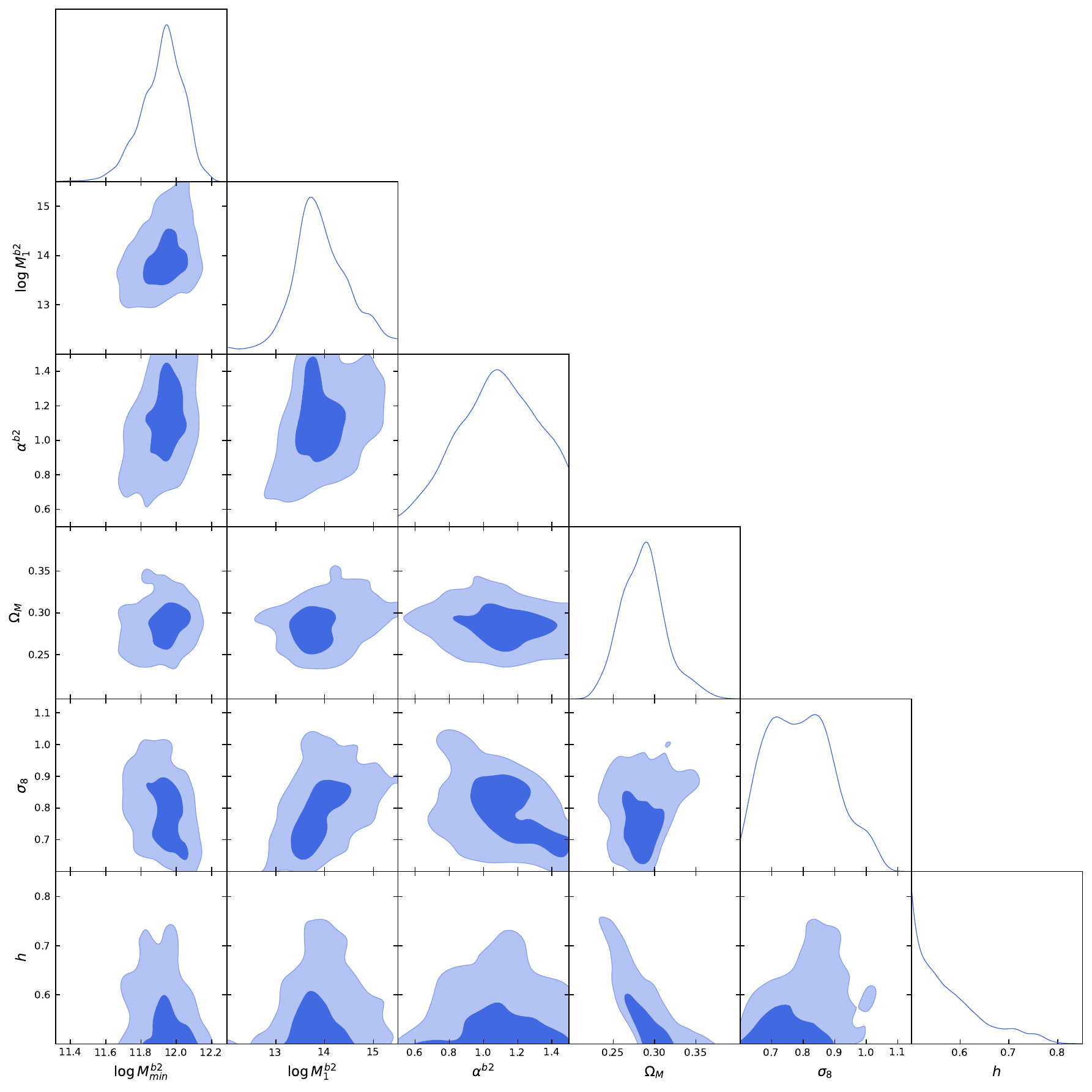}
\includegraphics[width=0.43\textwidth]{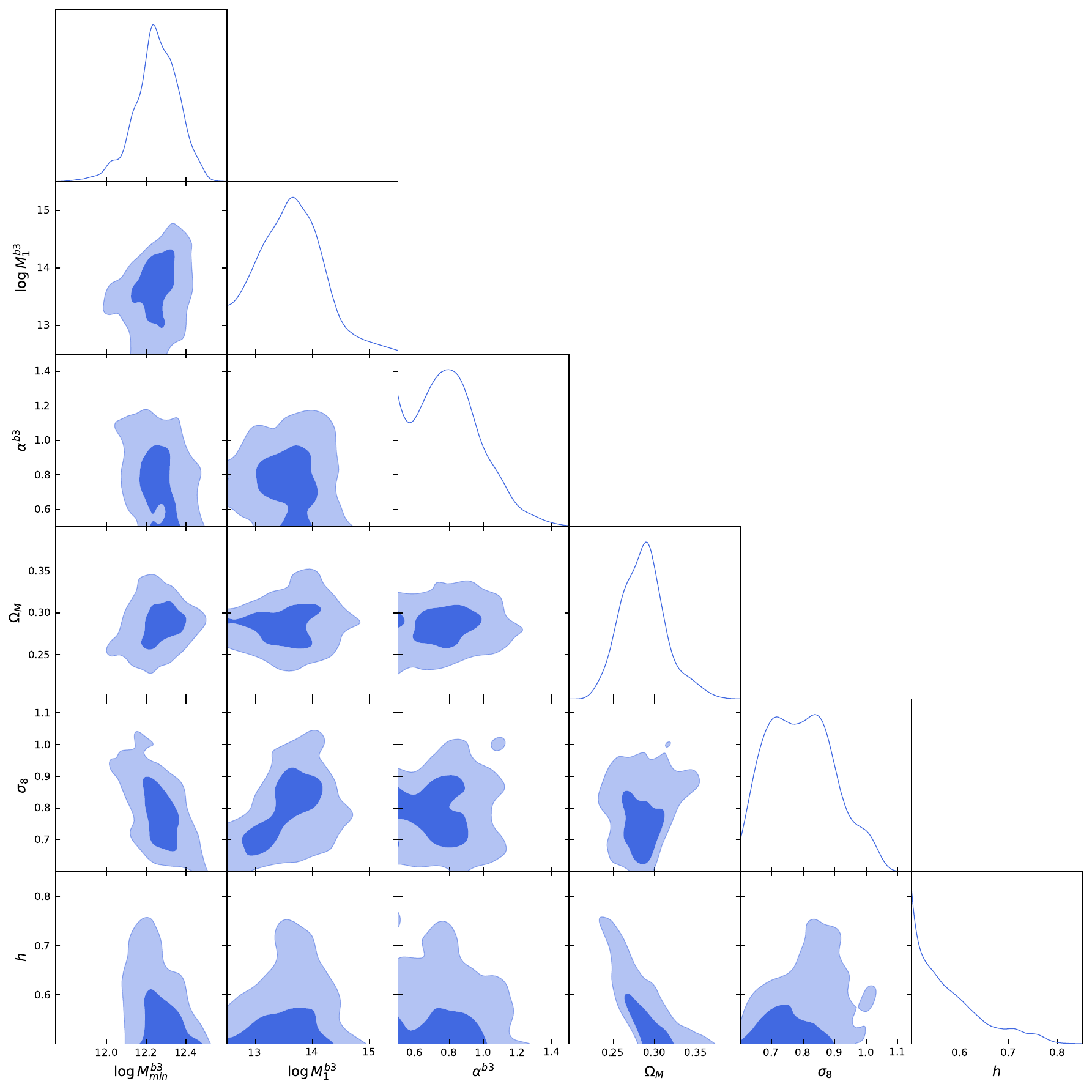}\\
\includegraphics[width=0.43\textwidth]{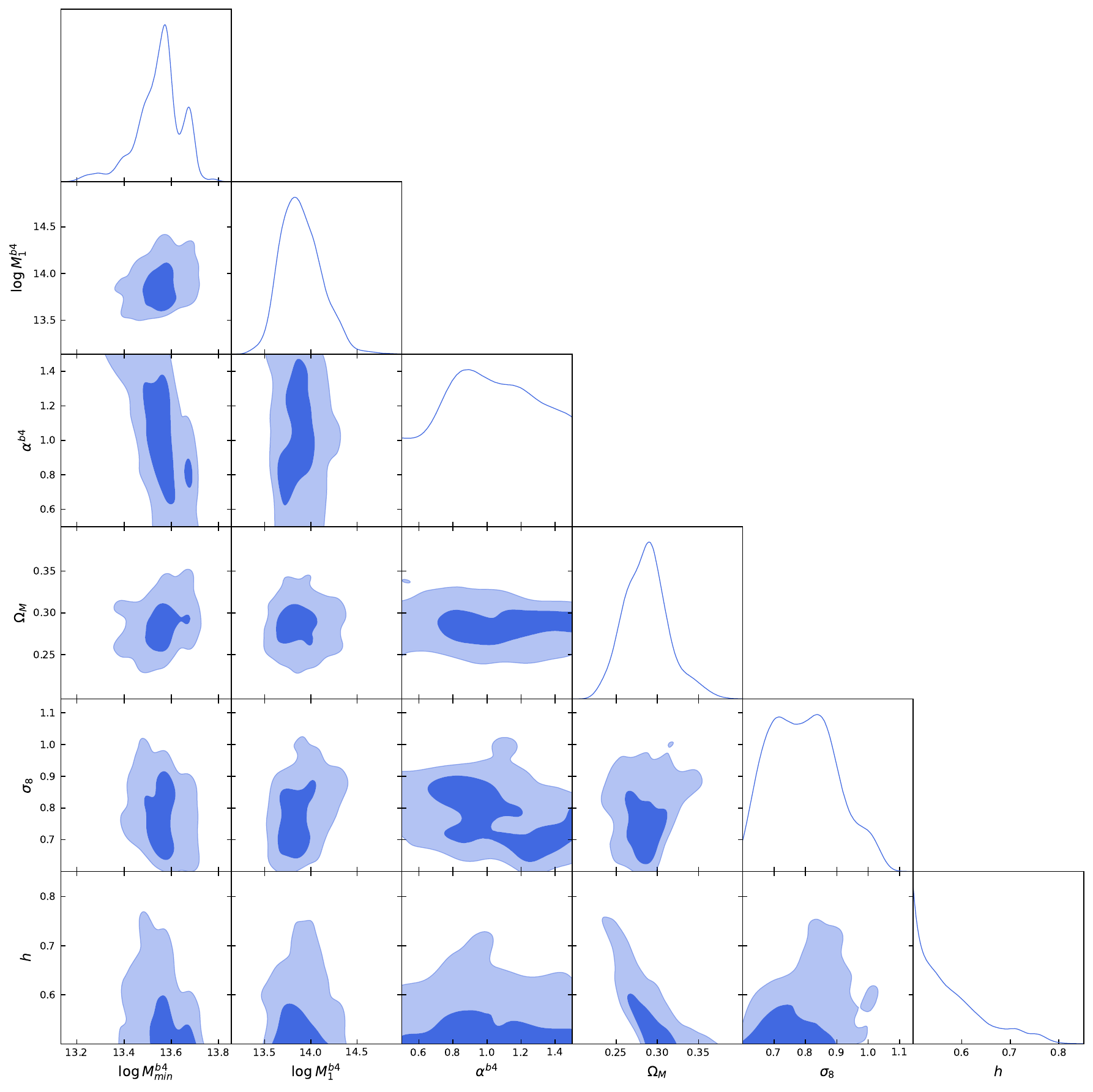}
\includegraphics[width=0.43\textwidth]{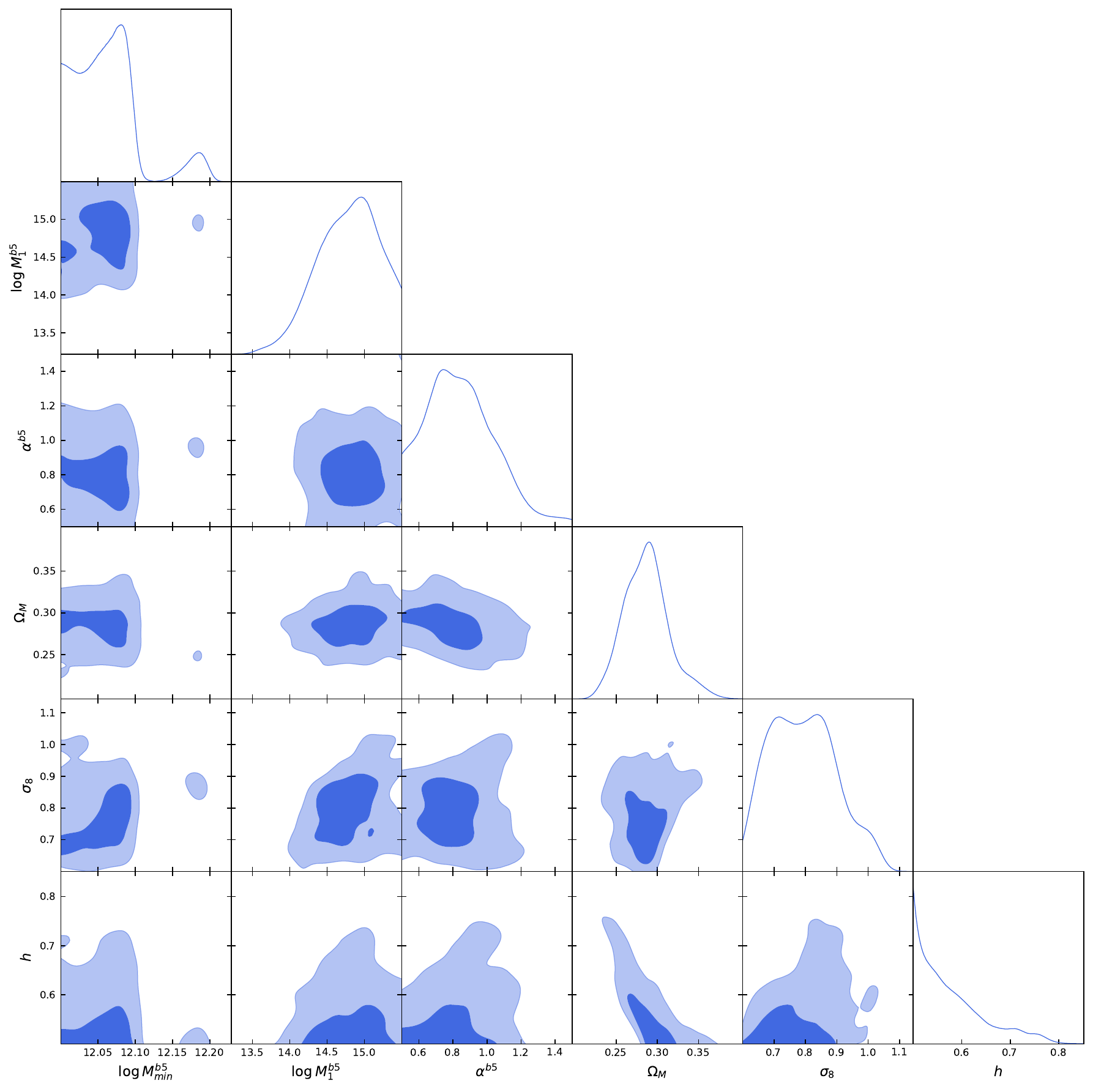}
 \caption{Marginalised posterior distribution and probability contours (set to 0.393 and 0.865) for the HOD and cosmological parameters obtained in the " six bins-WR" (i.e. $0.01<z<0.9$) case.
 }
 \label{fig:corner_tomo_6bins_09}
\end{figure*}

The" six bins" cases described in Sect. \ref{sec:6bins} are depicted in Figs.~\ref{fig:corner_tomo_6bins} and \ref{fig:corner_tomo_6bins_09}, while Fig. \ref{fig:corner_tomo_6bins} describes the findings of the "six bins" case, which are attained using bin 1, bin 2a and bin 2b (by splitting bin 2), bin 3a and bin 3b (by splitting bin 3), and bin 4. Figure~\ref{fig:corner_tomo_6bins_09} shows the "six bins-WR" case, obtained adding two extra bins to the "base case," that is:\ bin 0 (to the lower end of the original case, $0.01>z>0.1$), bin 1, bin 2, bin 3, bin 4 (but only up to z=0.7 to increase the statistics of bin 5), and bin 5 ($0.7>z>0.9$).

\end{appendix}

\end{document}